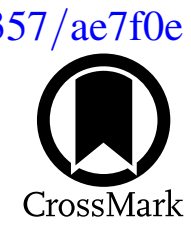

# Correlation between the Two-armed $V_R$ Spiral in the $Z$–$V_Z$ Plane and Moving Groups

Yan Xu[1], Chao Liu[2], Cheng Dong Li[3,4], Heidi Newberg[5], Hao Tian[2], Hua Jian Wang[1], Xiao Dian Chen[1], and Li Cai Deng[1]

[1] CAS Key Laboratory of Optical Astronomy, National Astronomical Observatories, Chinese Academy of Sciences, Beijing 100101, People's Republic of China; xuyan@bao.ac.cn
[2] Key Laboratory of Space Astronomy and Technology, National Astronomical Observatories, Chinese Academy of Sciences, Beijing 100101, People's Republic of China
[3] School of Astronomy and Space Science, Nanjing University, Nanjing 210093, People's Republic of China
[4] Key Laboratory of Modern Astronomy and Astrophysics, Nanjing University, Ministry of Education, Nanjing 210093, People's Republic of China
[5] Department of Physics, Applied Physics and Astronomy, Rensselaer Polytechnic Institute, Troy, NY 12180, USA

## Abstract

We use a crossmatched sample of 3.7 million stars from Gaia Data Release 3 and LAMOST Data Release 7 to investigate the velocity substructures in the Milky Way disk. The median radial velocity $V_R$ as a function of guiding-center radius $R_g$ exhibits alternating positive and negative stripes, which are strongly correlated with known moving groups. By examining the $V_R$ distribution in the $Z$–$V_Z$ phase space, we find that the D1, P2, D2, and P3 $V_R$ stripes display clear two-armed spirals. Among the moving groups embedded in these $V_R$ stripes, the Coma Berenices moving group in the P3 stripe exhibits the most pronounced two-armed spiral and serves as the primary contributor to the left arm of the overall P3 spiral. The angular momentum, eccentricities, orbital frequencies, and frequency ratios of its stars are consistent with either the corotation resonance of the spiral arms or the $m = 4$ inner Lindblad resonance of the bar. Test-particle simulations confirm that a bar with a varying pattern speed, together with static or transient spiral arms, can produce such two-armed spirals.

## 1. Introduction

The formation and evolution of the Milky Way have been influenced by numerous internal and external perturbations. For example, the formation and growth of spiral arms and the Galactic bar, as well as interactions between dwarf galaxies and the disk, all contribute to these perturbations. These asymmetric perturbations have created a multitude of substructures in the velocity and phase-space distributions of stars.

Large-scale surveys, such as the Gaia Data Release 3 (DR3; Gaia Collaboration et al. 2023), provide high-precision astrometric and radial velocity data, enabling researchers to study the distribution of stars in various dimensions. Recently, an increasing number of studies have discovered correlations between substructures in different dimensional spaces. L. M. Widrow et al. (2012) used the main-sequence star data from the Sloan Digital Sky Survey (SDSS) and found that the stellar number density in the vertical direction in the solar neighborhood exhibits a clear north–south asymmetry and features resembling wavelike perturbations. T. Antoja et al. (2018) used Gaia Data Release 2 (DR2) data to discover a single-armed spiral structure, while J. A. S. Hunt et al. (2022) used Gaia DR3 data to reveal a two-armed spiral structure in the $Z$–$V_Z$ phase space. Meanwhile, C. F. P. Laporte et al. (2019) analyzed the phase-space distribution of Gaia DR2 data and found that the spiral structures in the density distribution of stars near the Sun in the $Z$–$V_Z$ plane are consistent with the vertical density fluctuations observed by L. M. Widrow et al. (2012). Y. Xu et al. (2020) investigated 0.43 million K giants from the LAMOST Data Release 5 catalog. They found that by projecting the spiral features at different radii onto the $R$–$Z$ plane, these spiral features could well explain the observed asymmetric velocity substructures in the outer disk. The study clarified that the phase-space spiral is closely related to the overall velocity distribution of the Galactic outer disk. W. H. Trick et al. (2019) studied approximately 3.9 million stars in Gaia DR2 and found overdensities and ridges in the ($J_R$, $L_Z$) space. These features can be associated with the known moving groups in the solar neighborhood. The research by C. Cao et al. (2024) aimed to utilize the latest Gaia DR3 data to analyze the radial wave patterns discovered in the $L_Z$–$V_R$ space of the Galactic disk. They also aimed to distinguish the effects of different perturbations on these wave patterns through simulations. In the course of their study, they found that the ripple structures in the $L_Z$–$V_R$ space are projections of the ridges. S. Khoperskov & O. Gerhard (2022) focused on analyzing overdense regions in angular momentum space, traced their extensions in ($R$, $V_\phi$) space, and compared these with moving groups in the solar neighborhood. They utilized simulations to elucidate the impact of spiral arm and bar resonances on disk stars, examining these effects in both ($R$, $V_\phi$) and ($V_R$, $V_\phi$) dimensions. By utilizing multidimensional data, one can identify correlations between substructures in different spaces. For instance, velocity and density substructures in the $R-Z$ or $X-Y$ spaces, spiral structures in the $Z$–$V_Z$ phase space, ridge-like structures in the $R$–$V_\phi$ space, wave patterns in the $L_Z$–$V_R$ space, overdensities and ridges in the $J_R$–$L_Z$ space, and moving groups in the $V_R$–$V_\phi$ space are all interconnected. By analyzing the same structure across multiple spaces, one can obtain more information to characterize the substructures and better identify the sources of perturbations.

In this work, we utilize data from the crossmatch between Gaia DR3 and Large Sky Area Multi-Object Fiber Spectroscopic Telescope (LAMOST; X.-Q. Cui et al. 2012;

L.-C. Deng et al. 2012; G. Zhao et al. 2012) to investigate the velocity substructures of stars in different radial ranges ($R_g$) in the $Z$–$V_Z$ phase space, focusing on the $V_R$ component. We then compare these substructures with moving groups in the $R_g$–$V_R$ space. Through this comparison, we aim to study the perturbation origin of the two-armed spiral structure observed in the $V_R$ component of the $Z$–$V_Z$ phase space.

The sample adopted in this work is introduced in Section 2.1, and the $V_R$ distributions in the $R$–$R_g$ and $Z$–$V_Z$ planes are studied in Sections 2.3 and 2.4, respectively. Selection effects and their impact on the visibility of the phase spiral are discussed in Section 2.5. The correlation between the $V_R$ stripes along $R_g$ and moving groups is analyzed in Section 2.6, and the $V_R$ distribution of each moving group in the $Z$–$V_Z$ plane is examined in Section 2.7. The properties and possible resonant origins of the Coma Berenices moving group stars are investigated in Section 3, including action and frequency analysis, backward orbit integration, and test-particle simulations. A comparison with previous studies is given in Section 3.5, and the main conclusions are summarized in Section 4.

## 2. The Kinematic Distribution of the Observational Data

### 2.1. Sample

We crossmatched the data from Gaia DR3 and LAMOST Data Release 7 (DR7), removing duplicate observations and stars with radial velocity errors greater than 10 km s$^{-1}$. The systematic bias of 4.4 km s$^{-1}$ in LAMOST radial velocities (H. Wang et al. 2018) was corrected following the same procedure as in the previous work of (Y. Xu et al. 2020). We also required a signal-to-noise ratio greater than 15 in the $g$ band (as defined by the SDSS photometric system) and Gaia DR3 proper motion errors to be less than 0.2 mas yr$^{-1}$. After applying these criteria, we obtained a final sample of 3,711,823 stars. We calculated the cylindrical velocities ($V_R$, $V_\phi$, $V_Z$) using the same method as Y. Xu et al. (2020). The actions ($J_R$, $J_\phi$, $J_Z$), angles ($\theta_R$, $\theta_\phi$, $\theta_Z$), and frequencies ($\Omega_R$, $\Omega_\phi$, $\Omega_Z$; J. Binney & S. Tremaine 2008) were computed with the method of Y. Xu et al. (2023) implemented in the software package `galpy` (J. Bovy 2015), which employs the `MWPotential2014` and the Stackel approximation (J. Binney 2012). In `galpy`, the actions and frequencies were normalized such that a star on a circular orbit at the Galactic plane with $(R,\ V_\phi) = (8.34\,\mathrm{kpc}, 238\,\mathrm{km\,s^{-1}})$ (R. Schönrich et al. 2010; M. J. Reid et al. 2014) has $J_\phi = 1$ kpc km s$^{-1}$, $J_R = 0$ kpc km s$^{-1}$, $J_Z = 0$ kpc km s$^{-1}$, and a vertical frequency $\Omega_Z = 1$ km s$^{-1}$ kpc$^{-1}$.

### 2.2. Scientific Objectives and Motivation

From previous studies, we know that organizing stars in the Galactic disk in different ways, such as by different ranges of $R$ or different values of $L_Z$, can reveal different kinds of $V_R$ phase spiral in $Z-V_Z$ within the subsets of data (C. F. P. Laporte et al. 2019; Y. Xu et al. 2020; J. A. S. Hunt et al. 2022; T. Antoja et al. 2023). This suggests that sorting stars by $R$ or by $L_Z$ reveals structures that may be created by different types of perturbation. The two-armed $V_R$ phase spiral is more pronounced when sorting stars by $L_Z$. This may suggest that the response of stars to this type of perturbation is more closely related to their angular momentum rather than to the relative position of the perturbation sources. This evokes the characteristics of resonances.

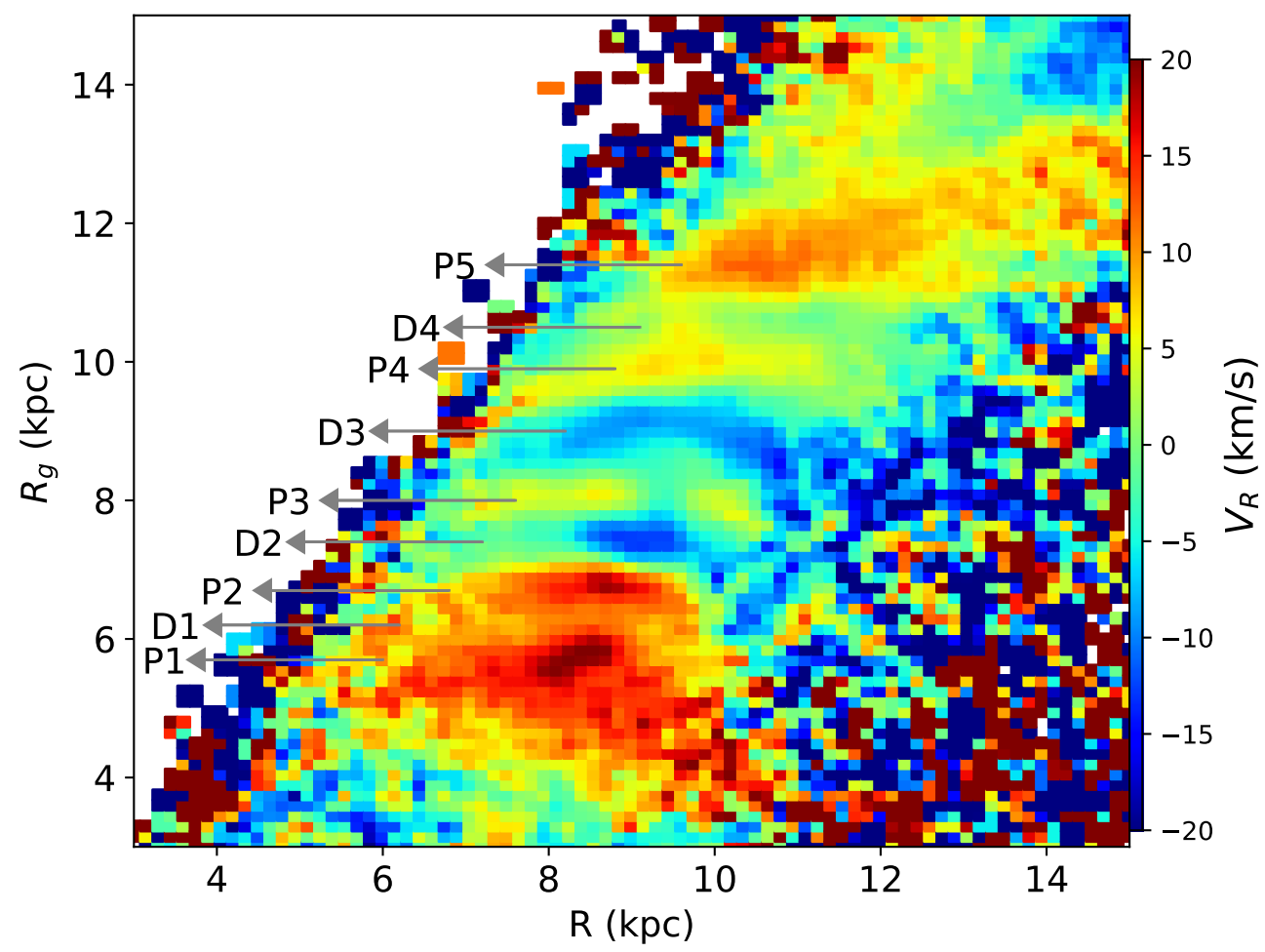

**Figure 1.** $V_R$ distribution of LAMOST DR7 data in the $R$ and $R_g$ plane. Positive and negative-$V_R$ stripes, roughly parallel to $R$ and perpendicular to $R_g$, are clearly visible. These $V_R$ stripes are labeled as peaks ("P") or dips ("D") of the $V_R$ distribution, with sequential numbering. The names are indicated on the plot.

In previous studies, fluctuations in the distribution of $V_R$ along $L_Z$ have been observed (C. Cao et al. 2024). The $V_R$ wavelike pattern along $L_Z$ may be associated with spiral arms and bar resonance and may correspond to the dominant moving group within that range of $L_Z$ (S. Khoperskov & O. Gerhard 2022). Different moving groups, in turn, are likely dominated by different perturbations. Therefore, we aim to segment the data based on the fluctuations of $V_R$ along $L_Z$, examining the $V_R$ distribution of these stars in phase space. Furthermore, we also aim to investigate the relationship between moving groups and the two-armed spiral structures in $Z-V_Z$ phase space, thereby seeking clues to the origin of these phase spirals.

### 2.3. The $V_R$ Stripes along $R_g$

We analyzed the kinematic pattern of our sample in the $R$ and $R_g$ plane. Figure 1 shows the radial velocity ($V_R$) distribution in this plane. The velocity stripes are more clearly defined when traced against $R_g$ because using $R_g$ avoids blurring due to orbital phase variations. As shown in Figure 1, the alternating positive and negative-$V_R$ stripes are roughly parallel to the $R$ direction and perpendicular to the $R_g$ direction. Such $V_R$ stripes have been attributed to internal and external perturbations in previous works (C. Cao et al. 2024). Figure 2 displays the median $V_R$ as a function of $R_g$. In the range $5 < R_g < 16$ kpc, the distribution exhibits five distinct peaks and four dips. We denote these $V_R$ stripes as peaks ("P") and dips ("D") with sequence numbers, as labeled in Figures 1 and 2. The boundaries of the stripes are defined by the midpoints between adjacent peaks and dips. The $R_g$ values of the starting and ending points, the corresponding median $L_Z$ values, and the number of stars in each $V_R$ stripe are listed in Table 1.

### 2.4. The $V_R$ Distribution in the $Z-V_Z$ Phase Space of Each $V_R$ Stripe

In this subsection, we divide the sample according to the distinct $V_R$ stripes and examine the $V_R$ distribution of stars in the $Z-V_Z$ phase space for each subsample.

**Table 1**
$R_g$ and $L_Z$ Ranges and Star Counts for Each $V_R$ Stripe

| $V_R$ Stripe | Starting $R_g$ (kpc) | Ending $R_g$ (kpc) | Starting $L_Z$ (kpc km s$^{-1}$) | Ending $L_Z$ (kpc km s$^{-1}$) | Number of Stars |
|---|---|---|---|---|---|
| P1 | 5.55 | 6.05 | 1354.5 | 1473.7 | 83,955 |
| D1 | 6.05 | 6.55 | 1473.7 | 1589.7 | 117,607 |
| P2 | 6.55 | 7.15 | 1589.7 | 1725.3 | 218,706 |
| D2 | 7.15 | 7.8 | 1725.3 | 1868.4 | 332,497 |
| P3 | 7.8 | 8.6 | 1868.4 | 2040.4 | 656,780 |
| D3 | 8.6 | 9.55 | 2040.4 | 2239.9 | 672,327 |
| P4 | 9.55 | 10.3 | 2239.9 | 2394.7 | 239,160 |
| D4 | 10.3 | 11.05 | 2394.7 | 2547.5 | 142,743 |
| P5 | 11.05 | 11.95 | 2547.5 | 2729.1 | 63,379 |

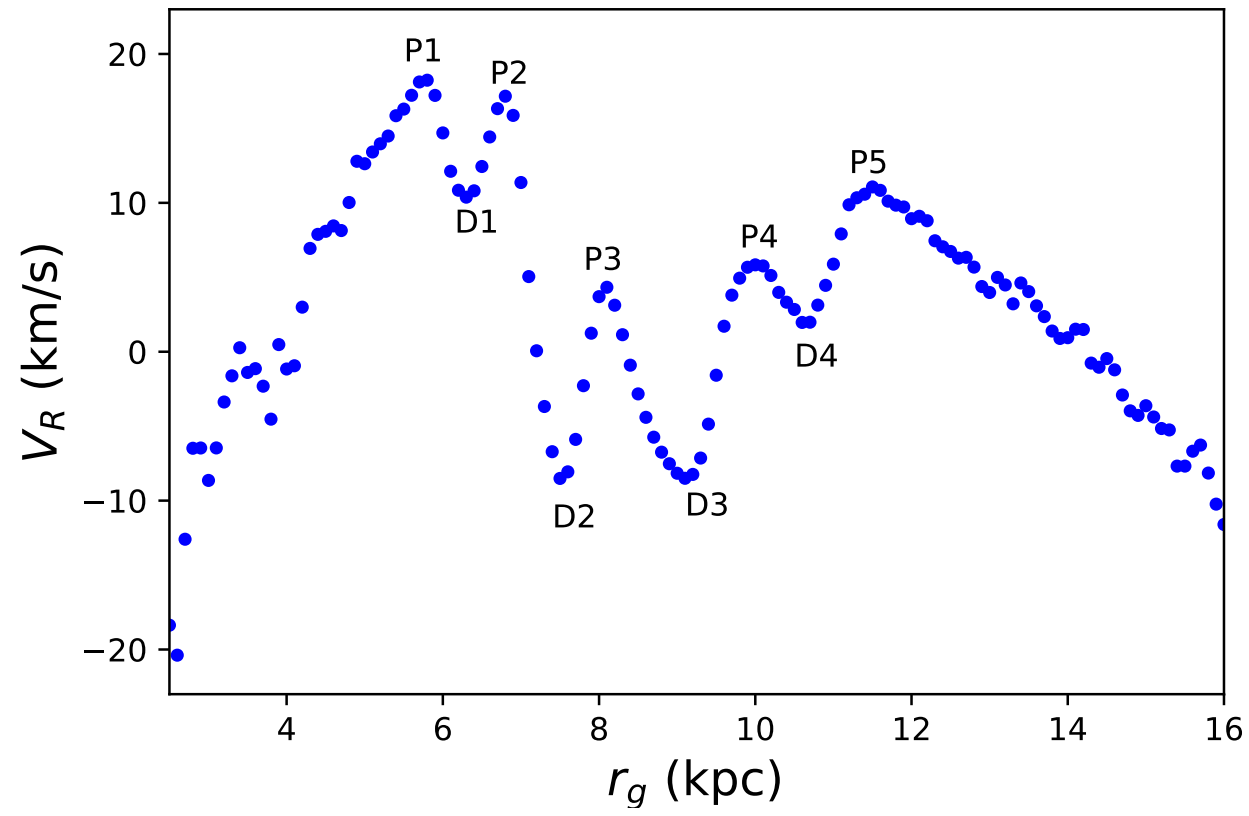

**Figure 2.** Median $V_R$ as a function of $R_g$ corresponding to the data shown in Figure 1. The $R_g$ and $L_Z$ values of the midpoints between adjacent peaks and dips, which define the boundaries of the $V_R$ stripes, are listed in Table 1.

Figure 3 shows $V_R$ distribution in $Z$–$V_Z$ phase space for each $V_R$ stripe. In our data, clear two-armed $V_R$ spirals are evident in the $Z$–$V_Z$ phase space of the D1, P2, D2, and P3 $V_R$ stripes. However, stars with larger $R_g$ ($R_g > 8.6$ kpc) do not exhibit clear two-armed $V_R$ spirals in the $Z$–$V_Z$ phase space. Instead, stars belonging to the D3, P4, D4, and P5 $V_R$ stripes display a single-armed $V_R$ spiral in the $Z$–$V_Z$ plane.

Stars in our sample have different orbital phases (here we use the conjugate angle of the radial action, $\theta_R$, to represent the orbital phase; Y. Xu et al. 2023) and eccentricities. The full maps of the $V_R$ distribution in the $Z$–$V_Z$ plane for each $V_R$ stripe, separated into different eccentricity ranges and orbital phases, are shown in Figures A1–A4 in Appendix A. The eccentricity and orbital phase ranges corresponding to the data in Figures A1–A4 are listed in Table 4. In these figures, eccentricity is divided into $e > 0.2$ and $e < 0.2$. The radial angle $\theta_R$ is divided into $\pi/2 < \theta_R < 3\pi/2$ and its complementary range ($\theta_R < \pi/2$ or $\theta_R > 3\pi/2$). (The interval $\pi/2 < \theta_R < 3\pi/2$ roughly corresponds to the region from apocenter to the guiding center radius, while $\theta_R < \pi/2$ or $\theta_R > 3\pi/2$ roughly corresponds to the region from the pericenter to the guiding center radius; Y. Xu et al. 2023.) When we subdivide the stars in each $V_R$ stripe by their orbital phase and eccentricity, we find that the two-armed $V_R$ spirals of the D1, P2, D2, and P3 $V_R$ stripes do not appear across all eccentricity ranges and orbital phases. As shown in Figures A1–A4, the two-armed $V_R$ spirals in the $Z$–$V_Z$ plane of the D1 and P2 stripes only appear when $e > 0.2$ and $\pi/2 < \theta_R < 3\pi/2$. The two-armed $V_R$ spirals of the D2 and P3 stripes only appear when $e < 0.2$ and $\pi/2 < \theta_R < 3\pi/2$. Figure 4 shows the distribution of stars from each $V_R$ stripe in the eccentricity versus $\theta_R$ plane. The subsets of stars in the D1, P2, D2, and P3 $V_R$ stripes that exhibit a two-armed phase-space spiral are highlighted with black boxes in Figure 4.

It is important to note that our data are limited to a specific range of $R$, and the stars in our sample have varying $R_g$, eccentricities, and orbital phases. The restriction on $R$ may introduce selection effects on these other parameters, which could also influence the $V_R$ distribution of stars in each $V_R$ stripe within the $Z$–$V_Z$ phase space. We will discuss the impact of selection effects on the $V_R$ distribution in the $Z$–$V_Z$ space in detail in Section 2.5 and select data that are not severely affected by these effects for the subsequent analysis.

### 2.5. Selection Effects and Their Impact on the $V_R$ Distribution in the $Z$–$V_Z$ Space

As shown in Figure 5, the $R$ values of our data are predominantly concentrated within 7.557 to 9.937 kpc (this range corresponds to $\pm 1\sigma$ from the median of the $R$ distribution). Meanwhile, the disk stars we observe have different angular momenta (or $R_g$), eccentricities, and orbital phases. Therefore, restricting $R$ to a certain range may lead to the following effect: stars with $R_g$ within this ($\pm 1\sigma$) range are preferentially low-eccentricity stars, while stars with $R_g < 7.557$ kpc or $R_g > 9.937$ kpc tend to have high eccentricities. Among these, stars with $R_g < 7.557$ kpc are more likely to be observed near their apocenter, whereas stars with $R_g > 9.937$ kpc are preferentially sampled near their pericenter.

Specifically, the coverage of data with different $R$ and eccentricities in the $R_g$ versus $\theta_R$ space is shown in Figure 6, and the details are summarized in Tables 2 and 3. In this figure and in both tables, the $R_g$ intervals are distinguished by different $V_R$ stripes. The radial angle $\theta_R$ is divided into two intervals: $\pi/2 < \theta_R < 3\pi/2$, and $\theta_R < \pi/2$ or $\theta_R > 3\pi/2$. Within each $R_g$–$\theta_R$ interval, $R_g$ is binned with a step size of 0.015, and $\theta_R$ is binned with a step size of 0.05. The coverage of each $R_g$ versus $\theta_R$ interval is defined as the proportion of bins that contain at least one star relative to the total number of bins in that interval. The coverage rates for stars with different $R$ and eccentricity in each $R_g$–$\theta_R$ interval are listed in Tables 2 and 3.

Below we discuss how the selection effects described above influence the appearance of the two-armed spiral in the $Z$–$V_Z$ plane. First, as noted above, due to selection effects the typical eccentricity of the sampled stars varies with $R_g$, which in turn leads to differences in the observed morphology of the two-armed spiral in the $Z$–$V_Z$ plane. From Figures 3 we can see that for the D2 and P3 $V_R$ stripes the two-armed spiral in the $Z$–$V_Z$ diagram is confined to $|Z| \lesssim 0.75$ kpc and the spiral arms are very narrow. In contrast, for the D1 and P2 $V_R$ stripes the spiral extends to $|Z| \lesssim 1.5$ kpc and the arms are much broader. This difference likely arises because the $R_g$ ranges of D2 (7.15–7.8 kpc) and P3 (7.8–8.6 kpc) favor the sampling of lower-eccentricity stars, whereas the $R_g$ ranges of D1 (6.05–6.55 kpc) and P2 (6.55–7.15 kpc) favor higher-eccentricity stars. As shown in Figure 4, the stars that exhibit two-armed spirals in the D2 and P3 $V_R$ stripes mostly have eccentricities $e < 0.2$, and these stars possess smaller vertical amplitudes and velocity dispersions. Consequently, the spiral arms in the $Z$–$V_Z$ plane are thin and confined to a narrow $Z$ range. For the D1 and P2 $V_R$ stripes, the stars with two-armed spirals are predominantly those with $e > 0.2$, which have larger vertical amplitudes and velocity dispersions, leading to thicker spiral arms that span a wider $Z$ range.

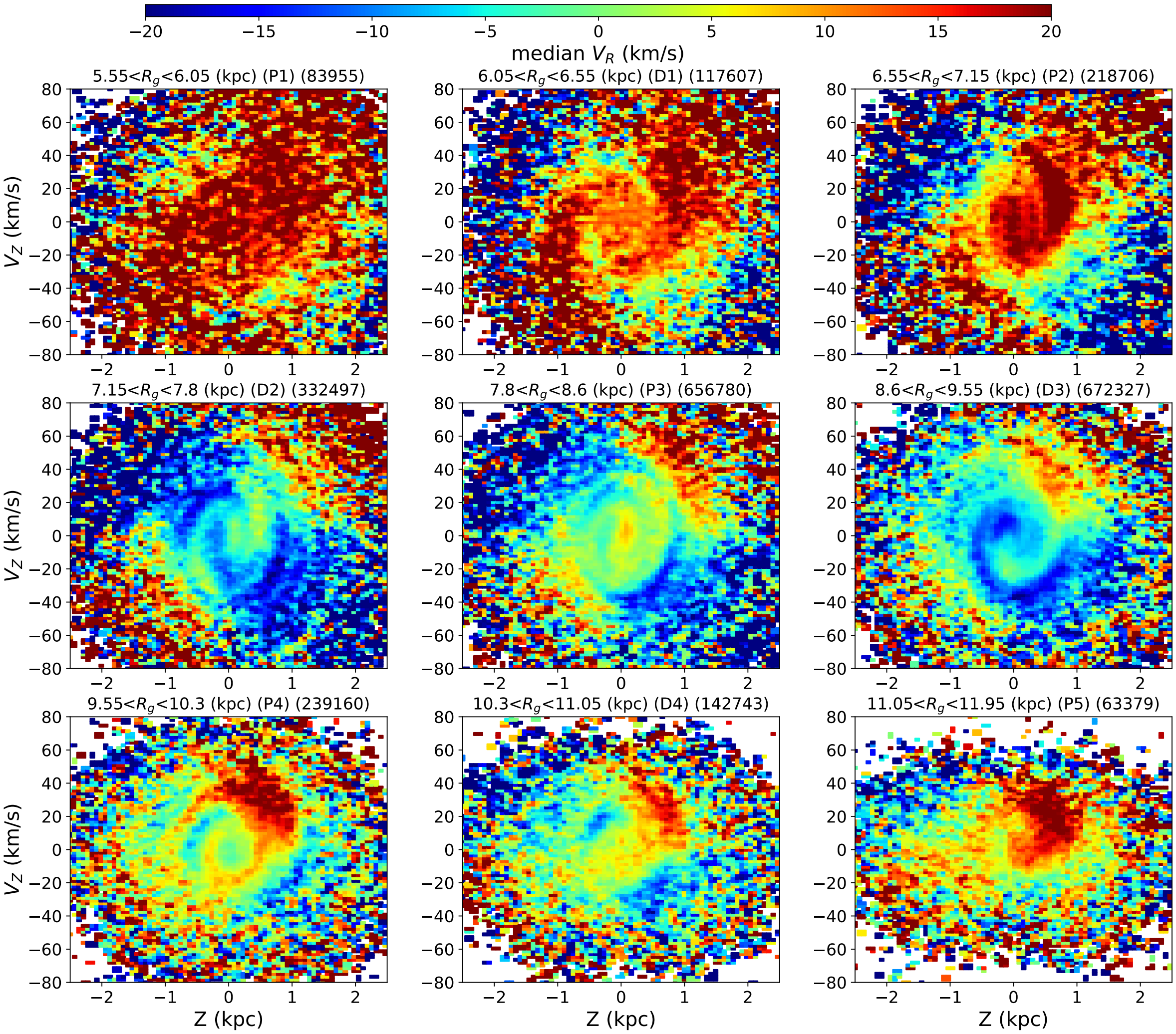

**Figure 3.** $V_R$ distribution in the $Z$–$V_Z$ phase space for each $V_R$ stripe shown in Figure 1. Each panel lists the $R_g$ range, the $V_R$ stripe name, and the number of stars. Two-armed $V_R$ spirals are clearly visible in the D1, P2, D2, and P3 $V_R$ stripes.

Furthermore, in Section 2.4 we noted that the two-armed spiral in the $Z$–$V_Z$ plane appears only in the orbital phase range $\pi/2 < \theta_R < 3\pi/2$. We wish to determine whether its absence in the complementary intervals ($\theta_R < \pi/2$ or $\theta_R > 3\pi/2$) is due to selection effects or whether it is genuinely absent. To this end, we performed test-particle simulations using a model that includes a bar and spiral arms. We integrated the orbits of the test particles and then resampled the simulation output according to the $R$ distribution of the observational data. We then examined how this resampling affects the two-armed spiral in $Z$–$V_Z$ plane. A complete description of the model and the simulation results is given in Appendix B. In the first row of Figure B2 we show the slices that exhibit clear two-armed $V_R$ spirals in the $Z$–$V_Z$ plane in the model. In the second row of Figure B2, the data are divided into two parts: $\pi/2 < \theta_R < 3\pi/2$ and ($\theta_R < \pi/2$ or $\theta_R > 3\pi/2$). In these two parts, the two-armed spiral is clear in one part and relatively blurred in the other. After resampling (see Figure B4 and Table B1), the quality of the originally clear spirals does not degrade significantly, whereas the quality of the originally blurred spirals deteriorates noticeably.

Specifically, we analyze the P3 $V_R$ stripe. The P3 $V_R$ stripe contains 412,300 stars with $\pi/2 < \theta_R < 3\pi/2$, and still as many as 244,480 stars with $\theta_R < \pi/2$ or $\theta_R > 3\pi/2$. Therefore, the absence of a two-armed $V_R$ spiral in the $\theta_R < \pi/2$ or $\theta_R > 3\pi/2$ interval for P3 cannot be attributed to insufficient sampling. According to our simulation results, there is no clear spiral structure in this interval ($\theta_R < \pi/2$ or $\theta_R > 3\pi/2$); as argued above, if a clear spiral structure were present, resampling according to the $R$ profile should not severely weaken it. Therefore, we conclude that for P3 there is genuinely no clear two-armed spiral in the $\theta_R < \pi/2$ or $\theta_R > 3\pi/2$ interval.

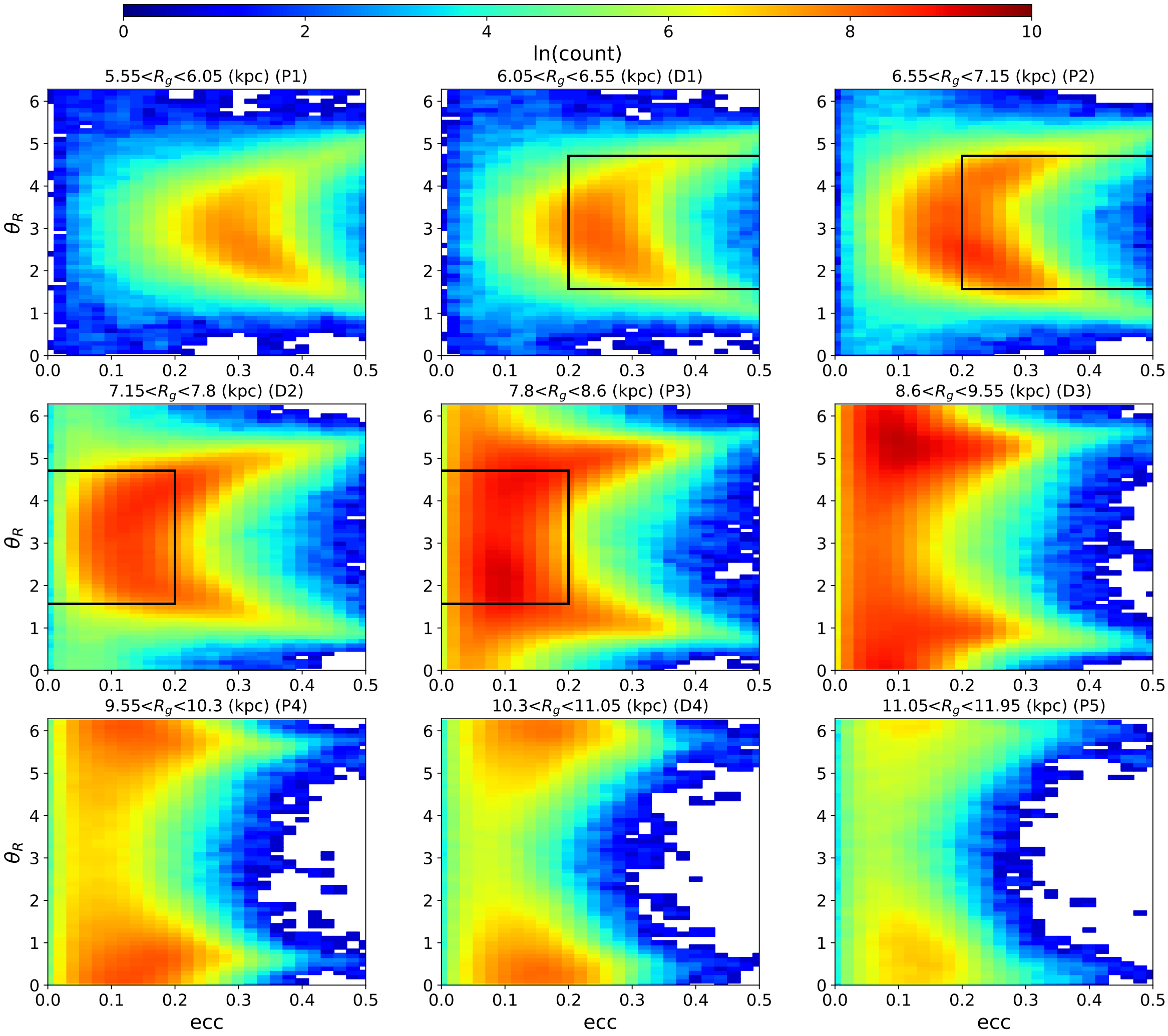

**Figure 4.** Number counts of stars from each $V_R$ stripe in the eccentricity vs. $\theta_R$ plane. Black rectangles mark the regions where stars exhibit a two-armed phase-space spiral in the $Z$–$V_Z$ plane.

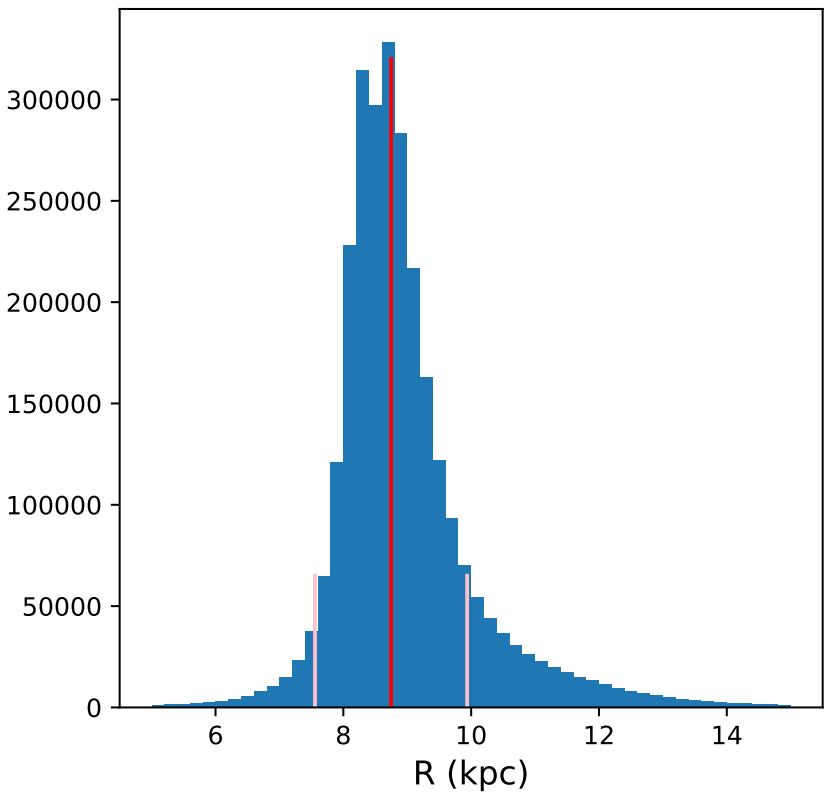

**Figure 5.** Histogram of the $R$ distribution of our sample. The red line marks the median ($R = 8.747$ kpc). The pink lines indicate $\pm 1\sigma$ from the median, at $R = 7.557$ and 9.937 kpc.

For the D1, P2, and D2 $V_R$ stripes, the coverage fraction in the $\theta_R < \pi/2$ or $\theta_R > 3\pi/2$ region is significantly lower than that in the $\pi/2 < \theta_R < 3\pi/2$ region, and the number of stars in these two phase intervals differs by at least an order of magnitude. Therefore, it is difficult for us to determine whether the absence of the two-armed phase spiral in the $\theta_R < \pi/2$ or $\theta_R > 3\pi/2$ interval for these $V_R$ stripes is due to a genuine lack of such a spiral or to selection effects.

Finally, based on the results of the above analysis, we determine a relatively complete analysis space. First, we only analyze the data with $\pi/2 < \theta_R < 3\pi/2$. Figure 7 shows the $V_R$ distribution in the $Z$–$V_Z$ plane for stars in each $V_R$ stripe, considering only the range $\pi/2 < \theta_R < 3\pi/2$. Comparing Figures 3 and 7, it can be seen that the two-armed spiral in the $Z$–$V_Z$ space for the $V_R$ stripes from D1 to P3 becomes clearer. Moreover, considering that most disk stars are on nearly circular orbits, the D2 and P3 $V_R$ stripes, which preferentially sample low-eccentricity stars, are able to capture the majority of stars in their respective $R_{\rm g}$ ranges. By contrast, D1 and P2, which preferentially sample high-eccentricity stars, miss a large fraction of the stars in their $R_{\rm g}$ ranges. Hence, the data with $\pi/2 < \theta_R < 3\pi/2$ in the D2 and P3 $V_R$ stripes constitute the most complete subset. In the subsequent analysis, we select this subset of data.

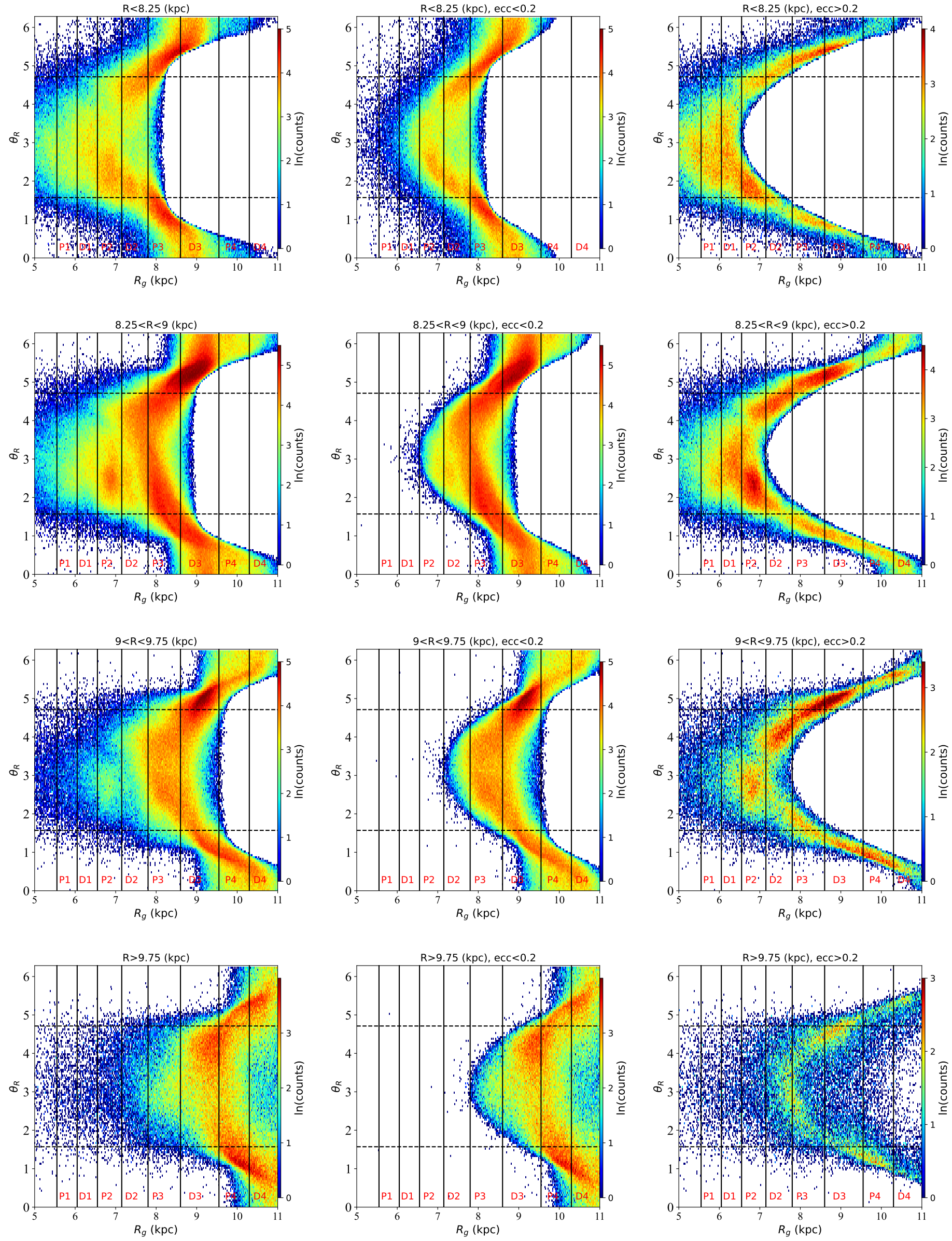

**Figure 6.** Star count distributions in the $R_g$–$\theta_R$ plane for different $R$ ranges: $R < 8.25$ kpc, $8.25 < R < 9$ kpc, $9 < R < 9.75$ kpc, and $R > 9.75$ kpc (top to bottom rows). To better show the substructures, only stars with $-1 < Z < 1$ kpc are plotted. The panels in the second column correspond to $e < 0.2$, and those in the third column to $e > 0.2$. The $V_R$ stripe names and their $R_g$ ranges are indicated in each panel.

**Table 2**
Coverage Rates of Stars in Different $R$ Ranges within Each $R_g$ and Orbital Phase Interval

| $R$ (kpc) | Orbital Phase (rad) | P1 (%) | D1 (%) | P2 (%) | D2 (%) | P3 (%) | D3 (%) | P4 (%) | D4 (%) |
|---|---|---|---|---|---|---|---|---|---|
| $R < 8.25$ | $\pi/2 < \theta_R < 3\pi/2$ | 100 | 100 | 100 | 100 | 45 | 0 | 0 | 0 |
| $R < 8.25$ | $\theta_R < \pi/2$ or $\theta_R > 3\pi/2$ | 63 | 70 | 82 | 96 | 90 | 52 | 32 | 11 |
| $8.25 < R < 9$ | $\pi/2 < \theta_R < 3\pi/2$ | 99 | 100 | 100 | 100 | 100 | 34 | 0 | 0 |
| $8.25 < R < 9$ | $\theta_R < \pi/2$ or $\theta_R > 3\pi/2$ | 30 | 36 | 43 | 49 | 79 | 85 | 51 | 34 |
| $9 < R < 9.75$ | $\pi/2 < \theta_R < 3\pi/2$ | 88 | 94 | 98 | 100 | 100 | 98 | 9 | 0 |
| $9 < R < 9.75$ | $\theta_R < \pi/2$ or $\theta_R > 3\pi/2$ | 14 | 19 | 26 | 33 | 38 | 79 | 81 | 100 |
| $R > 9.75$ | $\pi/2 < \theta_R < 3\pi/2$ | 62 | 74 | 84 | 96 | 98 | 100 | 100 | 100 |
| $R > 9.75$ | $\theta_R < \pi/2$ or $\theta_R > 3\pi/2$ | 11 | 14 | 16 | 20 | 24 | 33 | 80 | 100 |

**Note.** The $V_R$ stripe names denote the corresponding $R_g$ ranges.

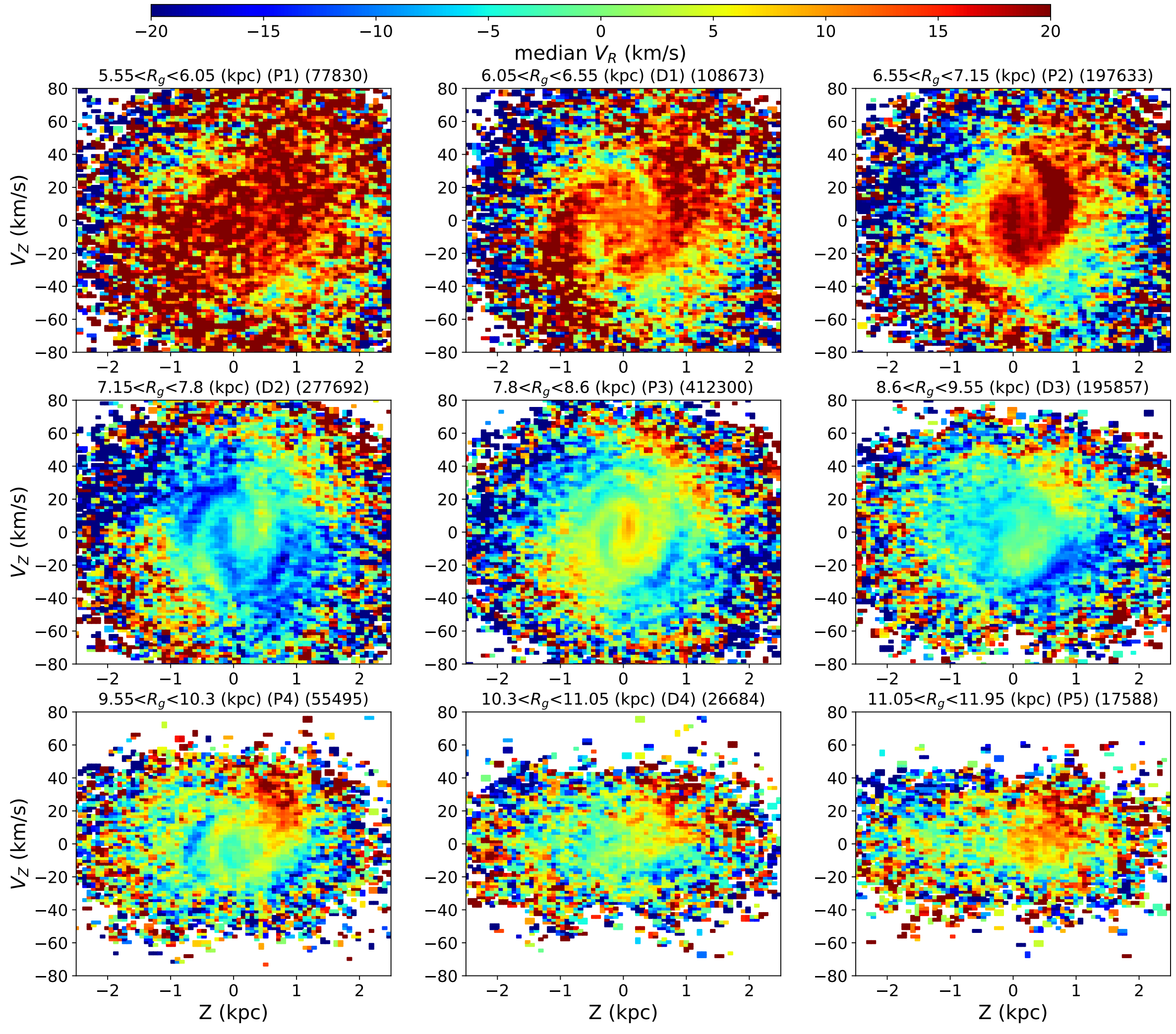

**Figure 7.** Replot of the $V_R$ distribution in the $Z$–$V_Z$ phase space for each $V_R$ stripe, restricted to stars with $\pi/2 < \theta_R < 3\pi/2$.

**Table 3**
Coverage Rates of Stars in Different $R$ and Eccentricity Bins within Each $R_g$ and Orbital Phase Interval

| $R$ (kpc) | Orbital Phase (rad) | $e$ | P1 (%) | D1 (%) | P2 (%) | D2 (%) | P3 (%) | D3 (%) | P4 (%) | D4 (%) |
|---|---|---|---|---|---|---|---|---|---|---|
| $R < 8.25$ | $\pi/2 < \theta_R < 3\pi/2$ | $e < 0.2$ | 89 | 97 | 100 | 100 | 45 | 0 | 0 | 0 |
| $R < 8.25$ | $\theta_R < \pi/2$ or $\theta_R > 3\pi/2$ | $e < 0.2$ | 27 | 41 | 65 | 94 | 90 | 49 | 9 | 0 |
| $R < 8.25$ | $\pi/2 < \theta_R < 3\pi/2$ | $e > 0.2$ | 100 | 100 | 54 | 10 | 0 | 0 | 0 | 0 |
| $R < 8.25$ | $\theta_R < \pi/2$ or $\theta_R > 3\pi/2$ | $e > 0.2$ | 53 | 59 | 66 | 74 | 63 | 45 | 32 | 11 |
| $8.25 < R < 9$ | $\pi/2 < \theta_R < 3\pi/2$ | $e < 0.2$ | 2 | 19 | 70 | 97 | 100 | 34 | 0 | 0 |
| $8.25 < R < 9$ | $\theta_R < \pi/2$ or $\theta_R > 3\pi/2$ | $e < 0.2$ | 0 | 0 | 0 | 11 | 71 | 85 | 48 | 16 |
| $8.25 < R < 9$ | $\pi/2 < \theta_R < 3\pi/2$ | $e > 0.2$ | 99 | 100 | 51 | 8 | 0 | 0 | 0 | 0 |
| $8.25 < R < 9$ | $\theta_R < \pi/2$ or $\theta_R > 3\pi/2$ | $e > 0.2$ | 30 | 36 | 43 | 48 | 51 | 41 | 45 | 34 |
| $9 < R < 9.75$ | $\pi/2 < \theta_R < 3\pi/2$ | $e < 0.2$ | 0 | 1 | 9 | 63 | 97 | 98 | 9 | 0 |
| $9 < R < 9.75$ | $\theta_R < \pi/2$ or $\theta_R > 3\pi/2$ | $e < 0.2$ | 0 | 0 | 0 | 1 | 9 | 75 | 81 | 50 |
| $9 < R < 9.75$ | $\pi/2 < \theta_R < 3\pi/2$ | $e > 0.2$ | 88 | 94 | 98 | 97 | 39 | 0 | 0 | 0 |
| $9 < R < 9.75$ | $\theta_R < \pi/2$ or $\theta_R > 3\pi/2$ | $e > 0.2$ | 14 | 19 | 26 | 33 | 37 | 42 | 33 | 39 |
| $R > 9.75$ | $\pi/2 < \theta_R < 3\pi/2$ | $e < 0.2$ | 0 | 0 | 1 | 7 | 63 | 97 | 94 | 100 |
| $R > 9.75$ | $\theta_R < \pi/2$ or $\theta_R > 3\pi/2$ | $e < 0.2$ | 0 | 0 | 0 | 0 | 0 | 13 | 77 | 100 |
| $R > 9.75$ | $\pi/2 < \theta_R < 3\pi/2$ | $e > 0.2$ | 62 | 74 | 84 | 96 | 95 | 16 | 6 | 4 |
| $R > 9.75$ | $\theta_R < \pi/2$ or $\theta_R > 3\pi/2$ | $e > 0.2$ | 11 | 14 | 16 | 20 | 24 | 30 | 42 | 48 |

**Note.** The $V_R$ stripe names denote the corresponding $R_g$ ranges.

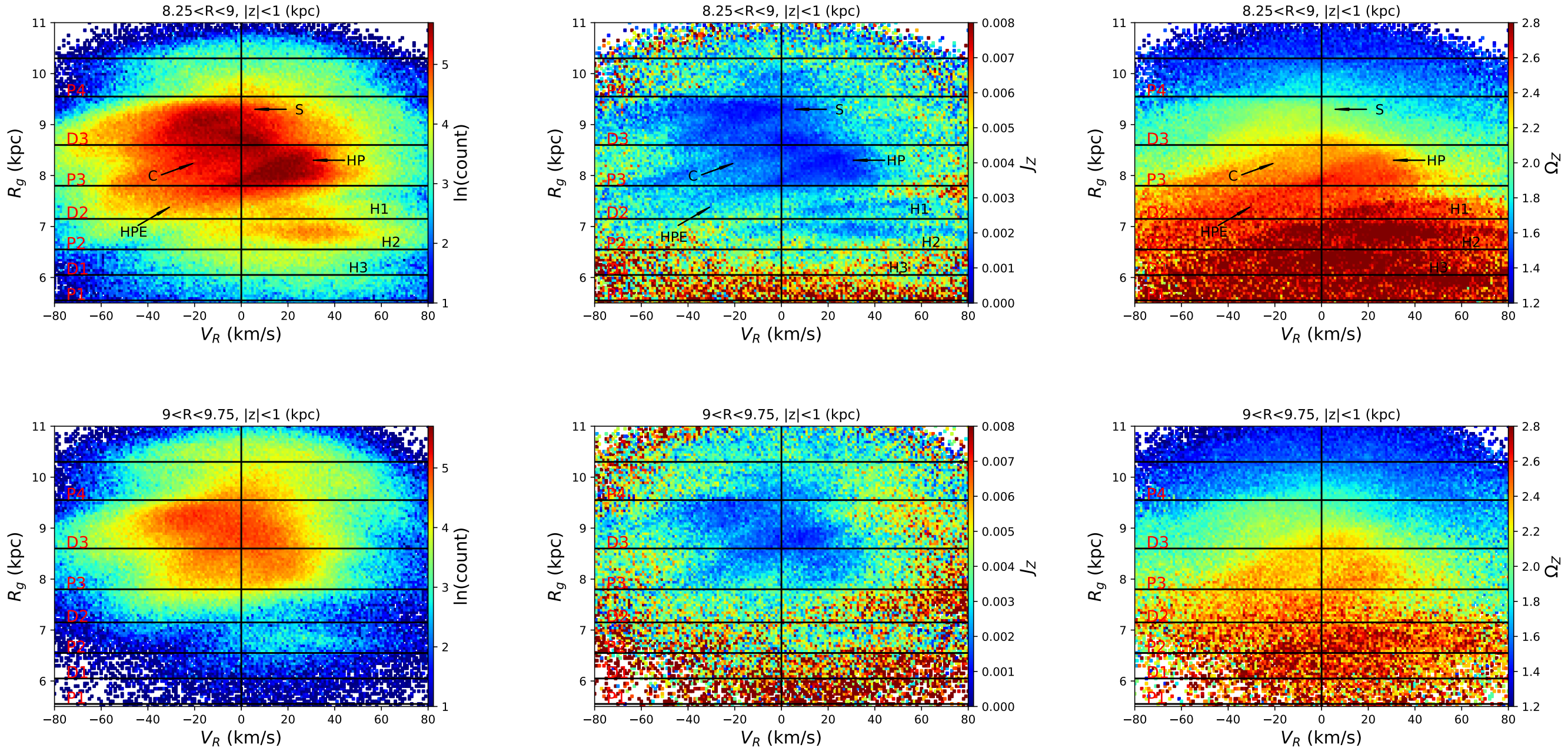

**Figure 8.** Distributions of stellar number counts (left panels), median $J_Z$ (middle panels), and median $\Omega_Z$ (right panels) in the $V_R$–$R_g$ plane. First row: $8.25 < R < 9$ kpc. Second row: $9 < R < 9.75$ kpc. In the first row, the arrows and labels indicate the main moving groups: Sirius ("S"), Coma Berenices ("C"), Hyades–Pleiades ("HP"), Hyades–Pleiades expansion tail ("HPE"), Hercules 1 ("H1"), Hercules 2 ("H2"), and Hercules 3 ("H3").

### 2.6. The Correlation between Moving Groups and $V_R$ Stripes

In the solar neighborhood, prominent moving groups such as the Hyades–Pleiades, Sirius, Hercules, and Coma Berenices are clearly visible in the $V_R$ versus $V_\phi$ plane. The $V_R$ distribution as a function of $L_Z$ (or $R_g$) is significantly shaped by these moving groups, which are known to extend several kiloparsecs from the Sun (S. Lucchini et al. 2023). They are likely important contributors to the alternating positive and negative-$V_R$ stripes shown in Figure 1 (S. Khoperskov & O. Gerhard 2022). In this subsection we discuss the correlation between the moving groups and the $V_R$ stripes.

The azimuthal velocity ($V_\phi$) of moving groups varies significantly with $R$. Although moving groups share similar angular momentum ($L_Z$) in the range $8 < R < 10$ kpc, they deviate from a curve of constant $L_Z$ at larger Galactocentric radii (M. Bernet et al. 2022). Consequently, stars belonging to a given moving group but located at different $R$ tend to occupy similar positions in the $V_R$–$R_g$ plane, particularly for stars within $8 < R < 10$ kpc.

In the left panels of Figure 8, we show the number counts in the $V_R$–$R_g$ plane for two wide $R$ ranges: $8.25 < R < 9$ kpc and $9 < R < 9.75$ kpc. The $R_g$ boundaries of the $V_R$ stripes are indicated in the figure. The main moving groups—Hyades–Pleiades, Sirius, Hercules, and Coma Berenices—are still distinguishable in the number count maps, although they appear more blurred in the second row (larger $R$). The distribution of these moving groups is clearly aligned with the alternating positive and negative-$V_R$ stripes. For instance, the D2 $V_R$ stripe encompasses the Sirius moving group and part of the Coma Berenices moving group. Stars in the Sirius moving group predominantly have $V_R < 0\,\mathrm{km\,s^{-1}}$, while stars in the Coma Berenices moving group cluster around $V_R \approx 0\,\mathrm{km\,s^{-1}}$. This is consistent with the D2 stripe being a negative-$V_R$ structure. Similarly, the P3 $V_R$ stripe mainly includes the Hyades–Pleiades moving group and part of the Coma Berenices moving group. The Hyades–Pleiades stars generally exhibit $V_R > 0\,\mathrm{km\,s^{-1}}$, whereas the Coma Berenices group contains stars with both positive and negative $V_R$. This matches the fact that the P3 stripe corresponds to a positive-$V_R$ structure.

In addition to examining the traditional number count distribution in the $V_R$–$R_g$ diagram, we also investigated the distribution of vertical actions and frequencies ($J_Z$ and $\Omega_Z$) to better delineate the boundaries of moving groups. This is because we expect that stars belonging to a moving group dominated by a particular perturbation may exhibit similar characteristics in action and frequency space. The normalization of $J_Z$ and $\Omega_Z$ was introduced in Section 2.1.

The distributions of the moving groups, colored by median $\Omega_Z$ and median $J_Z$, are shown in the second and third columns of Figure 8. These maps exhibit oblique arc-like structures similar to those seen in the number count panels. In these panels, the regions associated with moving groups are characterized by relatively low median $J_Z$ values. Moreover, the boundaries between different moving groups are clearly delineated in the median $\Omega_Z$ maps.

Notably, a sharp boundary separating the Coma Berenices and Sirius groups extends across the ($V_R$, $R_g$) plane from approximately ($-30\,\mathrm{km\,s^{-1}}$, 8.6 kpc) to ($18\,\mathrm{km\,s^{-1}}$, 9.55 kpc). The median $\Omega_Z$ values on either side of this boundary differ significantly. In the panel with $8.25 < R < 9$ kpc (third panel, first row), the median $\Omega_Z$ is about $2.04\,\mathrm{km\,s^{-1}\,kpc^{-1}}$ in the Sirius region and around $2.43\,\mathrm{km\,s^{-1}\,kpc^{-1}}$ in the Coma Berenices region. Similarly, in the panel with $9 < R < 9.75$ kpc (third panel, second row), the median $\Omega_Z$ is about $1.98\,\mathrm{km\,s^{-1}\,kpc^{-1}}$ in the Sirius region and about $2.30\,\mathrm{km\,s^{-1}\,kpc^{-1}}$ in the Coma Berenices region.

A similar $\Omega_Z$ boundary exists between the Coma Berenices and Hyades–Pleiades groups within $7.8 < R_g < 8.6$ kpc, running from approximately ($20\,\mathrm{km\,s^{-1}}$, 8.6 kpc) to ($-20\,\mathrm{km\,s^{-1}}$, 7.8 kpc) in the ($V_R$, $R_g$) plane. For the Hyades–Pleiades region, the median $\Omega_Z$ is approximately $2.53\,\mathrm{km\,s^{-1}\,kpc^{-1}}$ in the $8.25 < R < 9$ kpc panel and $2.35\,\mathrm{km\,s^{-1}\,kpc^{-1}}$ in the $9 < R < 9.75$ kpc panel.

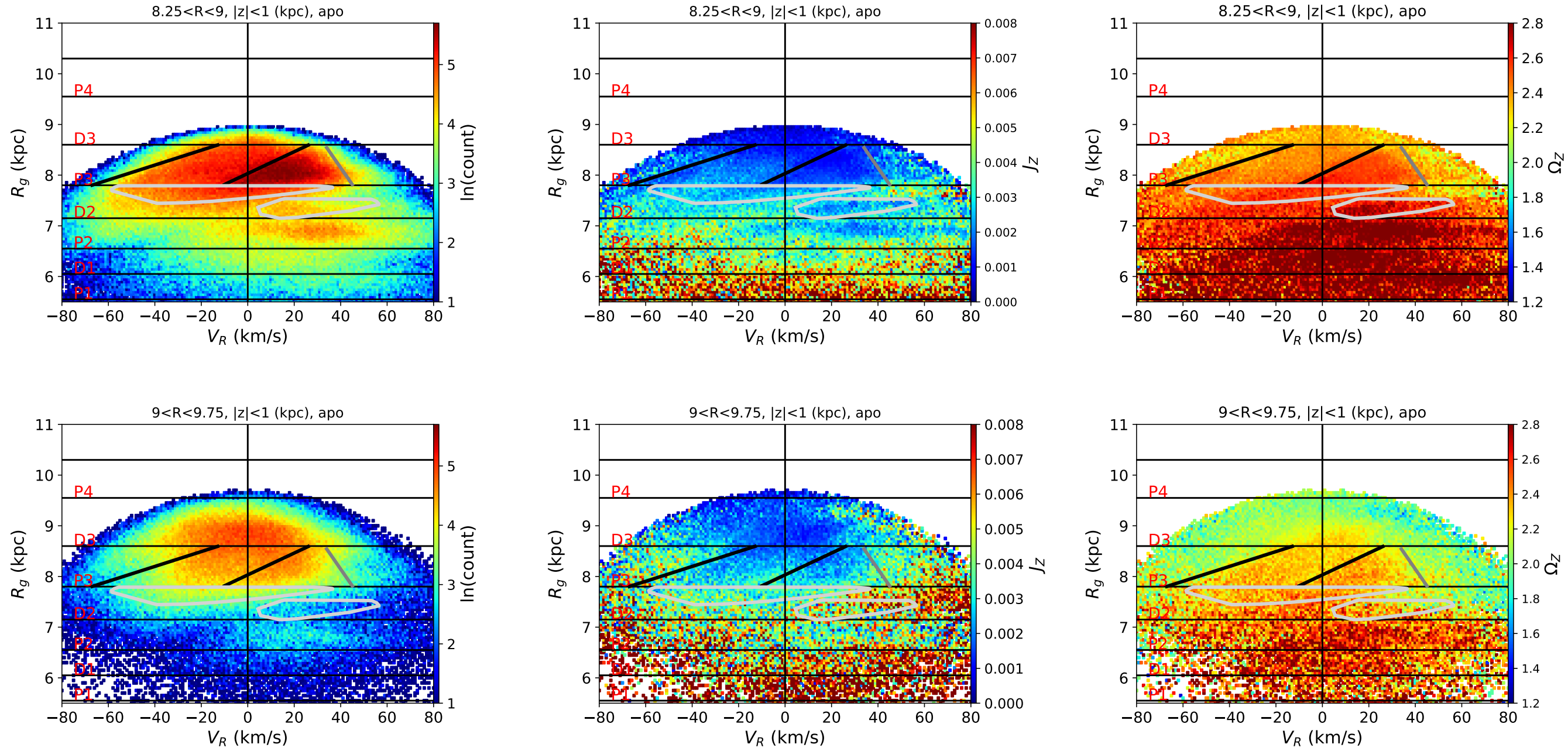

**Figure 9.** Distributions of stellar number counts (left panels), median $J_Z$ (middle panels), and median $\Omega_Z$ (right panels) for stars with $\pi/2 < \theta_R < 3\pi/2$ in the $V_R$–$R_g$ plane. First row: $8.25 < R < 9$ kpc. Second row: $9 < R < 9.75$ kpc. The outlined regions correspond to the Coma Berenices moving group in the P3 $V_R$ stripe (the region between the two black lines), the Hyades–Pleiades moving group in the P3 $V_R$ stripe (the region between the right black line and the gray line), the Hyades–Pleiades expansion tail in the D2 $V_R$ stripe (the left gray polygon), and the Hercules 1 moving group in the D2 $V_R$ stripe (the right gray polygon). The process of determining the boundaries of each moving group is described in Appendix C.

**Table 4**
$V_R$ Stripes and Their Associated Moving Groups

| $V_R$ stripes | Moving Groups | $V_R$ Distribution (km s$^{-1}$) |
|---|---|---|
| D1 | Hercules 3 | $V_R > 0$ |
| P2 | Hercules 2 | $V_R > 0$ |
| D2 | Hercules 1 | $V_R > 0$ |
| … | Hyades–Pleiades expansion tail | $V_R < 0$ |
| P3 | Hyades | $V_R > 0$ |
| … | Pleiades | $V_R > 0$ and $V_R < 0$ |
| … | Coma Berenices | $V_R > 0$ and $V_R < 0$ |
| D3 | Sirius | $V_R < 0$ |
| … | Coma Berenices | $V_R > 0$ and $V_R < 0$ |

Therefore, the $\Omega_Z$ map complements the number count map in distinguishing different moving groups in the $V_R$–$R_g$ plane. By comparing the number density distribution with the $\Omega_Z$ map, we find that although the Hyades–Pleiades moving group is predominantly located in the P3 $V_R$ stripe according to the stellar density, the characteristic $\Omega_Z$ pattern associated with Hyades–Pleiades extends from the $V_R > 0\,\mathrm{km\,s^{-1}}$ side of the P3 stripe into the $V_R < 0\,\mathrm{km\,s^{-1}}$ region of the D2 stripe. We refer to this $V_R < 0\,\mathrm{km\,s^{-1}}$ extension as the "Hyades–Pleiades expansion tail" in the following discussion.

Furthermore, a clear boundary is evident between Hyades–Pleiades and Hercules, and distinct boundaries are also observed among the Hercules subgroups (Hercules 1, Hercules 2, and Hercules 3) in both the median $J_Z$ and median $\Omega_Z$ maps.

The dominant moving groups in each $V_R$ stripe are listed in Table 4.

### 2.7. The $\mathrm{V_R}$ Distribution of Moving Groups in the $\mathrm{Z}$–$\mathrm{V_Z}$ Space

Section 2.4 demonstrates that the stars in the $V_R$ stripes D1, P2, D2, and P3 exhibit two-armed $V_R$ spirals in the $Z$–$V_Z$ phase space. Section 2.6 shows that moving groups are embedded in each of these $V_R$ stripes. This raises the question of how moving groups contribute to the two-armed $V_R$ spirals observed in these stripes. In this subsection, we select stars belonging to the moving groups in each $V_R$ stripe, and examine their $V_R$ distribution in the $Z$–$V_Z$ space. We investigate the extent to which these moving groups contribute to the two-armed $V_R$ spirals of their respective $V_R$ stripes.

In Section 2.5, we noted that only stars from the D2 and P3 $V_R$ stripes with $\pi/2 < \theta_R < 3\pi/2$ are considered. Figure 9 shows the number counts, median $\Omega_Z$, and median $J_Z$ distributions, analogous to Figure 8, but restricted to stars within the $\pi/2 < \theta_R < 3\pi/2$ interval in the $V_R$–$R_g$ plane.

The boundaries of the main regions of the Coma Berenices, Hyades–Pleiades, Hyades–Pleiades expansion tail, and Hercules 1 moving groups within the D2 and P3 $V_R$ stripes are outlined by polygons in the $V_R$–$R_g$ plane in Figure 9. The moving group boundaries are determined from abrupt changes in the median $\Omega_Z$ distribution in the $V_R$–$R_g$ plane. For the two moving groups in the D2 $V_R$ stripe, the $J_Z$ distribution is also used as a reference. The detailed procedure for defining the moving group boundaries is described in Appendix C. For analytical convenience, the first- and second-row panels of Figure 9 adopt the same boundaries to separate the moving groups, although the moving group patterns appear more blurred in the second row.

Here, we examine the $V_R$ distribution of each moving group in the $Z$–$V_Z$ phase space and compare it with the $V_R$ distribution of the corresponding $V_R$ stripe.

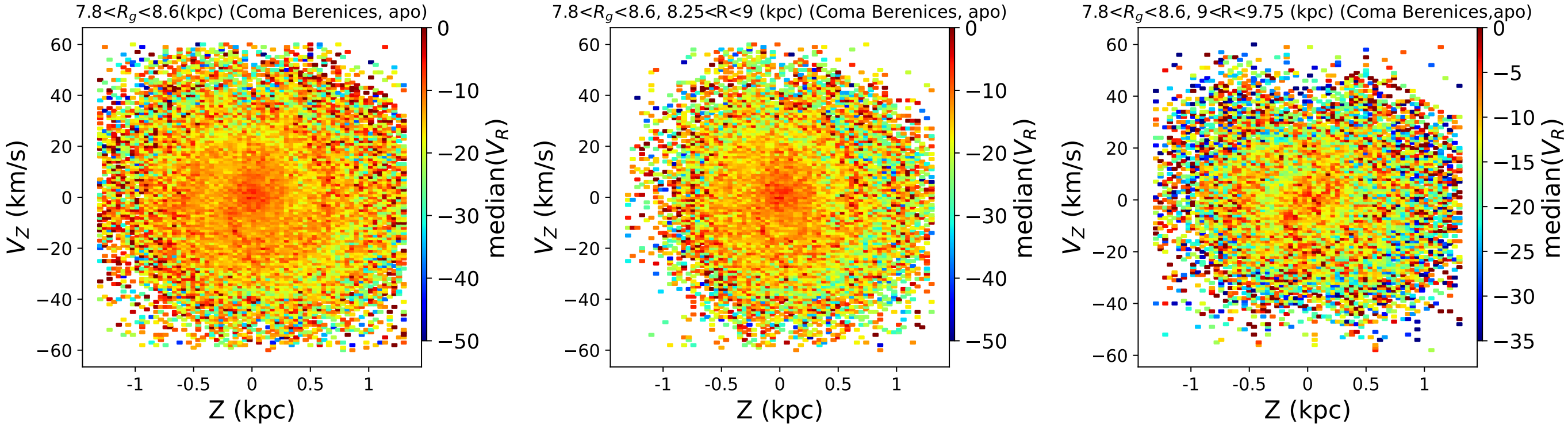

**Figure 10.** $V_R$ distribution in the $Z$–$V_Z$ plane for stars in the Coma Berenices moving group region (selected from Figure 9). Left: all stars in this region. Middle: stars with $8.25 < R < 9$ kpc. Right: stars with $9 < R < 9.75$ kpc.

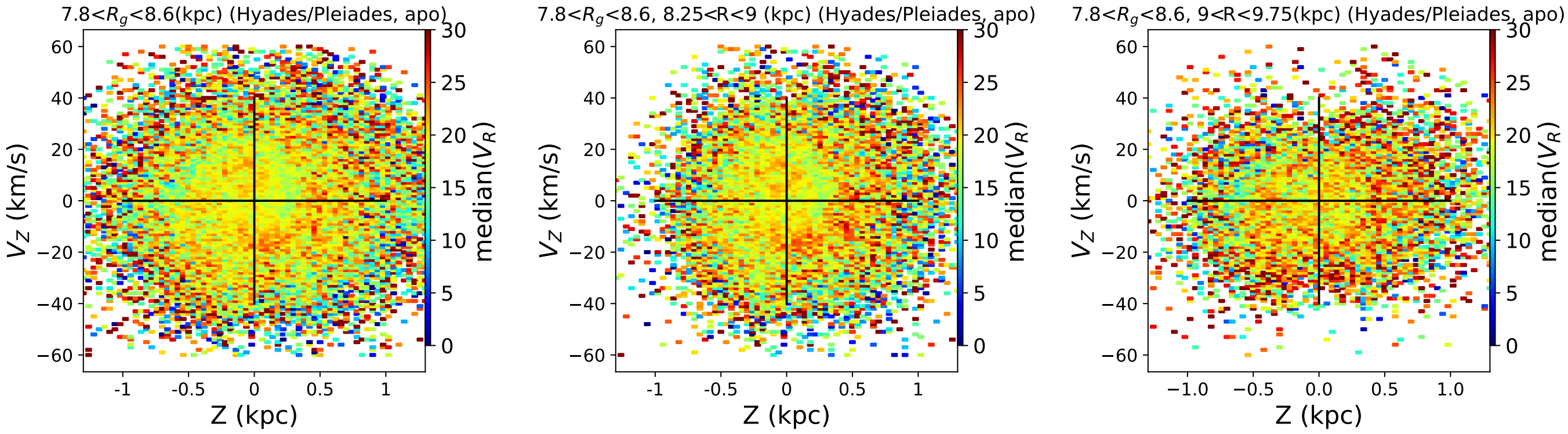

**Figure 11.** $V_R$ distribution in the $Z$–$V_Z$ plane for stars in the Hyades–Pleiades moving group region (selected from Figure 9). Left: all stars in this region. Middle: stars with $8.25 < R < 9$ kpc. Right: stars with $9 < R < 9.75$ kpc.

*2.7.1. The* $\mathrm{V_R}$ *Distribution of Moving Groups in the P3* $\mathrm{V_R}$ *Stripe in the* Z–$\mathrm{V_Z}$ *Plane and Their Contribution to Its Two-armed* $\mathrm{V_R}$ *Spiral*

From Figure 9, the Coma Berenices moving group is distributed around $V_R \approx 0\,\mathrm{km\,s^{-1}}$ and extends into the $V_R < 0\,\mathrm{km\,s^{-1}}$ region. In the $V_R$–$R_\mathrm{g}$ plane, the region of the Coma Berenices moving group is enclosed by a closed polygon. Its upper and lower boundaries are defined by the natural boundaries of the P3 $V_R$ stripe at $R_\mathrm{g} = 7.8$ and 8.6 kpc, respectively. The left and right boundaries are delineated by two black slanted lines, which are determined from abrupt changes in the median $\Omega_Z$ distribution and constrain the extent of the moving group in the $V_R$ direction; this region is outlined by a black polygon. The stars with $V_R > 0\,\mathrm{km\,s^{-1}}$ in the P3 $V_R$ stripe primarily belong to the Hyades–Pleiades moving groups, whose left and right boundaries are marked by the black slanted line and the gray slanted line, respectively. The $V_R$ distributions in the $Z$–$V_Z$ plane for stars within the Coma Berenices and Hyades–Pleiades moving group regions are shown in Figures 10 and 11, respectively. In these figures, we also display the $V_R$ distribution in the $Z$–$V_Z$ plane for two $R$ ranges, $8.25 < R < 9$ kpc and $9 < R < 9.75$ kpc, to examine how the $V_R$ distribution of the moving groups depends on $R$.

We first discuss the $V_R$ distribution of stars in the Coma Berenices region in the $Z$–$V_Z$ plane. Figure 10 shows the $V_R$ distribution in the $Z$–$V_Z$ plane for stars within the Coma Berenices region of the P3 $V_R$ stripe (the region between the two black slanted lines in Figure 9), and for stars in the radial ranges $8.25 < R < 9$ kpc and $9 < R < 9.75$ kpc. The left panel displays the $V_R$ distribution of stars over the full $R$ range, the middle panel shows the distribution for $8.25 < R < 9$ kpc, and the right panel illustrates the distribution for $9 < R < 9.75$ kpc.

The $V_R$ distribution for stars with $8.25 < R < 9$ kpc (middle panel of Figure 10) exhibits a two-armed $V_R$ spiral, with the $V_R$ values along the spiral biased toward zero and positive values compared to the surrounding regions. A prominent central concentration of the phase spiral is visible within $-0.15 < Z < 0.25$ kpc and $-10 < V_Z < 20\,\mathrm{km\,s^{-1}}$. The right arm extends from this central concentration into the second quadrant, passing approximately through $(Z,\ V_Z) = (0\,\mathrm{kpc}, -20\,\mathrm{km\,s^{-1}})$, $(0.66\,\mathrm{kpc}, 0\,\mathrm{km\,s^{-1}})$, and $(0\,\mathrm{kpc}, 35\,\mathrm{km\,s^{-1}})$. The left arm extends from the central concentration into the first quadrant; it appears highly diffuse in the first and second quadrants, but becomes relatively more distinct in the third and fourth quadrants, passing roughly through $(Z,\ V_Z) = (0\,\mathrm{kpc}, -44\,\mathrm{km\,s^{-1}})$ and $(1\,\mathrm{kpc}, 0\,\mathrm{km\,s^{-1}})$.

The right panel of Figure 10 displays a distinct two-armed $V_R$ phase spiral for stars in the Coma Berenices region within the radial range $9 < R < 9.75$ kpc. The two arms of the phase spiral are highly symmetric, and there is no prominent central concentration.

The left panel of Figure 10 shows a clear two-armed $V_R$ phase spiral, whose morphology combines the features seen in the middle and right panels.

Figure 11 shows the $V_R$ distribution in the $Z$–$V_Z$ plane for stars in the Hyades–Pleiades region of the P3 $V_R$ stripe (the region between the right black slanted line and the gray slanted line in Figure 9). The left panel displays the $V_R$ distribution for these stars over the full $R$ range. The middle panel shows the same region but restricted to $8.25 < R < 9$ kpc, and the right panel shows the distribution for $9 < R < 9.75$ kpc.

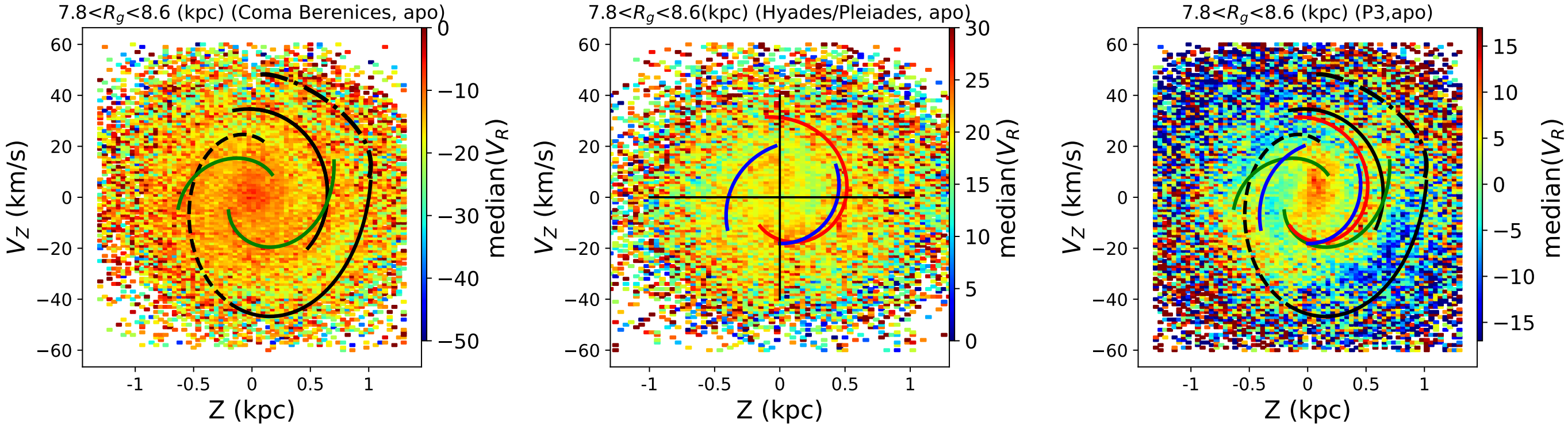

**Figure 12.** Left: $V_R$ distribution in the $Z$–$V_Z$ plane for stars in the Coma Berenices moving group region of the P3 $V_R$ stripe (as in the left panel of Figure 10). The fitted two-armed spirals for $8.25 < R < 9$ kpc and $9 < R < 9.75$ kpc from Figures D1 and D2 are overplotted as black and green curves, respectively. Middle: $V_R$ distribution for stars in the Hyades–Pleiades region of the P3 $V_R$ stripe (as in the left panel of Figure 11). The fitted spirals from Figures D3 (single-armed, red) and D4 (two-armed, blue) are overlaid. Right: $V_R$ distribution for all stars in the P3 $V_R$ stripe (as in the middle panel of the second row of Figure 7), with the fitting curves from the left and middle panels overlaid. For details of the fitting procedure, see Appendix D.

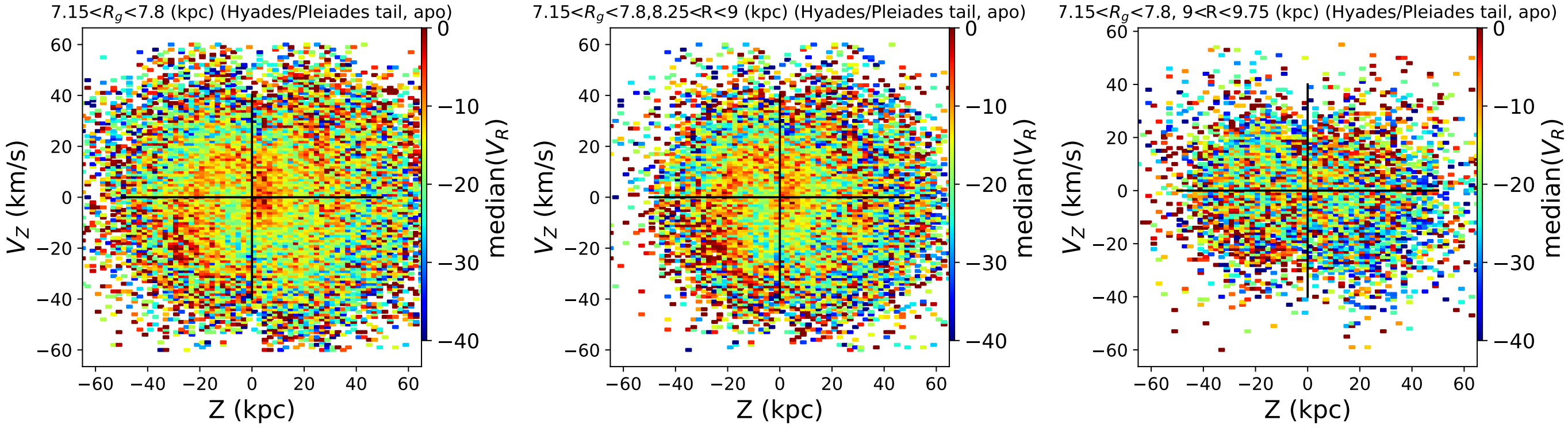

**Figure 13.** $V_R$ distribution in the $Z$–$V_Z$ plane for stars in the Hyades–Pleiades expansion tail region of the D2 $V_R$ stripe (selected from Figure 9). Left: all stars in this region. Middle: stars with $8.25 < R < 9$ kpc. Right: stars with $9 < R < 9.75$ kpc.

The $V_R$ distributions in the left and middle panels both exhibit a faint single-armed phase spiral in the $Z$–$V_Z$ plane. In contrast, the right panel ($9 < R < 9.75$ kpc) displays curved patterns that resemble a suspected two-armed spiral. In the following, we fit these patterns with a two-armed spiral.

From Figures 10 and 11, we can see the two-armed and one-armed $V_R$ spirals in the $Z$–$V_Z$ plane. The $V_R$ values along the spiral arms are biased toward positive values relative to the adjacent regions. We determine a threshold by comparing the median $V_R$ of the spiral arms with that of the surrounding areas, then select the bins that lie on the spiral arms and fit them with a logarithmic spiral. The detailed fitting procedure is described in Appendix D. The resulting fitted spirals are used directly in Figure 12.

Figure 12 compares the $V_R$ distributions of each moving group in the $Z$–$V_Z$ plane, as shown in Figures 11 and 10, with the $V_R$ distribution of all stars in the P3 $V_R$ stripe (Figure 7), in order to determine the contribution of the moving groups to the two-armed spiral in this stripe.

The left panel of Figure 12 shows the fitted two-armed spirals for stars in the Coma Berenices region: the black curve corresponds to $8.25 < R < 9$ kpc, and the green curve to $9 < R < 9.75$ kpc. The middle panel displays the fitted curves for the Hyades–Pleiades region: a one-armed spiral for $8.25 < R < 9$ kpc (red curve) and a two-armed spiral for $9 < R < 9.75$ kpc (blue curve).

The fitted curves from the left and middle panels are overlaid on the $V_R$ distribution of the P3 $V_R$ stripe in the $Z$–$V_Z$ plane in the right panel of Figure 12. A comparison of these ridge curves with the full P3 $V_R$ phase spiral shows that the right arm of the P3 spiral is composed of three components: the tail of the single-armed spiral in the Hyades–Pleiades region for $8.25 < R < 9$ kpc (red curve), the right arm of the suspected two-armed spiral in the same region for $9 < R < 9.75$ kpc (blue curve), and the right arm of the Coma Berenices region in both $8.25 < R < 9$ kpc (black curve) and $9 < R < 9.75$ kpc (green curve). The left arm of the P3 $V_R$ phase spiral coincides with two contributions: the left arm of the Coma Berenices region in both $R$ ranges ($8.25 < R < 9$ kpc and $9 < R < 9.75$ kpc; black and green curves), and the left arm of the curved pattern of the Hyades–Pleiades region for $9 < R < 9.75$ kpc (blue curve).

Among the two main moving groups in the P3 $V_R$ stripe, the Coma Berenices moving group exhibits a more pronounced two-armed spiral than the Hyades–Pleiades, and is the primary contributor to the left arm of the two-armed spiral in this stripe.

#### 2.7.2. *The* $V_R$ *Distribution of Moving Groups in the D2* $V_R$ *Stripe in the* $Z$–$V_Z$ *Plane and Their Contribution to Its Two-armed* $V_R$ *Spiral*

According to Figure 9, stars with $V_R > 0$ km s$^{-1}$ in the D2 $V_R$ stripe consist predominantly of the Hercules 1 moving group. Stars with $V_R < 0$ km s$^{-1}$ in this stripe correspond mainly to the Hyades–Pleiades expansion tail, as defined in Section 2.6.

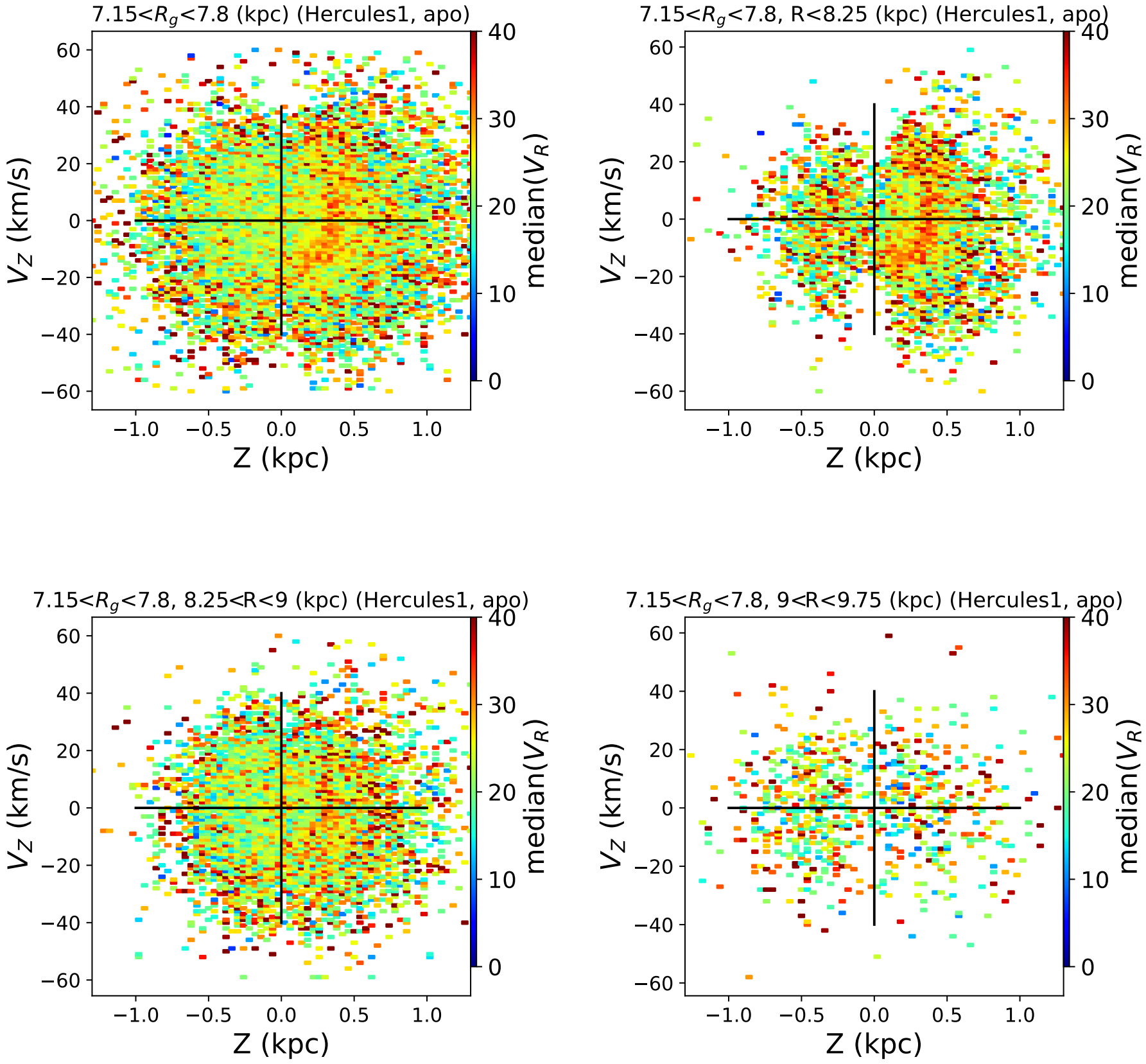

**Figure 14.** $V_R$ distribution in the $Z$–$V_Z$ plane for stars in the Hercules 1 moving group region of the D2 $V_R$ stripe (selected from Figure 9). Top left: all stars in this region. Top right: $R < 8.25$ kpc. Bottom left: $8.25 < R < 9$ kpc. Bottom right: $9 < R < 9.75$ kpc.

Figure 13 shows the $V_R$ distribution in the $Z$–$V_Z$ plane for stars within the Hyades–Pleiades expansion tail region (the left gray polygon in Figure 9) of the D2 $V_R$ stripe. The left panel displays the $V_R$ distribution over the full $R$ range; a single-armed phase spiral is clearly visible. The middle panel shows stars with $8.25 < R < 9$ kpc. The distribution closely resembles that in the left panel. The inner loop of the single-armed spiral passes approximately through $(Z, V_Z) = (0\,\mathrm{kpc}, 10\,\mathrm{km\,s^{-1}})$, $(-0.5\,\mathrm{kpc}, 0\,\mathrm{km\,s^{-1}})$, and $(0\,\mathrm{kpc}, -35\,\mathrm{km\,s^{-1}})$, while the outer tail passes roughly through $(0.7\,\mathrm{kpc}, 0\,\mathrm{km\,s^{-1}})$ and $(0\,\mathrm{kpc}, 40\,\mathrm{km\,s^{-1}})$. Both the inner loop and the tail are relatively clear, but the region connecting them appears faint. In Appendix D, we select bins with median $V_R$ above a threshold of $-10\,\mathrm{km\,s^{-1}}$. It can be seen that the inner loop and the tail tend to trace two spiral segments with different pitch angles, rather than forming a single one-armed spiral. The two spiral segments are fitted by logarithmic spirals in Figure D5. The right panel, which includes stars with $9 < R < 9.75$ kpc, does not show a clear phase spiral.

Figure 14 shows the $V_R$ distribution in the $Z$–$V_Z$ plane for stars in the Hercules 1 region (the right gray polygon of the D2 $V_R$ stripe in Figure 9). The top-left panel displays the $V_R$ distribution over the full $R$ range. A short $V_R$ phase spiral is present in the first and fourth quadrants of the $Z$–$V_Z$ plane. This short spiral passes roughly through $(Z, V_Z) = (0.27\,\mathrm{kpc}, 15\,\mathrm{km\,s^{-1}})$, $(0.35\,\mathrm{kpc}, 0\,\mathrm{km\,s^{-1}})$, and $(0.2\,\mathrm{kpc}, -15\,\mathrm{km\,s^{-1}})$. In the top-right, bottom-left, and bottom-right panels, we divide the stars in the Hercules 1 region into three $R$ subsets: $R < 8.25$ kpc, $8.25 < R < 9$ kpc, and $9 < R < 9.75$ kpc. It can be seen that the single-armed spiral in the Hercules 1 region is primarily contributed by stars with $R < 8.25$ kpc. In Appendix D (Figure D6), we fit this short spiral with a logarithmic spiral.

In Figure 15, we compare the two-armed $V_R$ spiral of the D2 $V_R$ stripe with the phase-spiral features in the Hyades–Pleiades expansion tail and the Hercules 1 region in the $Z$–$V_Z$ plane. The fitted inner loop and outer tail of the phase spiral in the Hyades–Pleiades expansion tail are overplotted as black curves in the left panel, while the fitted short $V_R$ arm of the Hercules 1 region is shown as a red curve in the middle panel. These curves are overlaid on the $V_R$ distribution of the D2 $V_R$ stripe in the right panel. Comparing the fitted curves with the two-armed spiral of the D2 stripe reveals that the left arm of the D2 spiral is contributed by the inner loop of the Hyades–Pleiades expansion tail phase spiral. The right arm of the D2 spiral is formed by a combination of the outer tail of that same phase spiral and the short $V_R$ arm of the Hercules 1 region. Thus, the two-armed $V_R$ phase spiral in the D2 stripe is not an independent structure; rather, it is composed of $V_R$ arm features from two distinct moving groups in the $Z$–$V_Z$ plane.

## 3. Properties and Possible Origin of Coma Berenices Moving Group Stars

In the last section, we investigated the $V_R$ distribution in the $Z$–$V_Z$ plane for stars in the four moving group regions of the P3 and D2 $V_R$ stripes. Among these, only stars in the Coma Berenices moving group region exhibit a clear two-armed spiral in the $Z$–$V_Z$ plane. Here, we further explore the properties and origin of this two-armed $V_R$ spiral.

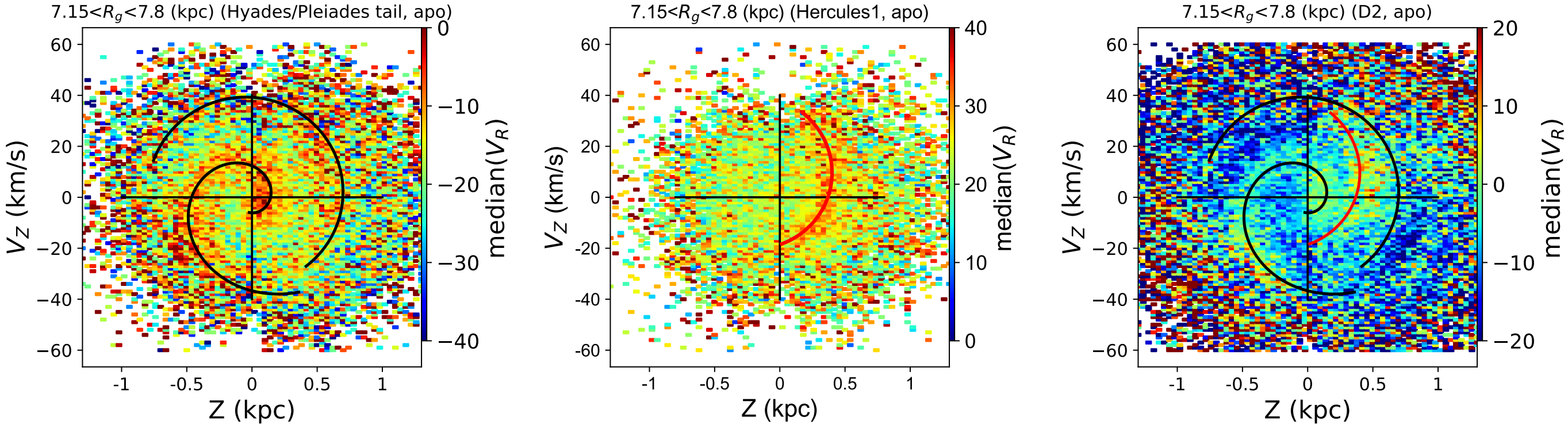

**Figure 15.** Left: $V_R$ distribution in the $Z$–$V_Z$ plane for stars in the Hyades–Pleiades expansion tail region of the D2 $V_R$ stripe (as in the left panel of Figure 13). The ridge of the single-armed $V_R$ spiral from Figure D5 is overplotted as a blue curve. Middle: $V_R$ distribution for stars in the Hercules 1 moving group region of the D2 $V_R$ stripe (as in the left panel of Figure 14). The ridge of the short $V_R$ arm from Figure D6 is overplotted as a black curve. Right: $V_R$ distribution for stars in the D2 $V_R$ stripe (as in the left panel of the second row of Figure 7), with the blue and black curves from the left and middle panels overlaid. For details of the fitting procedure, see Appendix D.

### 3.1. Properties of the Two-armed $V_R$ Spiral of Coma Berenices Moving Group Stars: Insights from $L_Z$ and $J_R$ Distributions

Taking the stars in the Coma Berenices moving group region with $8.25 < R < 9$ kpc as an example, we examine the distributions of $L_Z$ and $J_R$ in the $Z$–$V_Z$ plane in Figure 16. From this figure, stars located on the spiral arms exhibit relatively higher median $L_Z$ values compared to those in the adjacent regions. At the same time, these stars have smaller $J_R$ values, indicating more circular orbits relative to their surroundings. This observational result is highly consistent with a conclusion from previous work on the origin of Coma Berenices, which suggests that this moving group originates from the corotation resonance of spiral arms (T. A. Michtchenko et al. 2018). It is also consistent with the description in Section 3 of J. Binney & S. Tremaine (2008), where it is noted that nonaxisymmetric gravitational perturbations can increase the angular momentum of stars. The increase in angular momentum reduces the radial amplitude of the orbit, making the orbit more circular. This effect is particularly pronounced near the corotation resonance, where the azimuthal velocity of the stars matches the pattern speed of the nonaxisymmetric gravitational potential, further circularizing their orbits.

### 3.2. Identifying Resonances through Stellar Orbital Frequencies of Coma Berenices Moving Group Stars

Figure 17 shows the scatterplots of stars on the phase-spiral arms in the $Z$–$V_Z$ plane for the Coma Berenices moving group region, selected by applying a $V_R$ threshold over the radial ranges $8.25 < R < 9$ kpc and $9 < R < 9.75$ kpc within the P3 $V_R$ stripe (see also Figures D1 and D2 in Appendix D). In this subsection, we explore the frequency distribution of stars on the phase-spiral arms and the corresponding possible resonances.

The orange histograms in Figure 18 show the distribution of frequencies and frequency ratios for the red and blue arms identified in Figure 17. For comparison, the blue histograms display the same distributions for all stars within the same $R$ range as those in the orange histograms. The median and standard deviation of each orange histogram are listed in Table 5. To better assess the resonance conditions, the units of all frequencies in Table 5 have been converted to physical units here.

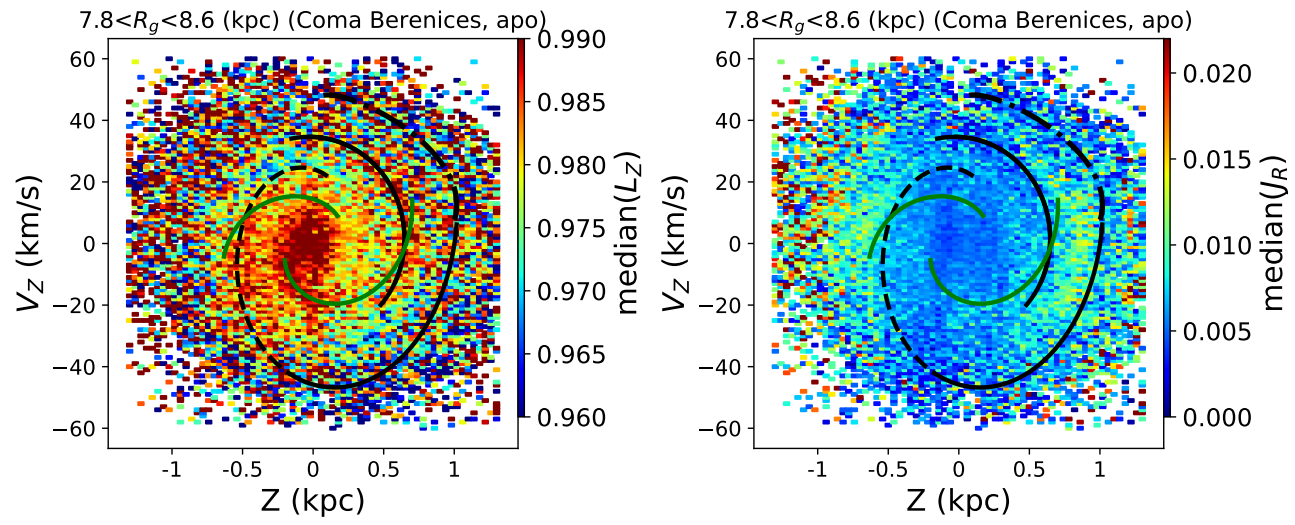

**Figure 16.** Distributions of median $L_Z$ (left) and median $J_R$ (right) in the $Z$–$V_Z$ plane for stars in the Coma Berenices moving group region with $8.25 < R < 9$ kpc. The fitted two-armed $V_R$ spiral curves from Figure 12 are overplotted on both panels.

In this table, the most notable feature is that the median $\Omega_\phi$ of the red and blue arms in Figure 17 is around 28 km s$^{-1}$ kpc$^{-1}$, with a standard deviation of only 0.6 km s$^{-1}$ kpc$^{-1}$. This median value is very close to the pattern speed of the spiral arms (W. S. Dias et al. 2019). Therefore, based on their $\Omega_\phi$ distribution, the stars in these arms are very likely located at the corotation resonance of the spiral arms.

Meanwhile, the $\Omega_R$ distribution is also relatively narrow, with a median value around 38 km s$^{-1}$ kpc$^{-1}$. Consequently, $\Omega_R/\Omega_\phi \approx 4/3$, suggesting that these stars are on orbits near the 4:3 radial–azimuthal resonance. Considering the possible pattern speeds of the spiral arms and the bar—12 km s$^{-1}$ kpc$^{-1}$ (A.-C. Eilers et al. 2020) and 28 km s$^{-1}$ kpc$^{-1}$ (W. S. Dias et al. 2019) for the spiral arms, and 37.5 km s$^{-1}$ kpc$^{-1}$ (J. A. S. Hunt et al. 2022), 41 km s$^{-1}$ kpc$^{-1}$ (M. Bernet et al. 2024), and 52 km s$^{-1}$ kpc$^{-1}$ (A. R. Pettitt et al. 2014) for the bar—we test each possible pattern speed $\Omega_P$ using the resonance condition $m(\Omega_\phi - \Omega_P) = \pm\,\Omega_R$, with azimuthal harmonics $m = 2$ and 4. For a star with $\Omega_\phi = 28$ and $\Omega_R = 38$ km s$^{-1}$ kpc$^{-1}$, the condition is satisfied when $\Omega_{bar} = 37.5$ km s$^{-1}$ kpc$^{-1}$ and $m = 4$: $4 \times (28–37.5) = -38 = -\Omega_R$. This corresponds to an $m = 4$ inner Lindblad resonance. Hence, taking into account the distributions of both $\Omega_\phi$ and $\Omega_R$, the $m = 4$ inner Lindblad resonance with the bar (pattern speed $\Omega_P = 37.5$ km s$^{-1}$ kpc$^{-1}$) can also explain the frequency distribution observed on the arms.

In Table 5, the median $\Omega_Z$ values differ between the red and blue arms because of their different locations in the $Z$–$V_Z$ plane (namely, 50, 57, 65, and 70 km s$^{-1}$ kpc$^{-1}$). Using the full

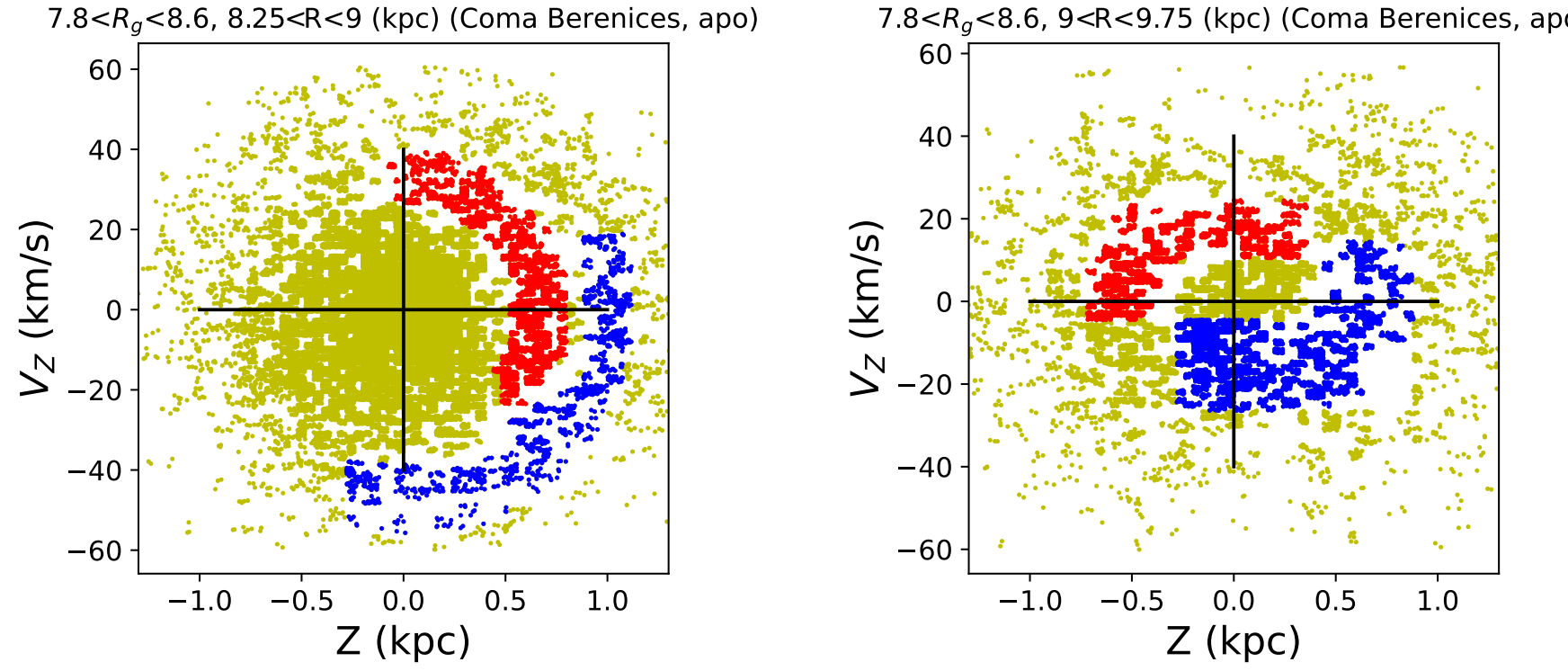

**Figure 17.** Left: stars in the Coma Berenices moving group region of the P3 $V_R$ stripe, within $8.25 < R < 9$ kpc, whose bins exceed the $V_R$ threshold in the $Z$–$V_Z$ plane (see also Figure D1 in Appendix D). The red and blue arms trace distinct, well-defined portions of the two-armed spiral. The red arm contains 3669 stars; the blue arm contains 1227 stars. Right: same as the left panel, but for $9 < R < 9.75$ kpc (see also Figure D2 in Appendix D). The red arm contains 3643 stars; the blue arm contains 5867 stars.

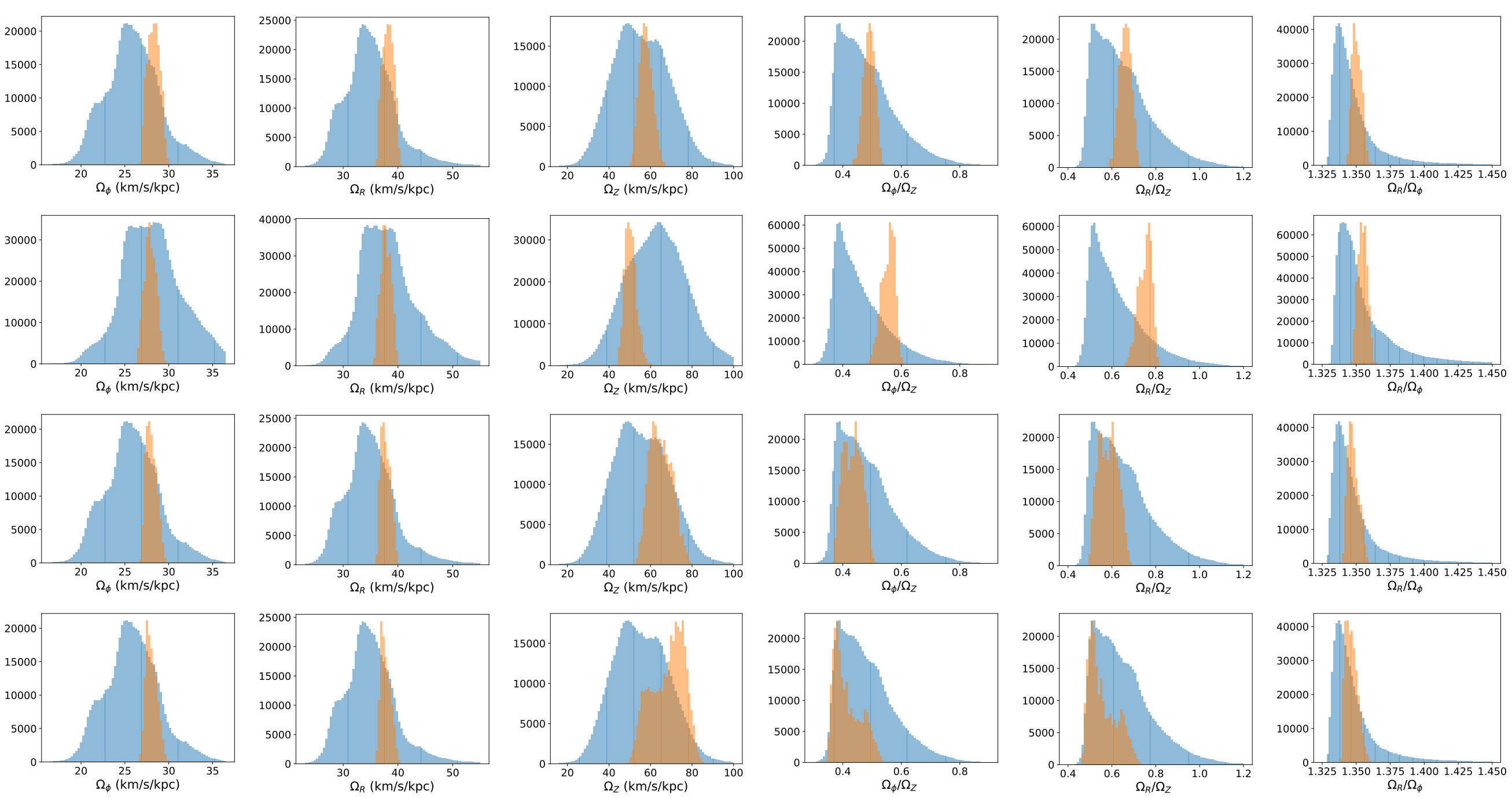

**Figure 18.** Orange histograms show the distributions of $\Omega_\phi$, $\Omega_R$, $\Omega_Z$, $\Omega_\phi/\Omega_Z$, $\Omega_R/\Omega_Z$, and $\Omega_R/\Omega_\phi$ for stars on the red (right) and blue (left) arms of the two-armed $V_R$ spiral in the Coma Berenices moving group region (see Figure 17). Blue histograms display the same quantities for all stars in the corresponding radial range, shown for comparison. First and second rows: $8.25 < R < 9$ kpc, with orange corresponding to the red arm (first row) and blue arm (second row). Third and fourth rows: $9 < R < 9.75$ kpc, similarly for the red and blue arms. All orange histograms refer to stars with $7.8 < R_g < 8.6$ kpc (the P3 $V_R$ stripe).

**Table 5**
Median and Standard Deviation of the Frequencies and Frequency Ratios of Stars in the Red Arm and Blue Arm of Figure 17

| | Median (std., 8r) (km $s^{-1}$ $kpc^{-1}$) | Median (std., 8b) (km $s^{-1}$ $kpc^{-1}$) | Median (std., 9r) (km $s^{-1}$ $kpc^{-1}$) | Median (std., 9b) (km $s^{-1}$ $kpc^{-1}$) |
|---|---|---|---|---|
| $\Omega_R$ (km $s^{-1}$ $kpc^{-1}$) | 38.2 (0.99) | 37.76 (0.96) | 37.68 (0.88) | 37.64 (0.9) |
| $\Omega_\phi$ (km $s^{-1}$ $kpc^{-1}$) | 28.28 (0.68) | 27.87 (0.66) | 27.96 (0.6) | 27.97 (0.62) |
| $\Omega_Z$ (km $s^{-1}$ $kpc^{-1}$) | 57.67 (3.1) | 50.15 (2.77) | 64.63 (5.3) | 69.85 (7.8) |
| $\Omega_R/\Omega_Z$ | 0.66 (0.026) | 0.757 (0.03) | 0.585 (0.04) | 0.54 (0.06) |
| $\Omega_R/\Omega_\phi$ | 1.35 (0.0034) | 1.35 (0.003) | 1.347 (0.004) | 1.35 (0.004) |
| $\Omega_\phi/\Omega_Z$ | 0.49 (0.019) | 0.56 (0.02) | 0.434 (0.03) | 0.4 (0.046) |

**Notes.** The frequencies have been converted to physical units for comparison with the pattern speeds of the spiral arms and the bar.
In this table, (8r) and (8b) refer to the red and blue arms in Figure 17 for the radial range $8.25 < R < 9$ kpc, while (9r) and (9b) denote the red and blue arms for $9 < R < 9.75$ kpc. The frequencies are given in physical units.

resonance condition $m(\Omega_\phi - \Omega_P) + l\Omega_Z + n\Omega_R = 0$, we test the candidate pattern speeds $\Omega_P = 12$ and $28\ \mathrm{km\,s^{-1}\,kpc^{-1}}$ for the spiral arms, and 37.5, 41, and $52\ \mathrm{km\,s^{-1}\,kpc^{-1}}$ for the bar, restricting to small integers $|m| \leqslant 4$, $|l| \leqslant 2$, and $|n| \leqslant 1$ to identify vertical resonances. For $\Omega_Z = 57\ \mathrm{km\,s^{-1}\,kpc^{-1}}$, the resonance condition is satisfied for the bar (pattern speed $\Omega_P = 37.5\ \mathrm{km\,s^{-1}\,kpc^{-1}}$) with $(m, l, n) = (2, 1, -1)$. For $\Omega_Z = 70\ \mathrm{km\,s^{-1}\,kpc^{-1}}$, the resonance condition is satisfied for the spiral arm (pattern speed $\Omega_P = 12\ \mathrm{km\,s^{-1}\,kpc^{-1}}$) with $(m, l, n) = (2, -1, 1)$.

### *3.3. Backward Orbit Integration of Stars in the Coma Berenices Moving Group Region with 9 < R < 9.75 kpc*

Among the moving groups in the P3 and D2 $V_R$ stripes we studied, the stars in the Coma Berenices moving group region are the only ones that exhibit a clear two-armed $V_R$ spiral in the $Z$–$V_Z$ plane. In particular, within the radial range $9 < R < 9.75$ kpc, the stars display a symmetric two-armed spiral that is more continuous and simpler than the spiral observed for $8.25 < R < 9$ kpc. Therefore, we selected this subset of data to integrate the orbits of these stars backward in time, in order to further understand the formation mechanism of the two-armed $V_R$ spiral in the Coma Berenices moving group region.

We performed the orbit integration using the gravitational potential from the `galpy` package (J. Bovy 2015). Since we only integrate backward for 100 Myr, we adopt approximately static models for the bar and spiral arms in the gravitational potential. The total potential includes the Galaxy-like background potential (`MWPotential2014`), a spiral arm potential (`SpiralarmPotential`; D. P. Cox & G. C. Gómez 2002), and a bar potential (`DehnenBarPotential`; W. Dehnen 2000; G. Monari et al. 2016). The bar's major axis makes an angle of $\phi_{\rm bar} = 25^\circ$ with the Galactic center–Sun line, lying to the left of that line (i.e., in the direction of increasing Galactic longitude, which is also the direction of Galactic rotation). The bar pattern speed is set to $37.5\ \mathrm{km\,s^{-1}\,kpc^{-1}}$, and its length to 5 kpc. For the spiral arms, we use a four-armed model and adjust the parameters of `SpiralArmsPotential` so that the simulated arms best overlap with the Sagittarius–Carina, Local, Perseus, and Outer arms in the Milky Way. The adopted parameters are `amp` = 1, `N` = 4, `alpha` = –0.15, `Cs` = [1.0, 0.5, 0.2], `Rs` = 0.8, and `H` = 0.3. Here, `amp` is the amplitude of the spiral arm density perturbation, `N` is the number of arms, `alpha` is the pitch angle, `Cs` are the constants multiplying the $\cos(n\gamma)$ terms in the potential function, `Rs` is the radial scale length of the density perturbation, and `H` is its vertical scale height. A schematic diagram of the model spiral arm potential is shown in Figure 19.

We integrated the orbits of the stars in the Coma Berenices region with $9 < R < 9.75$ kpc backward in time over the past 100 Myr, recording positions at intervals of 25 Myr. Figure 20 shows the distribution of these stars in the $X$–$Y$ plane at each time step. We also compared the positions of these stars relative to the Sagittarius–Carina Arm, Local Arm, and Perseus Arm, adopting a common pattern speed of $28\ \mathrm{km\,s^{-1}\,kpc^{-1}}$ for the spiral arms (W. S. Dias et al. 2019). For each look-back time, Table 6 lists the median Galactocentric radius $R$, as well as the median and standard deviation of the distances from these stars to the three spiral arms.

From Figure 20 and Table 6, the current orbital phases of these stars have just passed the apocenter, with a median

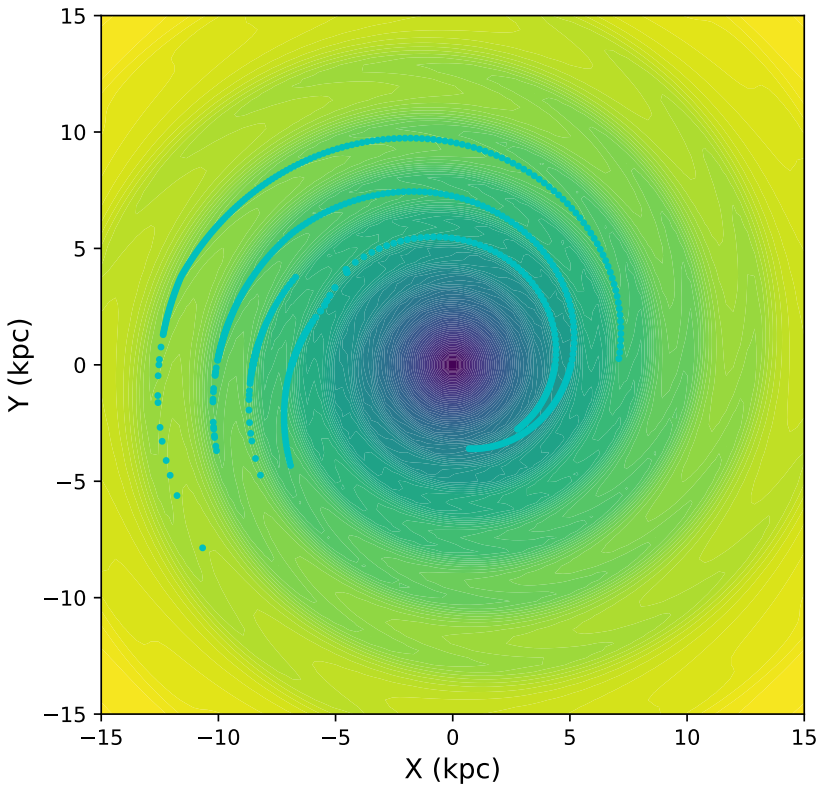

**Figure 19.** Schematic diagram of the `galpy` spiral arm potential (`SpiralarmPotential`), with the amplitude (`amp`) increased to 15 to make the spiral arms more distinct. The cyan curves, labeled from inside to outside, trace the Sagittarius–Carina Arm, Local Arm, Perseus Arm, and Outer Arm from M. J. Reid et al. (2019). The model spiral arms roughly coincide with these observed arms.

Galactocentric radius of $R = 9.24$ kpc. Relative to the spiral arms (M. J. Reid et al. 2019), they are located approximately between the Perseus Arm and the Local Arm. After integrating backward for 100 Myr, the stars are near the pericenter, with a median Galactocentric radius of $R = 7.3$ kpc, and lie close to the Sagittarius Arm. At look-back times of 25, 50, and 75 Myr, they are situated between pericenter and apocenter, with median Galactocentric radii of 9.2, 8.46, and 7.43 kpc, respectively. Therefore, as these stars move between pericenter and apocenter, they repeatedly cross the corotation radius of the spiral arms.

In Table 6, we compared the relative positions of this subset of the data with the spiral arms from M. J. Reid et al. (2019). In addition, M. Monguió et al. (2014) found that the Perseus Arm is located at a distance of approximately $1.6 \pm 0.2$ kpc from the Sun, based on the density and kinematic features of B4–A1-type stars in the anticenter direction. This distance is slightly smaller than the 2.0 kpc derived by M. J. Reid et al. (2014) from studies of star-forming regions. Moreover, Y. Xu et al. (2023) used masers and OB stars as tracers and found that the Perseus Arm has a Galactocentric distance of about 9.5–9.9 kpc near the anticenter. Adopting this location for the Perseus Arm, the stars in the Coma Berenices region with $9 < R < 9.75$ kpc are closer to the Perseus Arm than to the Local Arm.

### *3.4. Galactic Potential Models*

In our data, stars in the Coma Berenices moving group region within the P3 $V_R$ stripe exhibit a two-armed spiral in the $Z$–$V_Z$ plane. Such a symmetric vertical perturbation could be caused by either the bar or the spiral arms (J. A. S. Hunt et al. 2022; L. M. Widrow 2023). To investigate the influence of the bar and spiral arms on disk stars, we constructed five models of the Galactic gravitational potential based on the model of C. Li et al. (2023). All five models share the same axisymmetric background potential, built with the AGAMA self-consistent modeling framework (E. Vasiliev 2019). This background comprises a thin stellar disk, a thick stellar disk, a gaseous disk, a classical bulge, and a dark matter halo. Their density profiles are initially taken from parametric forms (exponential disks, a Plummer-like bulge, and a Navarro–Frenk–White halo) and then iteratively adjusted to achieve dynamical self-consistency with the corresponding distribution

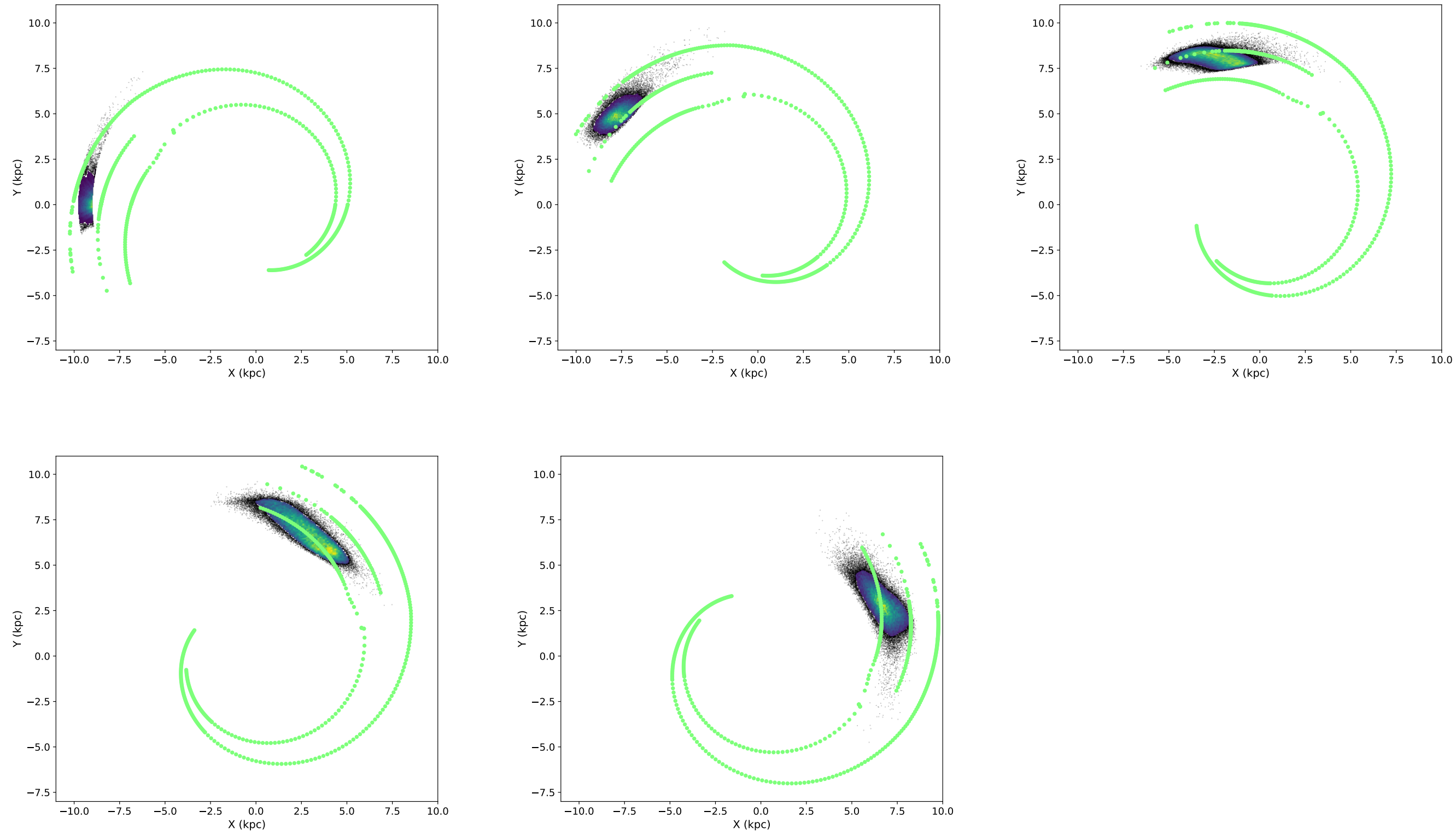

**Figure 20.** Distribution of stars in the Coma Berenices moving group region with $9 < R < 9.75$ kpc (see Figure 9) in the $X$–$Y$ plane, obtained by integrating their orbits backward in time. From left to right and top to bottom, the panels correspond to the present day and 25, 50, 75, and 100 Myr ago. The Sagittarius–Carina Arm, Local Arm, and Perseus Arm are also labeled. The spiral arm pattern speed is set to 28 km s$^{-1}$ kpc$^{-1}$.

**Table 6**
Median Distances and Standard Deviations From the Sagittarius–Carina, Local, and Perseus Arms, as Well as Median Galactocentric Radius $R$ and Orbital Phase $\theta_R$, for Stars in the Coma Berenices Region with $9 < R < 9.75$ kpc

| Look-back time (Myr) | Med. Dis. to Sag. (kpc) | Std. Dev. of Dis. to Sag. (kpc) | Med. Dis. to Loc. (kpc) | Std. Dev. of Dis. to Loc. (kpc) | Med. Dis. to Per. (kpc) | Std. Dev. of Dis. to Per. (kpc) | Med. of $\theta_R$ (rad) | Std. Dev. of $\theta_R$ (rad) | Med. of $R$ (kpc) | Std. Dev. of $R$ (kpc) |
|---|---|---|---|---|---|---|---|---|---|---|
| 0 | 2.33 | 0.24 | 0.74 | 0.23 | −0.78 | 0.23 | 3.56 | 0.47 | 9.24 | 0.2 |
| 25 | 1.93 | 0.38 | 0.4 | 0.32 | −1.139 | 0.32 | 2.62 | 0.47 | 9.2 | 0.356 |
| 50 | 1.22 | 0.316 | −0.31 | 0.34 | −1.85 | 0.32 | 2.02 | 0.7 | 8.46 | 0.48 |
| 75 | 0.2 | 0.413 | −1.21 | 0.37 | −2.78 | 0.41 | 0.55 | 0.47 | 7.43 | 0.54 |
| 100 | −0.7 | 0.52 | −1.42 | 0.57 | −2.95 | 0.59 | −0.09 | 0.565 | 7.3 | 0.44 |

**Notes.** Orbits were integrated backward at intervals of 25 Myr over the past 100 Myr; spiral arm positions are from M. J. Reid et al. (2019).
Negative median distances indicate that the stars are located on the inner side of the corresponding spiral arm; positive values indicate the outer side.

functions (quasi isothermal for the disks, and double-power law for the bulge and halo). The final axisymmetric potential is a cylindrical-spline representation that accurately reproduces the rotation curve and vertical force profile of a Milky Way–like galaxy.

On top of this common background, we superimposed different nonaxisymmetric perturbations (a bar and/or spiral arms) to investigate how stellar orbits respond to various dynamical disturbances. The simulations were integrated forward for 8 Gyr, and the trajectories of all test particles were recorded. The nonaxisymmetric components of each model are listed below.

*Model 1: bar only*. The nonaxisymmetric component consists only of an elongated, two-component bar, with no spiral arms. The bar evolves as follows: its pattern speed starts at $\Omega_{\rm bar} = 88\,{\rm km\,s^{-1}\,kpc^{-1}}$ and decreases nonlinearly to $37.5\,{\rm km\,s^{-1}\,kpc^{-1}}$ over the first 6 Gyr. Its strength and size grow simultaneously: the mass multiplication factor increases from 1 to 2.5, and the radial multiplication factor from 1 to 1.5, after which they remain constant beyond 6 Gyr.

*Model 2: static spiral*. The nonaxisymmetric component consists only of a four-armed logarithmic spiral (D. P. Cox & G. C. Gómez 2002), without a bar. Its pattern speed is constant at $\Omega_{\rm sp} = 28\,{\rm km\,s^{-1}\,kpc^{-1}}$, giving a corotation radius of about 8.65 kpc, close to the solar orbit. The amplitude of the surface density perturbation is set to 50% of the background disk surface density. The spiral persists throughout the entire integration period (0–8 Gyr) with constant strength and constant rotation rate.

*Model 3: bar + static spiral*. The nonaxisymmetric components consist of an evolving bar (identical to Model 1) and a static four-armed logarithmic spiral (identical to Model 2).

*Model 4: transient spiral*. The nonaxisymmetric component consists only of a four-armed logarithmic spiral (D. P. Cox &

**Table 7**
Overview of the Five Dynamical Models

| Model | Nonaxisymmetric Component | Bar Prescription (Incl. Pattern Speed and Evolution) | Spiral Prescription (Incl. Pattern Speed and Evolution) |
|---|---|---|---|
| 1 | Bar only | Dual long bar (C. Wegg et al. 2015), $\Omega_{\rm p}$: $-88 \rightarrow -37.5$ (0–6 Gyr), mass × 2.5, scale × 1.5 (0–6 Gyr), $\rho_{\rm bar} \propto {\rm sech}^2\left(\frac{z}{h_z}\right)\exp\left[-\left(\left(\frac{\lvert x\rvert}{x_0}\right)^{c_\perp}+\left(\frac{\lvert y\rvert}{y_0}\right)^{c_\perp}\right)^{c_\parallel/c_\perp} - \left(\frac{R}{R_{\rm out}}\right)^{s_{\rm out}} - \left(\frac{R_{\rm in}}{R}\right)^{s_{\rm in}}\right]$ | ⋯ |
| 2 | Spiral only (steady) | ⋯ | Four-arm logarithmic spiral (D. P. Cox & G. C. Gómez 2002), $\Omega_{\rm sp} = -28$ km s$^{-1}$ kpc$^{-1}$ (constant), constant amplitude, $\Phi_{\rm sp} = -4\pi G\Sigma_0 e^{-R/R_d}\sum_{n=1}^{3}\frac{C_n}{D_n K_n}\cos(n\gamma)\,{\rm sech}^{B_n}\left(\frac{K_n z}{B_n}\right)$ with $\gamma = m\left[\phi - \frac{\ln(R/R_d)}{\tan i_p} - \phi_0\right]$, $m=4$, $i_{\rm p} = \pi/8$, $\Sigma_0 = 0.5\Sigma_{\rm disk}$, $\Sigma_{\rm disk} = 3\times10^9\,M_\odot\,{\rm pc}^{-2}$ |
| 3 | Bar + static spiral | Same as Model 1 | Same as Model 2 |
| 4 | Spiral only (transient) | ⋯ | Same analytic form as Model 2, but: $\Sigma_0 = 0.4\Sigma_{\rm disk}$; transient envelope $A(t)$: $A(t) = \begin{cases} t/2 & 0 \leqslant t < 2\ {\rm Gyr} \\ 1-(t-2)/2 & 2 \leqslant t < 4\ {\rm Gyr} \\ 0 & t \geqslant 4\ {\rm Gyr}\end{cases}$ applied as $\Phi_{\rm sp}(t) = A(t)\cdot\Phi_{\rm sp,static}$ |
| 5 | Bar + transient spiral | Same as Model 1 | Same as Model 4 |

**Note.** "Bar" and "Spiral" columns indicate the presence of each component; the corresponding prescriptions, pattern speeds and temporal evolutions are given within the same cells.

G. C. Gómez 2002), without a bar. The spiral has a constant pattern speed $\Omega_{\rm sp} = 28\,{\rm km\,s^{-1}\,kpc^{-1}}$, giving a corotation radius of about 8.65 kpc, close to the solar orbit. The surface density perturbation amplitude is set to 40% of the background disk surface density. The transient envelope is defined as follows: a linear rise from zero to its maximum amplitude over 0–2 Gyr, a linear decay back to zero over 2–4 Gyr, after which the spiral vanishes completely.

*Model 5: bar + transient spiral*. The nonaxisymmetric components include an evolving bar (identical to Model 1) and a transient four-armed spiral (same amplitude as Model 4, but with the time window shifted later). The spiral's time window is as follows: linear growth from 2 to 4 Gyr, linear decay from 4 to 6 Gyr, after which it disappears completely. The total potential consists of the axisymmetric background, the evolving bar, and the transient spiral.

The formulas and parameters of the spiral arms and bar used in Models 1–5 are listed in Table 7.

For Models 1–5, we bin the results in azimuthal angle $\phi$ with a bin width of 40° and in angular momentum $L_Z$ with a bin width of 200 kpc km s$^{-1}$, and examine the distribution of radial velocity $V_R$ in the $Z$–$V_Z$ plane. Figure 21 shows typical $V_R$ distributions in the $Z$–$V_Z$ plane obtained from the orbital integrations for each model.

The top-left panel of Figure 21 shows the result for Model 1 (bar only). The stellar particles were initially located near the midplane. This panel displays a snapshot after 3 Gyr of integration: instead of remaining confined near $Z \sim 0$ kpc with small vertical oscillations, a significant fraction of stars are "kicked" to larger $|Z|$ and $|V_Z|$, forming two distinct clumps in the $Z$–$V_Z$ plane. This is a clear signature of vertical resonant heating or buckling instability induced by the bar. Furthermore, the vertical motions of stars captured by the bar's resonance are roughly synchronized, lacking a systematic phase variation with $Z$. Therefore, the bar produces localized clumps in the $Z$–$V_Z$ plane, rather than a continuous smooth spiral.

The top middle panel shows a typical result for Model 2 (static spiral only). This model does not exhibit a two-armed spiral in the $Z$–$V_Z$ plane over the 8 Gyr integration period.

The top-right panel shows a result for Model 3 (bar + static spiral). In this example after 3 Gyr of integration, the bar excites vertical resonances while the static spiral enhances phase mixing, leading to a two-armed spiral in the $V_R$ distribution in the $Z$–$V_Z$ plane.

The bottom-left panel shows the result for Model 4 (transient spiral only). The transient spiral grows from 0 Gyr, reaching its maximum amplitude at 2 Gyr. This panel displays a snapshot at 2 Gyr. Even a purely transient spiral can generate a two-armed $V_R$ spiral in the $Z$–$V_Z$ plane. This occurs because the rotating spiral potential imposes a periodic radial force that varies with vertical height $Z$ (through the ${\rm sech}^{B_n}(K_n Z/B_n)$ term). When a star crosses a spiral arm, the radial velocity impulse it receives depends on its current vertical position $Z$ and vertical velocity $V_Z$. Consequently, $V_R$ develops a well-defined pattern in the $Z$–$V_Z$ plane, imprinting a phase-dependent structure onto the vertical phase space.

The bottom-right panel shows the result for Model 5 (transient spiral + bar). Even with both components superposed, a clear two-armed $V_R$ spiral remains visible in the $Z$–$V_Z$ plane. From Model 4 (bottom-left panel), we see that the transient spiral alone can drive such a spiral, while Model 1 (top-left panel) shows that the bar provides vertical kicks that eject stars from the midplane. The superposition of the two perturbations still produces a distinct two-armed $V_R$ spiral.

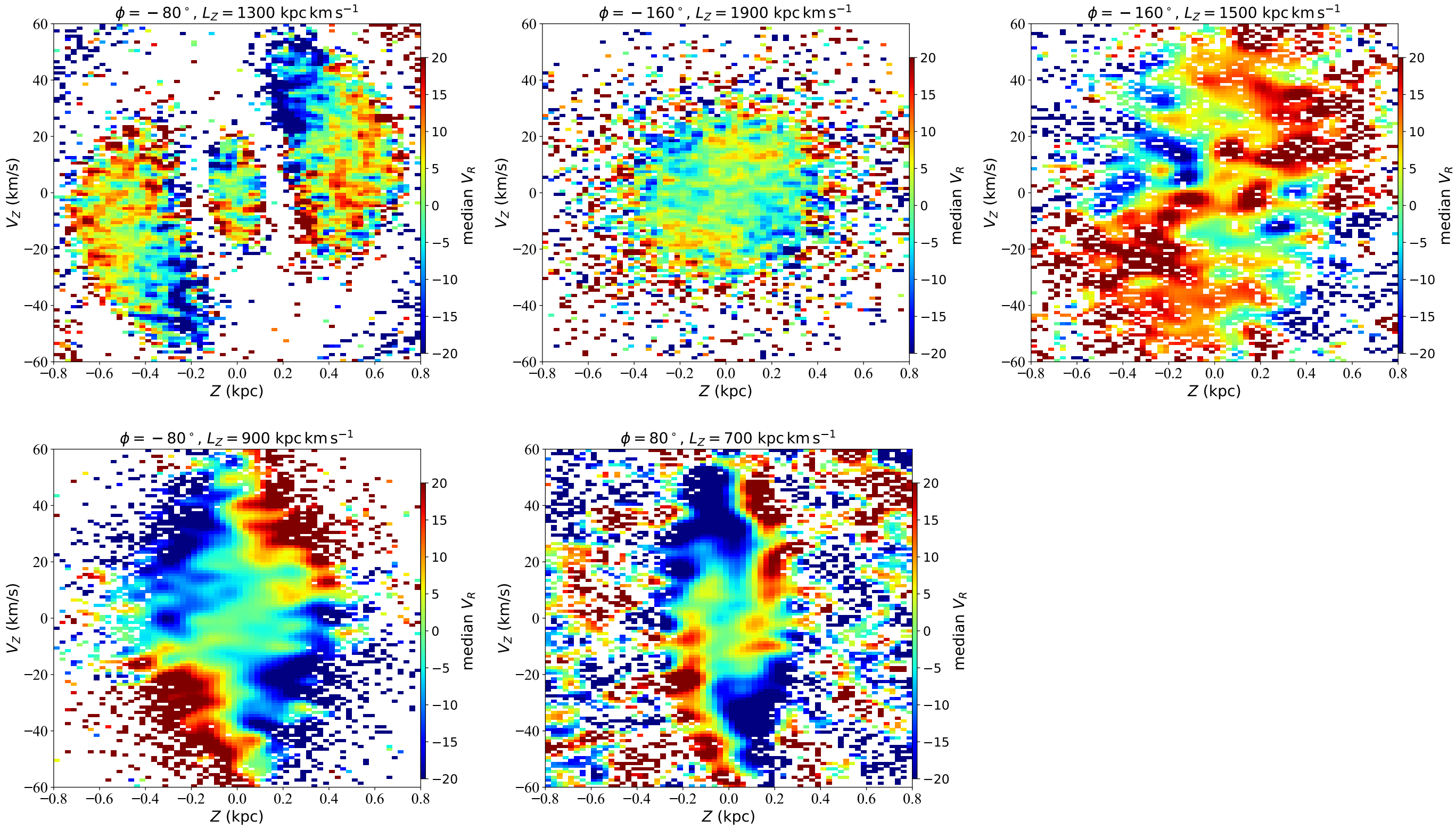

**Figure 21.** $V_R$ distributions in the $Z$–$V_Z$ plane from the orbital integrations. Top left: Model 1 (bar only) at 3 Gyr, for a slice with $-100^\circ < \phi < -60^\circ$ and $L_Z$ = 1200–1400 kpc km s$^{-1}$. Top middle: Model 2 (static spiral) at 3 Gyr, with $-180^\circ < \phi < -140^\circ$ and $L_Z$ = 1800–2000 kpc km s$^{-1}$. Top right: Model 3 (bar + static spiral) at 3 Gyr, with $-180^\circ < \phi < -140^\circ$ and $L_Z$ = 1400–1600 kpc km s$^{-1}$. Bottom left: Model 4 (transient spiral) at 2 Gyr, with $-100^\circ < \phi < -60^\circ$ and $L_Z$ = 800–1000 kpc km s$^{-1}$. Bottom right: Model 5 (bar + transient spiral) at 4 Gyr, with $60^\circ < \phi < 100^\circ$ and $L_Z$ = 600–800 kpc km s$^{-1}$.

### *3.5. Comparing with Previous Studies*

In a previous study, C. Li et al. (2023) found that among 33 million stars with radial velocity measurements from Gaia DR3, two-armed phase spirals are only observed when $L_Z < 1800$ kpc km s$^{-1}$. In contrast, in our data, stars within the P3 $V_R$ stripe, which have $1868 < L_Z < 2040$ kpc km s$^{-1}$, still exhibit a clear two-armed $V_R$ spiral in the $Z$–$V_Z$ plane. There are several possible reasons for these differences. First, we used different samples. Our sample was selected from the crossmatch between LAMOST DR7 and Gaia DR3, whereas the sample of C. Li et al. (2023) was drawn from the Gaia RVS catalog. Second, we used different radial velocities: we employed those measured by LAMOST, while they used those provided by Gaia RVS. Additionally, we adopted distances based on parallaxes from C. A. L. Bailer-Jones et al. (2018, 2023) in Gaia DR3, while C. Li et al. (2023) used distances based on the GSP-Phot Aeneas best-fit values (R. Andrae et al. 2023). These differences in sample selection, radial velocity, and distance estimation may account for the distinct kinematic features.

From our data, the Coma Berenices moving group is currently the only one that exhibits a clear two-armed $V_R$ spiral in the $Z$–$V_Z$ phase space. Therefore, in this section we explore and compare previous studies on the origin of the two-armed spiral structure in the $Z$–$V_Z$ phase space and on the origin of the Coma Berenices moving group.

The two-armed spiral in phase space typically corresponds to the breathing mode of disk stars (L. M. Widrow 2023). In simulations, several types of perturbations can generate a breathing mode, such as the rapid one-way passage of a dwarf galaxy through the Galactic disk. The simulations of T. Asano et al. (2025) show that the perturbation from the Sagittarius dwarf galaxy not only directly excites the bending mode in the Galactic disk but also further excites the breathing mode by inducing the formation of spiral arms. These breathing modes manifest as two-armed phase spirals in the inner Galaxy. Perturbations from the bar and spiral arms can also generate a breathing mode. In various simulations, both density-wave spiral arms and transient spiral arms can induce a breathing mode and can partially explain the observed kinematic distribution. V. P. Debattista (2014) used *N*-body simulations to investigate the three-dimensional density of density-wave spiral arms and the vertical velocities of stars. Inside the corotation radius, vertical motions are compressive on the trailing side of the spiral arm, while outside this radius, they are compressive on the leading side. S. Ghosh et al. (2022) found that the amplitude of breathing motions excited by spiral structures increases with height above the midplane and decreases with stellar age at a given height; these findings align with Gaia DR2 data and suggest a driving role for spiral density waves. T. Khachaturyants et al. (2022)'s numerical simulations revealed that the pattern speed of the breathing waves is consistent with that of the spiral density waves, suggesting that the breathing waves may be driven by the spiral arm structure. A. Kumar et al. (2022) showed through $N$-body simulations that tidally induced spiral arms are the primary drivers of large-scale vertical breathing motions in galactic disks. Additionally, the kinematic characteristics of stars reveal that the Perseus Arm is in a dissipating phase with expansive vertical motion, whereas the Outer and Local Arms are in a growth phase with compressive vertical motion. These observations are well explained by transient spiral arms, as demonstrated in studies by J. Baba et al. (2018), A. Widmark et al. (2022), N. Funakoshi et al. (2024), T. Asano et al. (2024), and X. Liu et al. (2025).

The two-armed spiral is associated with the breathing mode, but the breathing mode alone is not a sufficient condition for the formation of a two-armed spiral in phase space (C. Li et al. 2023; S. Tremaine et al. 2023; R. Chiba et al. 2025). The internal perturber must grow or decay rapidly, on a timescale comparable to the vertical oscillation period of the stars (U. Banik et al. 2022, 2023).

C. Li et al. (2023) showed that a central bar and spiral arms with constant pattern speeds can generate nonzero mean radial velocities ($V_R$), but they do not produce two-armed phase spirals. When a central bar with a time-varying pattern speed was introduced, however, two-armed phase-spiral structures emerged, suggesting that changes in pattern speed may be the key factor in generating such spirals.

The simulations of J. A. S. Hunt et al. (2022) indicate that if the Gaia two-armed phase spiral is caused by a spiral arm or bar, these structures are likely transient and possess a symmetric vertical profile. At the same time, they note that the existence of such rapidly evolving spiral arms or bars requires further observational confirmation.

R. Chiba et al. (2025) pointed out that the period of the internal perturbation is longer than the vertical oscillation period of the stars. Nevertheless, even for a stable spiral arm, the resonance of the perturbation can break the adiabaticity of the inner disk, and diffusion driven by small-scale stochastic kicks prevents complete phase mixing, which can result in an open two-armed phase spiral.

Meanwhile, the perturbations that drive the breathing motion can also create arcs and clumps in the $V_R$–$V_\phi$ plane. Most previous studies have suggested that the formation of the Coma Berenices moving group is related to spiral arms. D. A. Barros et al. (2020) found that the stars of the Coma Berenices moving group mainly originated from the Local Arm or the Sagittarius–Carina Arm. Based on an analysis of the stellar angular momentum distribution, S. Khoperskov & O. Gerhard (2022) considered the Coma Berenices stream to be part of the large-scale stellar density structures in the Milky Way disk, likely associated with spiral arm or bar resonances. Gaia data further indicate that this moving group has significant radial extensions, suggesting a nonlocal origin. T. A. Michtchenko et al. (2018) employed a dynamical model constrained by observational data to analyze the stellar velocity distribution in the solar neighborhood and suggested that the formation mechanisms of the Coma Berenices and Pleiades moving groups may involve multiple factors, including the corotation resonance of the spiral arms.

G. Monari et al. (2018) combined RAVE and Gaia DR2 data to reveal incomplete vertical phase mixing in the vertical motion of the Coma Berenices moving group and explored its connection to the resonance effects of the Galactic bar. They also pointed out that this incomplete phase mixing might be caused by perturbations from the Sagittarius dwarf satellite galaxy. Moreover, they found that the Coma Berenices moving group formed no more than ∼150 Myr ago, a timescale that coincides with the peri-Galactic passage of the Sagittarius dwarf galaxy. This suggests that external perturbations have a significant impact on the vertical velocity distribution of the Coma Berenices moving group.

## 4. Summary

The most notable observational facts are summarized as follows.

1. The median $V_R$ as a function of $R_g$ alternates between positive and negative values, forming distinct $V_R$ stripes.
2. These positive and negative-$V_R$ stripes are strongly influenced by moving groups. Table 4 lists the moving groups embedded in each $V_R$ stripe.
3. The D1, P2, D2, and P3 $V_R$ stripes exhibit a two-armed phase spiral in the $Z$–$V_Z$ plane ($6.05 < R_g < 8.6$ kpc). In these stripes, the two-armed spiral is observed only for stars in the orbital phase range $\pi/2 < \theta_R < 3\pi/2$. Section 2.5 discusses the impact of selection effects on the $V_R$ distributions in the $Z$–$V_Z$ plane. We conclude that, for the P3 $V_R$ stripe, no clear two-armed spiral is present in the $\theta_R < \pi/2$ or $\theta_R > 3\pi/2$ intervals.
4. In the P3 $V_R$ stripe, the Coma Berenices moving group exhibits a clear two-armed spiral, which is the primary contributor to the left arm of the overall P3 spiral. The Hyades–Pleiades moving group shows a faint single-armed spiral in the inner radial bin ($8.25 < R < 9$ kpc) and a curled pattern resembling a two-armed spiral in the outer bin ($9 < R < 9.75$ kpc); together, these features form the right arm and part of the left arm of the P3 phase spiral.
5. In the D2 $V_R$ stripe, the two-armed phase spiral is not an independent structure but a composite: the inner loop of the Hyades–Pleiades expansion tail contributes to the left arm, while the outer tail of this expansion combined with the short arm from the Hercules 1 moving group forms the right arm.

Thus, the phase-space spirals observed in these $V_R$ stripes are the result of overlapping contributions from different moving groups, each with its own resonance origin. The arm features identified in the $Z$–$V_Z$ plane for each moving group region are summarized in Table 8.

Among the moving groups studied, only the Coma Berenices moving group exhibits a clear two-armed spiral in the $Z$–$V_Z$ plane. In Section 3, we further explored the properties of these stars.

1. They display a two-armed $V_R$ spiral in the $Z$–$V_Z$ plane; stars located on the arms have higher $L_Z$ and lower $J_R$, consistent with the corotation resonance of the spiral arms.
2. The median frequencies of these stars are $\Omega_\phi \approx 28\,\mathrm{km\,s^{-1}\,kpc^{-1}}$ and $\Omega_R/\Omega_\phi \approx 4/3$. These frequencies can be explained either by the spiral corotation resonance ($\Omega_{sp} = 28\,\mathrm{km\,s^{-1}\,kpc^{-1}}$) or by the $m = 4$ inner Lindblad resonance of the bar ($\Omega_{bar} = 37.5\,\mathrm{km\,s^{-1}\,kpc^{-1}}$).
3. Backward orbit integration shows that these stars move between apocenter and pericenter, repeatedly crossing

**Table 8**
Arm Features of the Moving Groups in the $Z$–$V_Z$ Plane

| Moving Group | Full $R$ Range | $8.25 < R < 9$ kpc | $9 < R < 9.75$ kpc |
|---|---|---|---|
| Hyades–Pleiades (P3) | Single-armed spiral | Single-armed spiral | No clear feature |
| Coma Berenices (P3) | Two-armed spiral | Two-armed spiral | Two-armed spiral |
| Hercules 1 (D2) | Short single-armed spiral | $V_R$ blob | No clear feature |
| Hyades–Pleiades expansion tail (D2) | Single-armed spiral | Single-armed spiral | Suspected two-armed feature |

the spiral corotation radius, which reinforces the resonance origin of the phase spiral.

4. Test-particle simulations demonstrate that a bar with a time-varying pattern speed, combined with either a static or a transient spiral, can produce a two-armed spiral in the $Z$–$V_Z$ plane, consistent with the conclusions drawn from the frequency distributions of these stars.

## Acknowledgments

This work is supported by the National Natural Science Foundation of China (NSFC) under grant No. 12322306, and a science research grant from the China Manned Space Project with No. CMS-CSST-2021-B03. This work made use of the data from LAMOST (Large Sky Area Multi-Object Fiber Spectroscopic Telescope, also known as the Guoshoujing Telescope).[6] LAMOST is a Chinese national megascience facility, operated by National Astronomical Observatories, Chinese Academy of Sciences.

## Appendix A
## Analysis of the $V_R$ Distribution in the $Z$–$V_Z$ Phase Space as a Function of Eccentricity and Orbital Phase

Here, we divide the data according to eccentricity range and orbital phase into the following categories: $e > 0.2$, $e < 0.2$, $\pi/2 < \theta_R < 3\pi/2$, and $\theta_R < \pi/2$ or $\theta_R > 3\pi/2$, as summarized in Table A1. We then examine the $V_R$ distribution in the $Z$–$V_Z$ plane for each of these intervals; the results are shown in Figures A1, A2, A3, and A4.

**Table A1**
Eccentricity and Orbital Phase Ranges Corresponding to the Data Shown in Figures A1–A4

| Figure | Eccentricity Range | $\theta_R$ Range |
|---|---|---|
| Figure A1 | $e < 0.2$ | $\pi/2 < \theta_R < 3\pi/2$ |
| Figure A2 | $e > 0.2$ | $\pi/2 < \theta_R < 3\pi/2$ |
| Figure A3 | $e < 0.2$ | $\theta_R < \pi/2$ or $\theta_R > 3\pi/2$ |
| Figure A4 | $e > 0.2$ | $\theta_R < \pi/2$ or $\theta_R > 3\pi/2$ |

[6] https://cstr.cn/31118.02.LAMOST

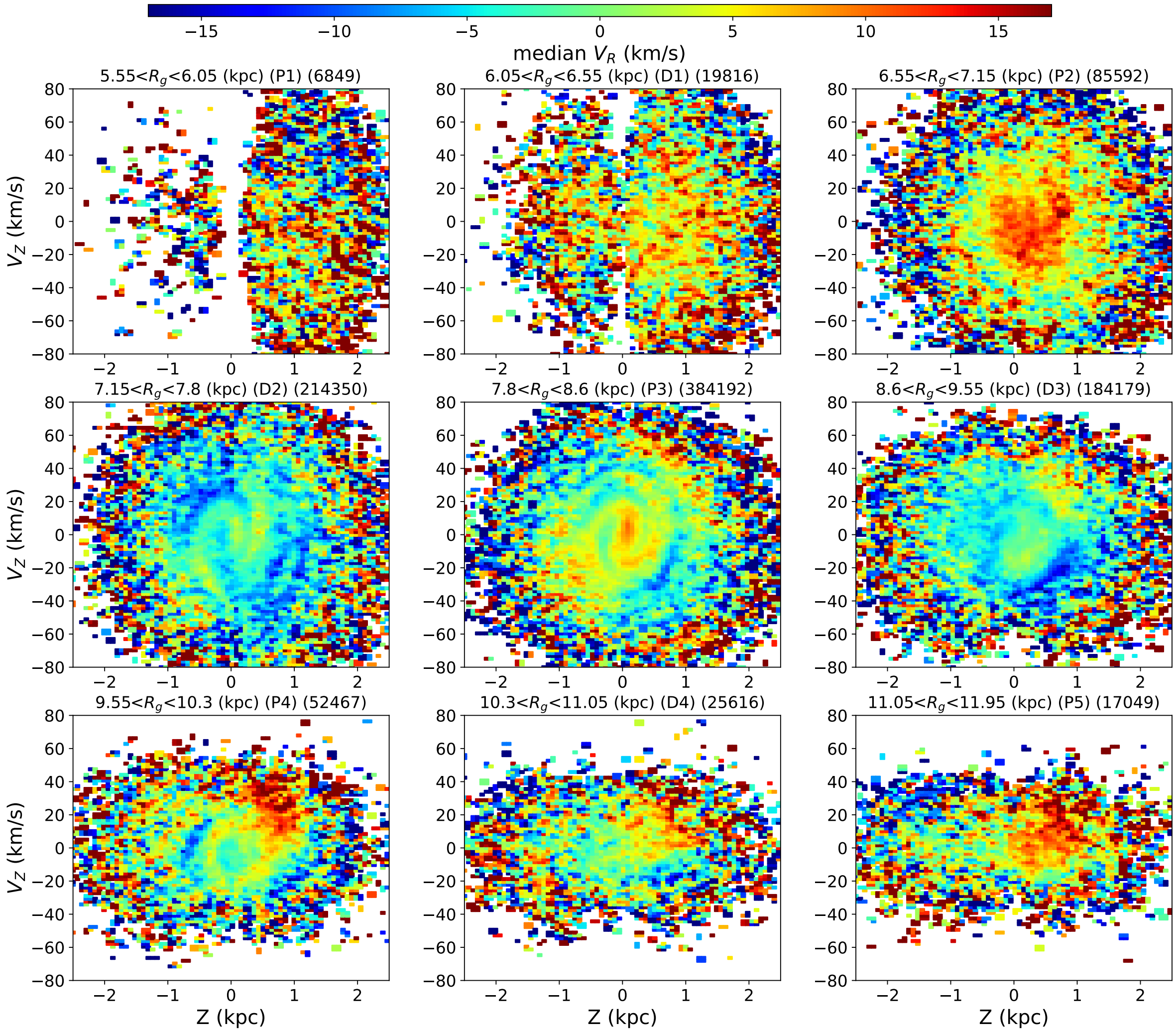

**Figure A1.** $V_R$ distribution in the $Z$–$V_Z$ phase space for each $V_R$ stripe shown in Figure 1, restricted to stars with $e < 0.2$ and $\pi/2 < \theta_R < 3\pi/2$. Each panel lists the $R_g$ range, the $V_R$ stripe name, and the number of stars. Two-armed $V_R$ spirals are visible in the D2 and P3 $V_R$ stripes.

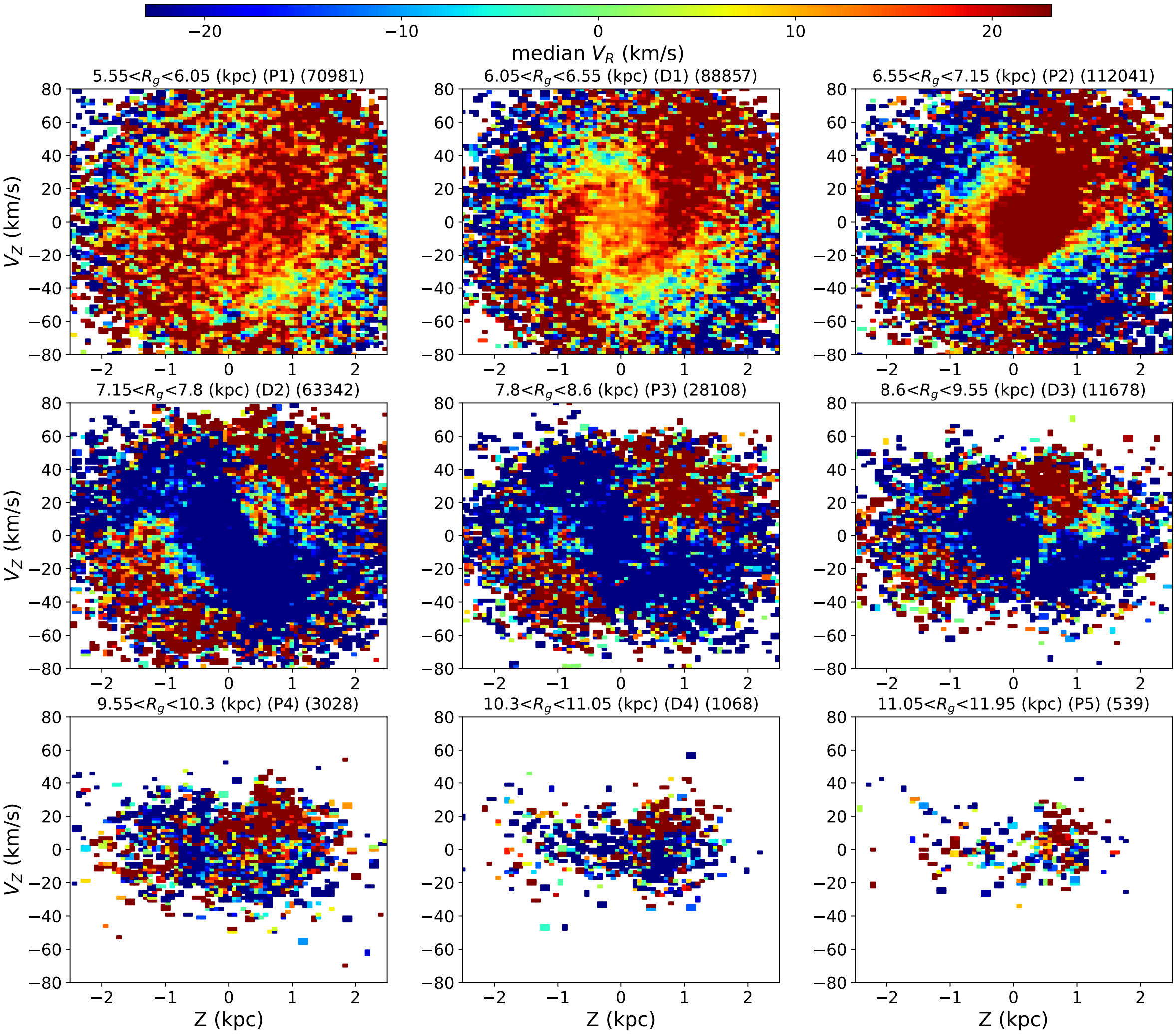

**Figure A2.** $V_R$ distribution in the $Z$–$V_Z$ plane for each $V_R$ stripe shown in Figure 1, restricted to stars with $e > 0.2$ and $\pi/2 < \theta_R < 3\pi/2$. Each panel lists the $R_g$ range, the $V_R$ stripe name, and the number of stars. Two-armed $V_R$ spirals are visible in the D1 and P2 $V_R$ stripes.

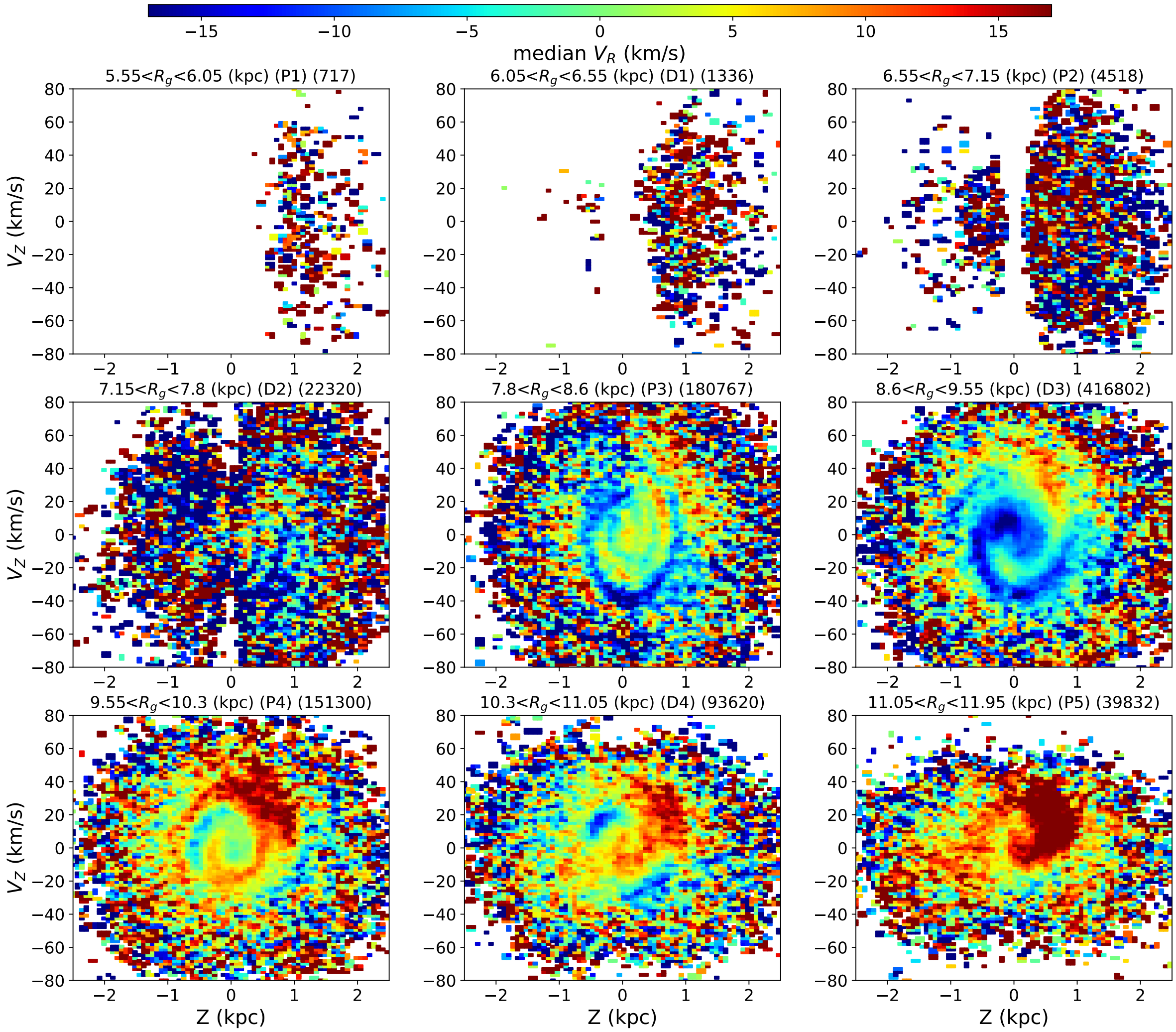

**Figure A3.** $V_R$ distribution in the $Z$–$V_Z$ plane for each $V_R$ stripe shown in Figure 1, restricted to stars with $e < 0.2$ and $\theta_R < \pi/2$ or $\theta_R > 3\pi/2$. Each panel lists the $R_g$ range, the $V_R$ stripe name, and the number of stars.

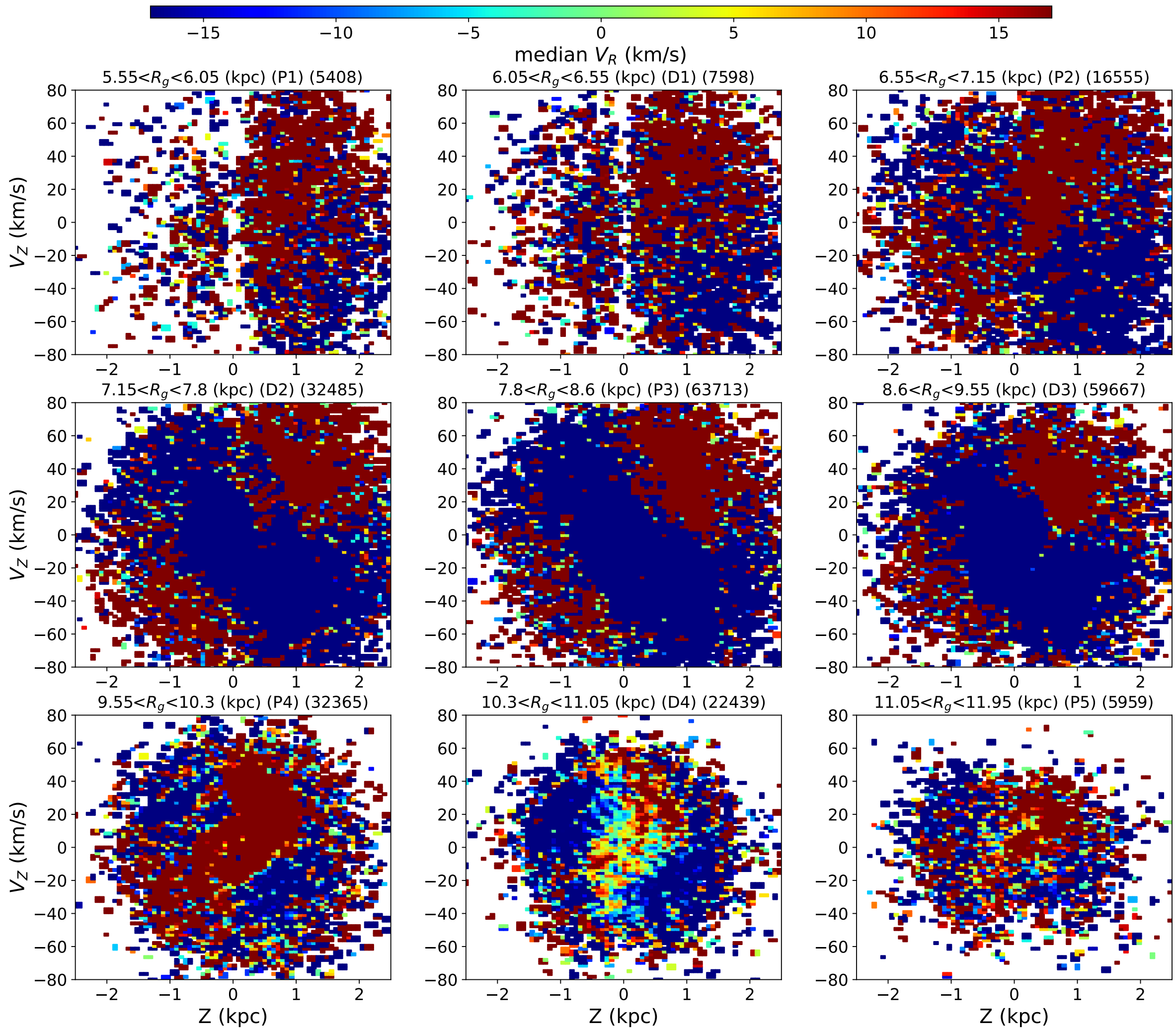

**Figure A4.** $V_R$ distribution in the $Z$–$V_Z$ plane for each $V_R$ stripe shown in Figure 1, restricted to stars with $e > 0.2$ and $\theta_R < \pi/2$ or $\theta_R > 3\pi/2$. Each panel lists the $R_g$ range, the $V_R$ stripe name, and the number of stars.

## Appendix B
## Investigating the Impact of Selection Effects on the Visibility of Two-armed Spirals Using Simulated Data

In this section, we perform orbit integrations using test-particle simulations to model the perturbations acting on disk stars. We then mimic the selection effects of the LAMOST survey by resampling the simulated data according to the observed $R$ distribution, in order to investigate whether and how this resampling process weakens the two-armed spiral in the $Z$–$V_Z$ plane.

We have described the various simulation models in detail in Section 3.4; here we use Model 3. Figure B1 shows the $V_R$ distribution in the ($R_g$, $Z$) plane for stars in a wedge of the simulation data at 3 Gyr of orbital integration. As shown, the $V_R$ distribution exhibits stripe-like patterns that alternate between positive and negative values with changing $R_g$, qualitatively similar to those found in the observational data.

In the wedge of the simulation data shown in Figure B1, a pronounced two-armed spiral appears in the $Z$–$V_Z$ phase space within the range $3.4 < R_g < 5$ kpc, centered at $R_g \approx 4.2$ kpc. The first row of Figure B2 shows the stellar number counts and the $V_R$ distribution in the $Z$–$V_Z$ plane for a slice with $3.4 < R_g < 5$ kpc from this wedge. In this figure, a two-armed phase spiral is clearly visible. For convenience, its ridge curve has been manually traced and is indicated by the yellow curve.

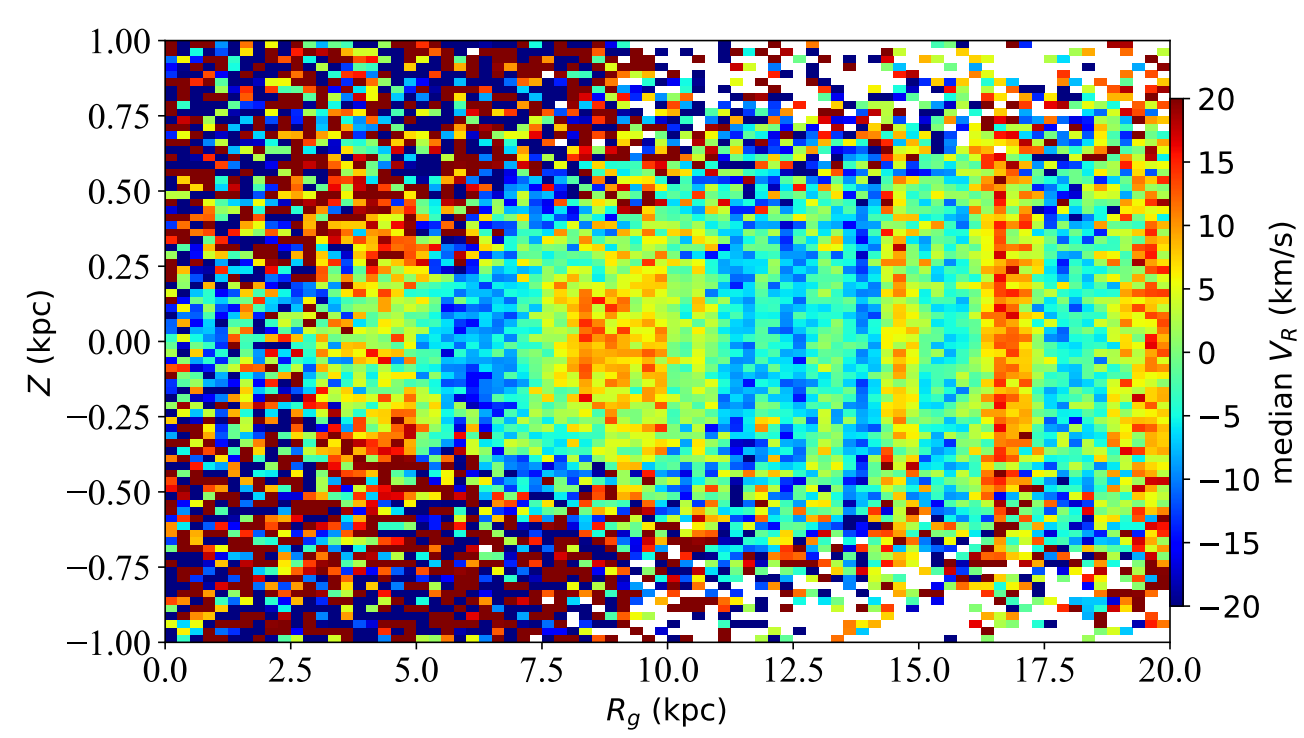

**Figure B1.** $V_R$ distribution in the ($R_g$, $Z$) plane for a Galactic wedge, from a snapshot of the simulation at 2 Gyr.

The second row of Figure B2 presents the $V_R$ distribution in the $Z$–$V_Z$ plane for the same model sample, divided by orbital phase into two subsamples: $\pi/2 < \theta_R < 3\pi/2$ and $\theta_R < \pi/2$ or $\theta_R > 3\pi/2$. It can be seen that the two-armed spiral in the lower-right panel is clear, whereas the two-armed spiral in the lower-left panel is faint.

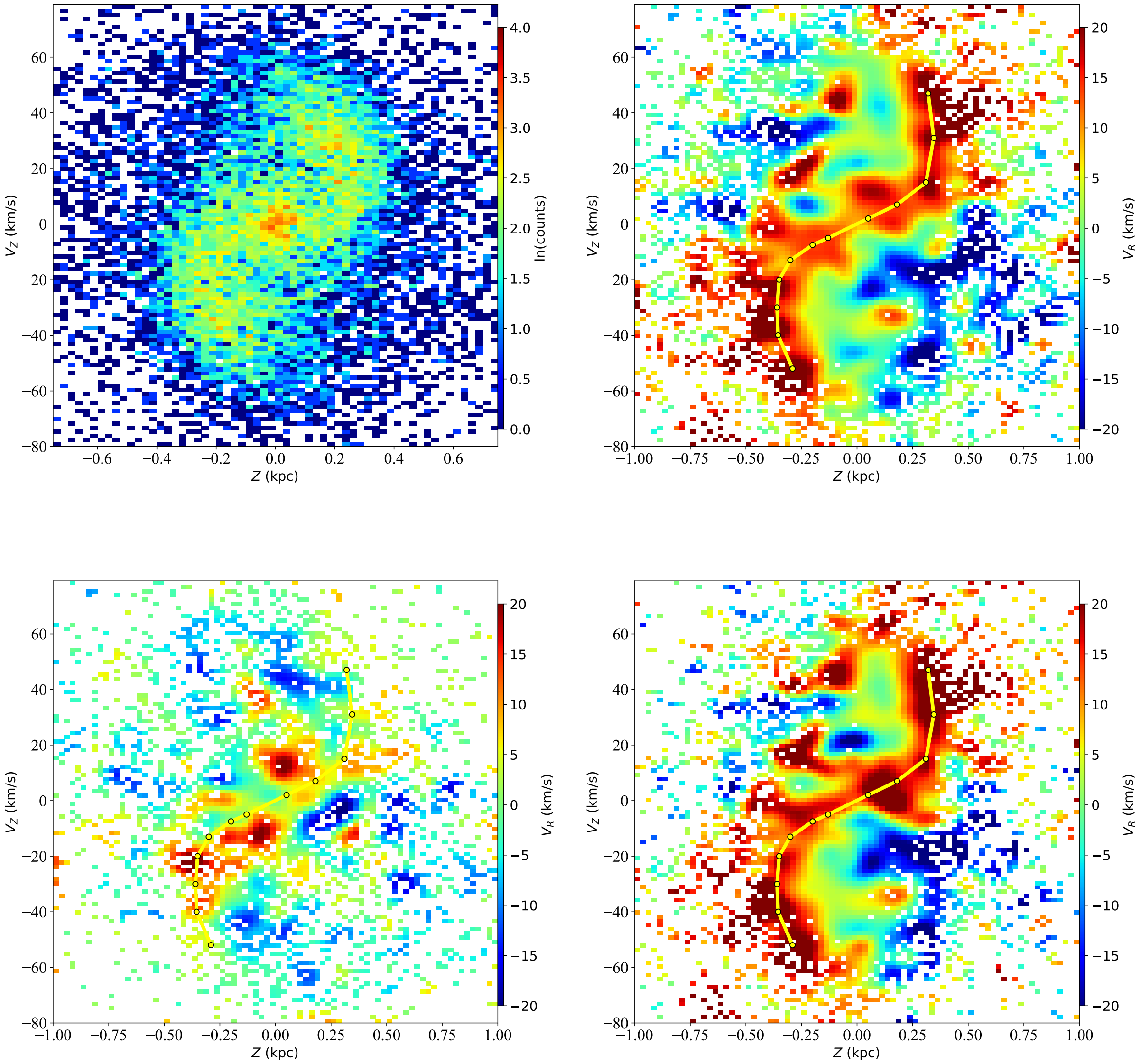

**Figure B2.** Stellar distributions for the wedge shown in Figure B1, restricted to $3.4 < R_g < 5$ kpc. Top row: stellar number counts (left) and $V_R$ distribution (right) in the $Z$–$V_Z$ plane. The $V_R$ distribution exhibits a clear two-armed spiral, with the ridge curves manually traced in yellow. Bottom row: $V_R$ distributions in the $Z$–$V_Z$ plane for the same stars, separated by orbital phase $\pi/2 < \theta_R < 3\pi/2$ (left) and $\theta_R < \pi/2$ or $\theta_R > 3\pi/2$ (right). The ridge curves from the top-right panel are overplotted for reference.

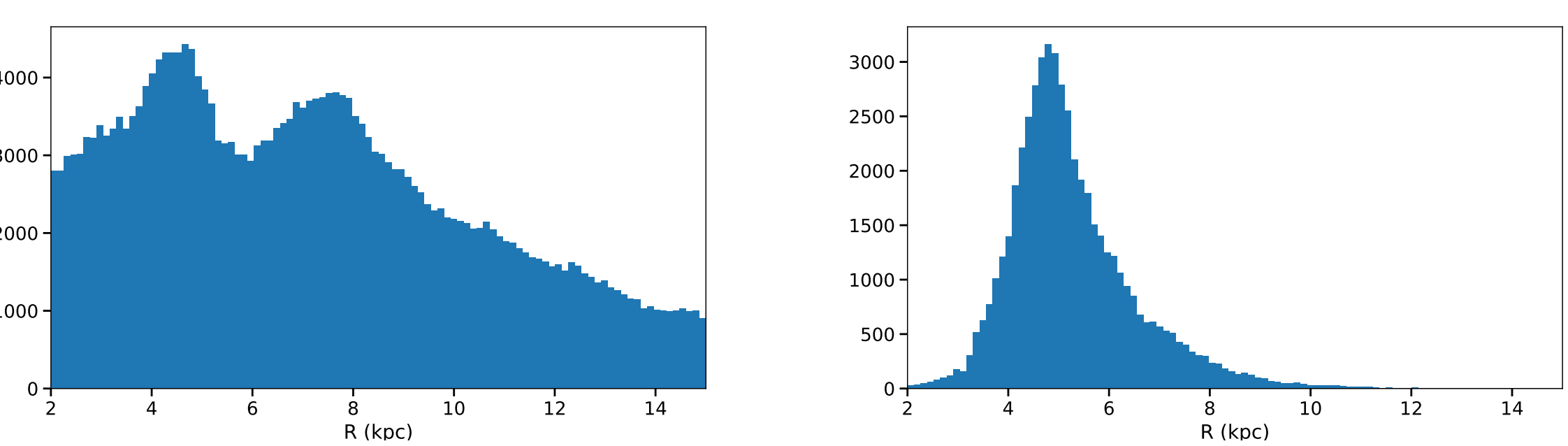

**Figure B3.** Stellar number distribution as a function of $R$ for the wedge shown in Figure B1. Left: original simulation data. Right: after resampling to match the observed $R$ distribution of LAMOST (Figure 5).

We then resample the simulation data according to the observed radial distribution to mimic the observational selection effects. The left panel of Figure B3 shows the stellar distribution as a function of $R$ for the wedge of simulation data shown in Figure B1. In the observational data, we focus primarily on how selection effects influence the two-armed spiral in the $Z$–$V_Z$ plane. Take the P3 $V_R$ stripe as an example, which spans $7.8 < R_g < 8.6$ kpc and is centered at $R_g = 8.2$ kpc. The $R$ distribution of the same observational data peaks at $R = 8.747$ kpc (Figure 5), which is offset by 0.547 kpc from the $R_g$ center of the P3 stripe. As described earlier, in the simulation wedge the two-armed spiral in the $Z$–$V_Z$ plane appears centered at $R_g \approx 4.2$ kpc. Correspondingly, we shift the peak of the observed $R$ profile to $4.2 + 0.547$ kpc and resample the data in Figure B1 accordingly. The right panel of Figure B3 shows the result of this resampling.

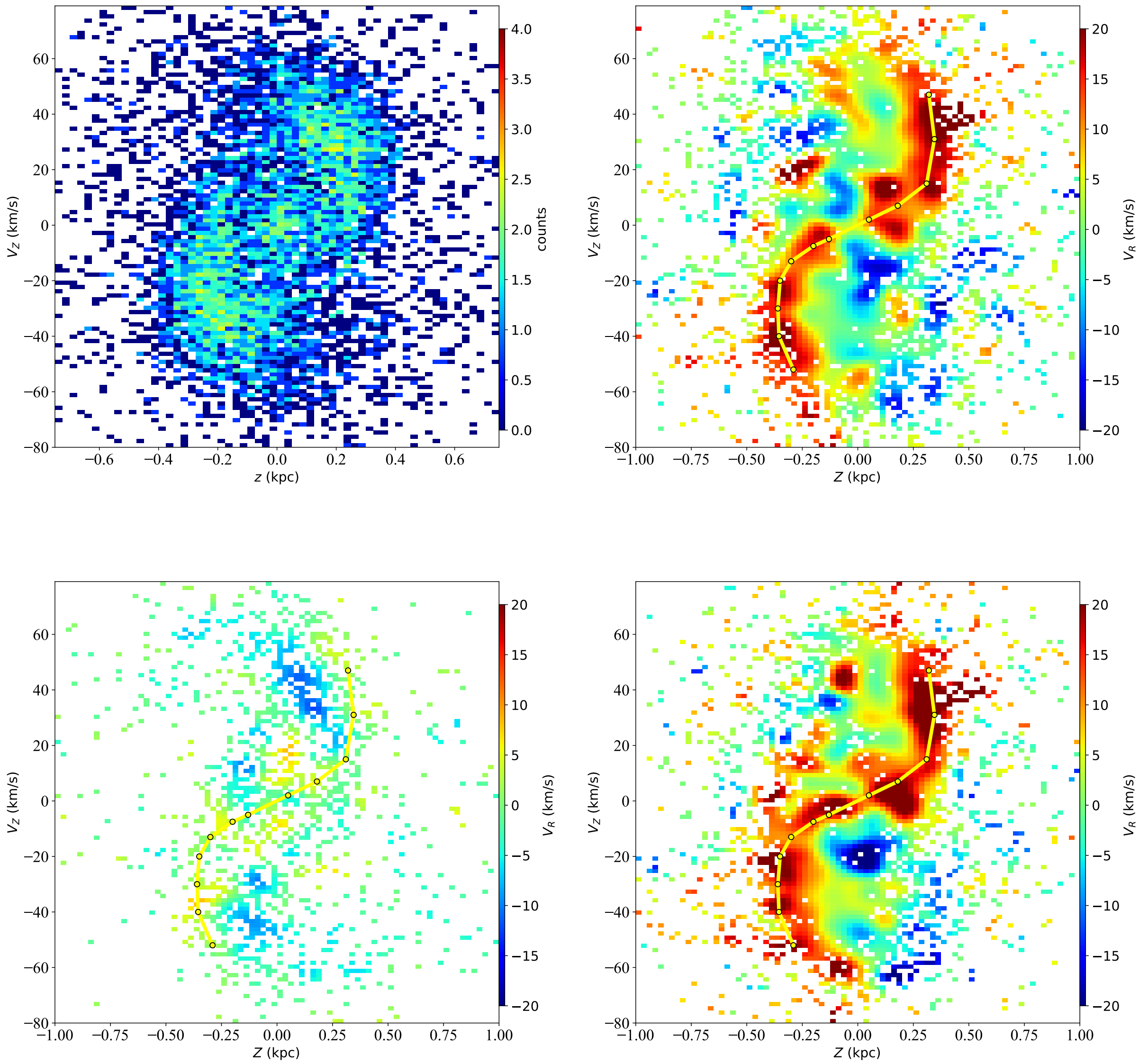

**Figure B4.** Stellar distributions for the wedge shown in Figure B1 within $3.4 < R_g < 5$ kpc, after resampling to match the observed $R$ distribution. Top row: number counts (left) and $V_R$ distribution (right) in the $Z$–$V_Z$ plane for the full resampled sample. Bottom row: $V_R$ distribution for the same stars, separated by orbital phase: $\pi/2 < \theta_R < 3\pi/2$ (left) and $\theta_R < \pi/2$ or $\theta_R > 3\pi/2$ (right). The manually traced ridge curves (yellow) from Figure B2 are overplotted on all $V_R$ panels.

For the resampled sample, we select the data within $3.4 < R_g < 5$ kpc and study the distribution in the $Z$–$V_Z$ plane. The first row of Figure B4 shows the stellar number counts (top-left panel) and the $V_R$ distribution (top-right panel) in the $Z$–$V_Z$ plane for the resampled sample in this $R_g$ range. The second row separates the data by orbital phase into $\pi/2 < \theta_R < 3\pi/2$ (bottom-left panel) and its complement, $\theta_R < \pi/2$ or $\theta_R > 3\pi/2$ (bottom-right panel). Comparing the panels corresponding to $\theta_R < \pi/2$ or $\theta_R > 3\pi/2$ in the two figures, a clear two-armed spiral is visible in Figure B2, and it remains clearly discernible in Figure B4. In contrast, the spiral pattern in the $\pi/2 < \theta_R < 3\pi/2$ panel of Figure B2, which was initially less distinct, becomes even weaker in Figure B4.

To assess the spatial similarity between the $V_R$ distribution and the ridge curve of the $V_R$ spiral, we calculated the normalized maximum cross-correlation coefficient via a Fourier transform. Table B1 lists, in the second column, the correlation coefficients between the $V_R$ distribution and the yellow ridge curve for each panel shown in Figures B2 and B4. The third column gives the ratio of each coefficient to the reference value of 0.7121 obtained from the full sample in $3.4 < R_g < 5$ kpc (first row of the table). Rows 2 and 3 list the correlation coefficients for the two phase subsamples ($\pi/2 < \theta_R < 3\pi/2$ and $\theta_R < \pi/2$ or $\theta_R > 3\pi/2$) as 0.46 and 0.68, corresponding to ratios of 64.8% and 95.5%, respectively. For the same phase subsamples after resampling (rows 5 and 6), the correlation coefficients become 0.167 and 0.65, with ratios of 23% and 91.2%, respectively.

**Table B1**
Correlation Coefficients between the $V_R$ Distribution and the Two-armed Spiral Ridge Curve for the Model Samples, and Their Ratios to the Reference Value (F1)

| Sample | Correlation Coefficient | Ratio to F1 |
|---|---|---|
| Full sample of $3.4 < R_g < 5$ kpc (F1) | 0.7121 | 100% |
| Phase I ($\pi/2 < \theta_R < 3\pi/2$) of F1 | 0.4611 | 64.8% |
| Phase II ($\theta_R < \pi/2$ or $\theta_R > 3\pi/2$) of F1 | 0.68 | 95.5% |
| Resampled sample of $3.4 < R_g < 5$ kpc (F2) | 0.726 | 102% |
| Phase I ($\pi/2 < \theta_R < 3\pi/2$) of F2 | 0.167 | 23% |
| Phase II ($\theta_R < \pi/2$ or $\theta_R > 3\pi/2$) of F2 | 0.65 | 91.2% |

From rows (3) and (6) of Table B1, it is evident that for the $\theta_R < \pi/2$ or $\theta_R > 3\pi/2$ subset, which originally showed a clear two-armed spiral, the spiral quality does not degrade significantly after resampling. In contrast, rows (2) and (5) show that for the $\pi/2 < \theta_R < 3\pi/2$ subset, which originally displayed only a faint spiral, the spiral quality is severely degraded.

These results demonstrate that resampling the simulation data to match the observed $R$ distribution preserves clear spiral features while strongly suppressing weaker ones.

## Appendix C Identification of Moving Group Boundaries Using $\Omega_Z$ and $J_Z$

Comparing the three columns of panels in Figure 8, we see that in the $V_R$–$R_g$ plane the distribution of number counts is very similar to that of $J_Z$. At the center of each substructure, the number counts peak and $J_Z$ is smallest. Within each substructure, $\Omega_Z$ is relatively uniform, whereas near the boundaries the $\Omega_Z$ value exhibits a clear jump. Therefore, we use the $\Omega_Z$ distribution to distinguish different moving groups.

To locate the $\Omega_Z$ jumps, we divide the right panel in the first row of Figure 8 (the $\Omega_Z$ distribution in the $V_R$–$R_g$ plane) into 30 bins along the $V_R$ axis and examine the variation of $\Omega_Z$ with $R_g$ within each bin. Figure C1 shows $\Omega_Z$ versus $R_g$ for each $V_R$ bin, with curves colored by the $V_R$ bin value. To more conveniently determine the positions of the $\Omega_Z$ jumps, we compute the mean $\Omega_Z$ as a function of $R_g$. In Figure C1, the black dots mark the mean $\Omega_Z$. We then subtract the linear fit (red line, $\Omega_Z = -0.3443R_g + 5.1033$) of the mean $\Omega_Z$ from the $\Omega_Z$–$R_g$ distribution in each bin.

The residuals obtained by subtracting the mean $\Omega_Z$ distribution from each $V_R$ bin in Figure C1 are shown in Figure C2. The gradient of the residuals is computed, and the locations of the minimum gradient (i.e., where $\Omega_Z$ changes most rapidly) within the $R_g$ range of each $V_R$ stripe are plotted in Figure C2 with colored markers.

The resulting $\Omega_Z$ jump positions are plotted as black dots in Figure C3. It can be seen that within the P3 $V_R$ stripe, the black dots form two distinct slanted lines, which clearly delineate the left boundary of the Coma Berenices moving group region and the boundary between the Coma Berenices and Hyades–Pleiades moving groups. We fitted these points with two straight lines, shown as black lines in Figure C3.

From the black dots in Figure C3, it is clear that the right boundary of the Hyades–Pleiades moving group region is not well defined by this method. This may be due to the differing slopes of the left and right boundaries. Therefore, we instead binned the data within $7.8 < R_g < 8.6$ kpc by $R_g$ and examined the $\Omega_Z$ distribution along $V_R$; the results are shown in Figure C4. In that figure, we also plot the gradient of $\Omega_Z$ with respect to $V_R$ for each $R_g$ bin and mark the gradient minima. The resulting $\Omega_Z$ jump positions (i.e., the locations of these minima) are then overplotted as gray dots in Figure C3. We fitted these points with a straight line, which is shown as a gray line in Figure C3.

In Figure C3, in addition to the moving groups within the P3 $V_R$ stripe, we are also interested in those within the D2 $V_R$ stripe. However, observing the black dots marking the $\Omega_Z$ jump positions, aside from the upper boundary of Hercules 1, the other boundaries of Hercules 1 and those of Hyades–Pleiades extension tail are not clearly defined. Since we observed in Figure 8 that the core regions of the moving groups have relatively small $J_z$, we plotted the $J_z$ distribution of the D2 $V_R$ stripe on the $V_R$–$R_g$ plane in Figure C5, and we drew contours for the $J_z$ distribution. The regions with relatively small $J_z$ values were then identified as polygons by selecting connected areas below a threshold of $J_z < 0.0028$ kpc km s$^{-1}$, applying Gaussian smoothing, and extracting the boundaries using contour detection. The resulting polygons clearly outline the core regions of the Hyades–Pleiades extension tail and Hercules 1 moving groups in the $V_R$–$R_g$ plane in Figure C3 and C5.

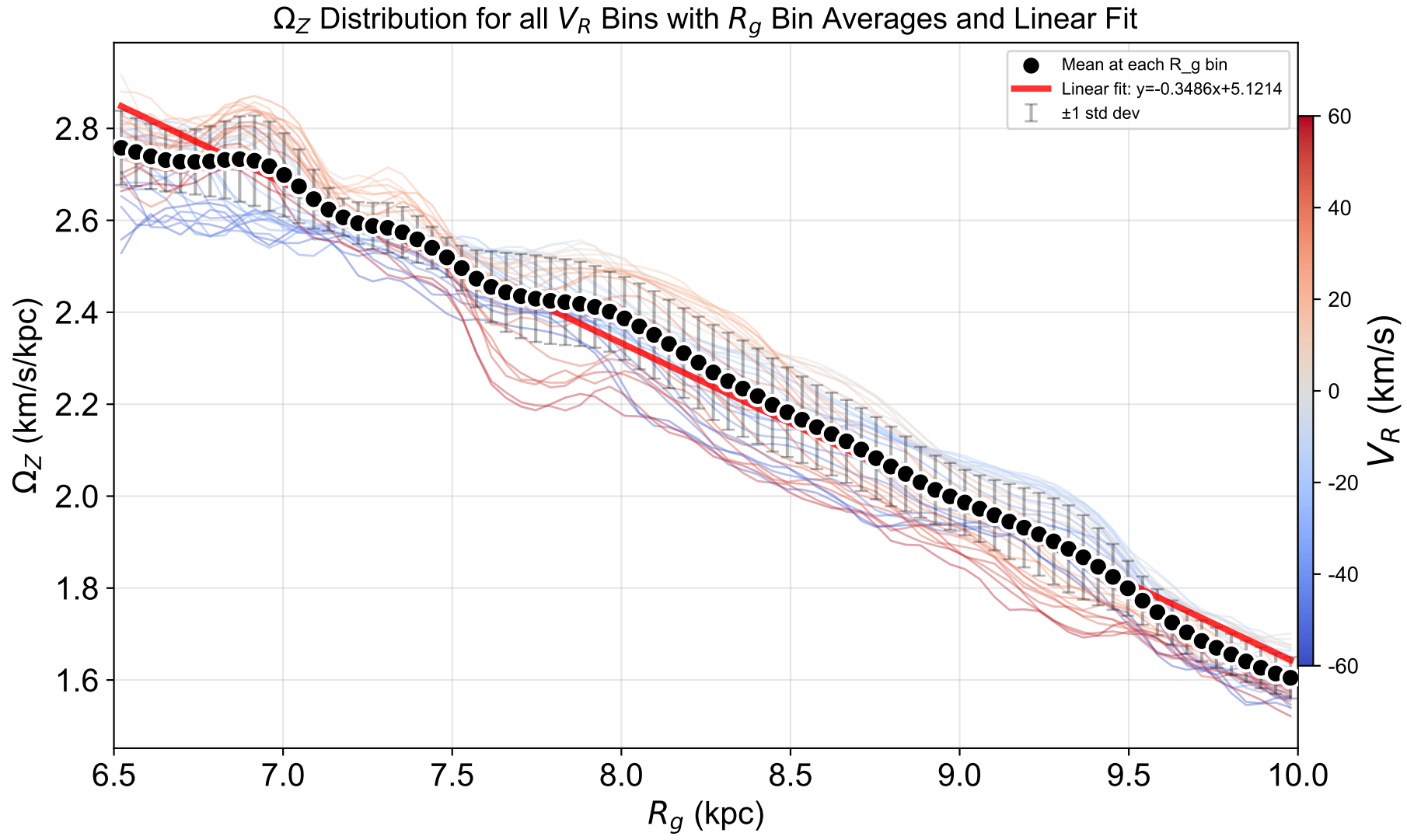

**Figure C1.** Stars in Figure 8 with Galactocentric radii $8.5 < R < 9$ kpc are divided into 30 bins in $V_R$. For each bin, $\Omega_Z$ as a function of $R_g$ is plotted as a colored curve, with the color representing the median $V_R$ of the bin (color bar range: −60 to 60 km s$^{-1}$). Black points show the median $\Omega_Z$ at each $R_g$ computed over all $V_R$ bins; error bars indicate the standard deviation. The red solid line is a linear fit to these median values ($\Omega_Z = -0.3443R_g + 5.1033$), illustrating the radial gradient of $\Omega_Z$ with $R_g$.

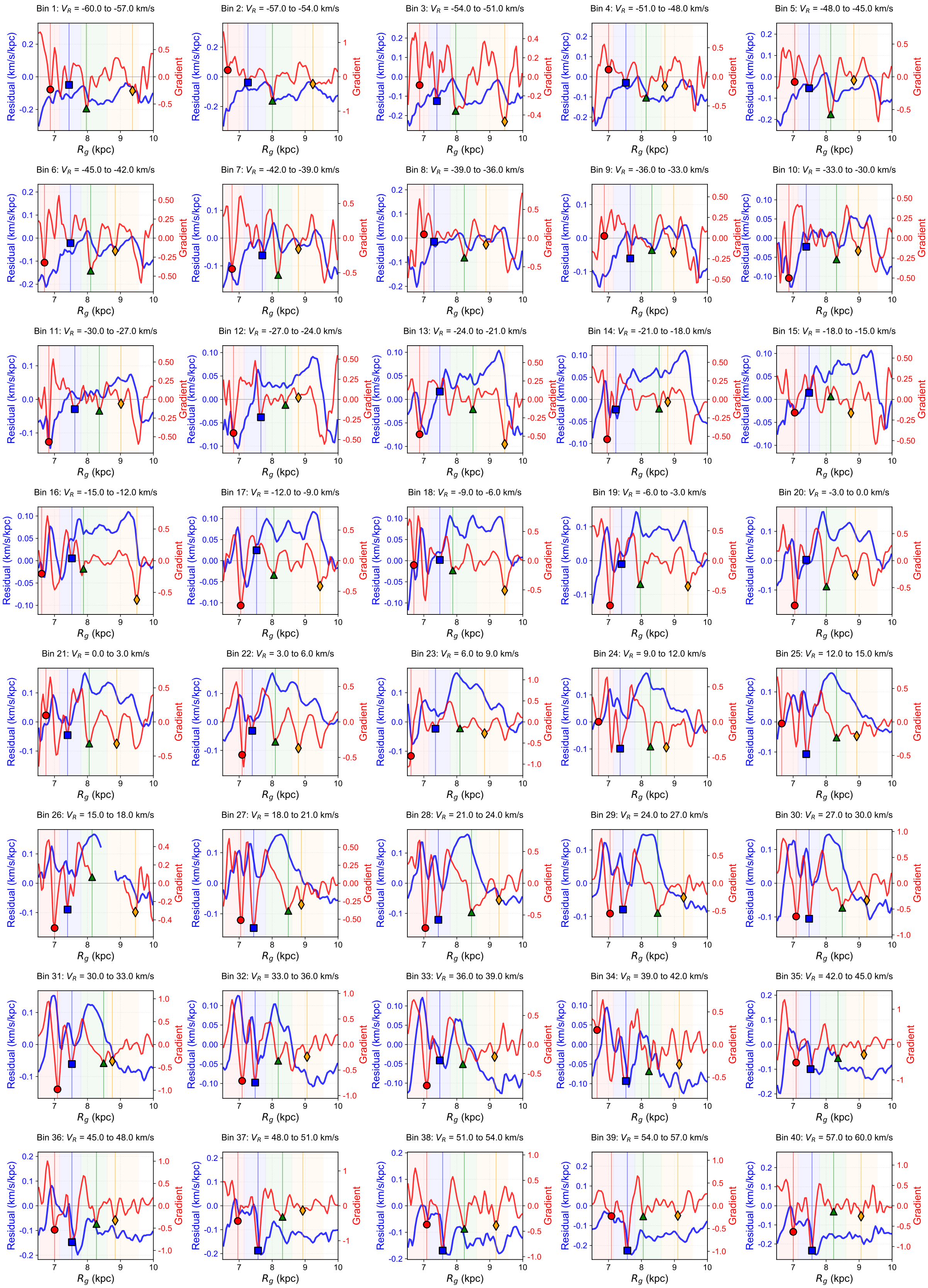

**Figure C2.** $\Omega_Z$ residuals and their gradients for stars in different $V_R$ bins. Blue curves show the $\Omega_Z$ residuals after subtracting the fitted linear trend $\Omega_Z = -0.3443 R_g + 5.1033$ (from Figure C1) from the $\Omega_Z$ vs. $R_g$ relation in each $V_R$ bin. Red curves show the gradients of these residuals. Colored markers indicate the positions of the gradient minima in four $R_g$ ranges corresponding to the $V_R$ stripes: P2 (6.55–7.15 kpc, red circles), D2 (7.15–7.8 kpc, blue squares), P3 (7.8–8.6 kpc, green triangles), and D3 (8.6–9.55 kpc, orange diamonds).

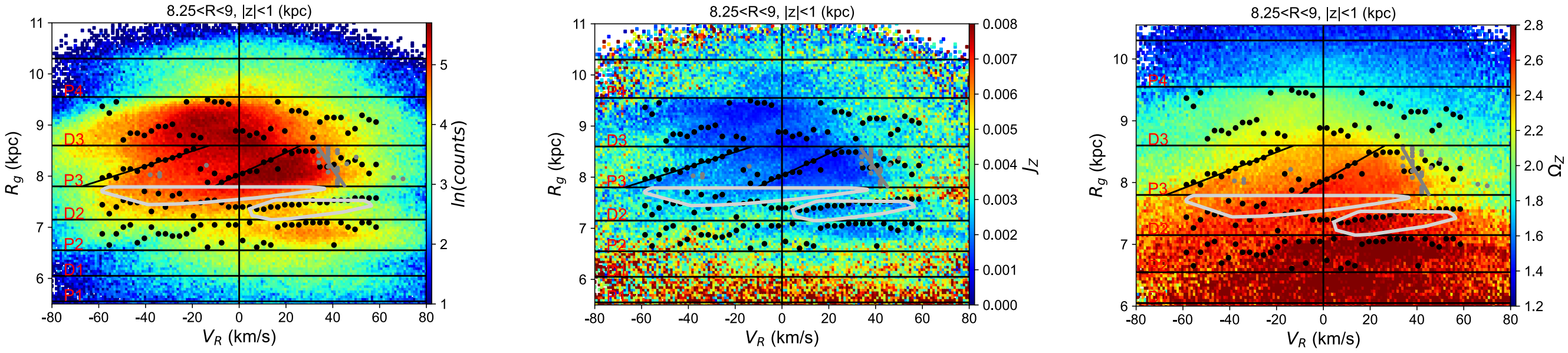

**Figure C3.** Distributions of number counts, $J_Z$, and $\Omega_Z$ in the $(V_R, R_g)$ plane for data with $8.25 < R < 9$ kpc. Black dots mark the $\Omega_Z$ gradient minima from Figure C2. Within $7.8 < R_g < 8.6$ kpc, these minima clearly delineate the left and right boundaries of the Coma Berenices moving group, which are fitted by the straight lines $R_g = 0.0146V_R + 8.784$ (left) and $R_g = 0.0215V_R + 8.031$ (right); the right boundary also separates the Coma Berenices and Hyades–Pleiades moving groups. Gray dots indicate the $\Omega_Z$ gradient minima that define the right boundary of the Hyades–Pleiades moving group, fitted by $R_g = 0.004V_R + 7.353$. In the range $7.15 < R_g < 7.8$ kpc, the $\Omega_Z$ minima do not reliably constrain the boundaries of the Hercules 1 and Hyades–Pleiades expansion tail moving groups; therefore, the $J_Z$ distributions in these regions are jointly examined. The gray polygon marks regions with median $J_Z < 0.0028$ kpc km s$^{-1}$ (derived from Figure C5), which are used to select stars likely belonging to these moving groups.

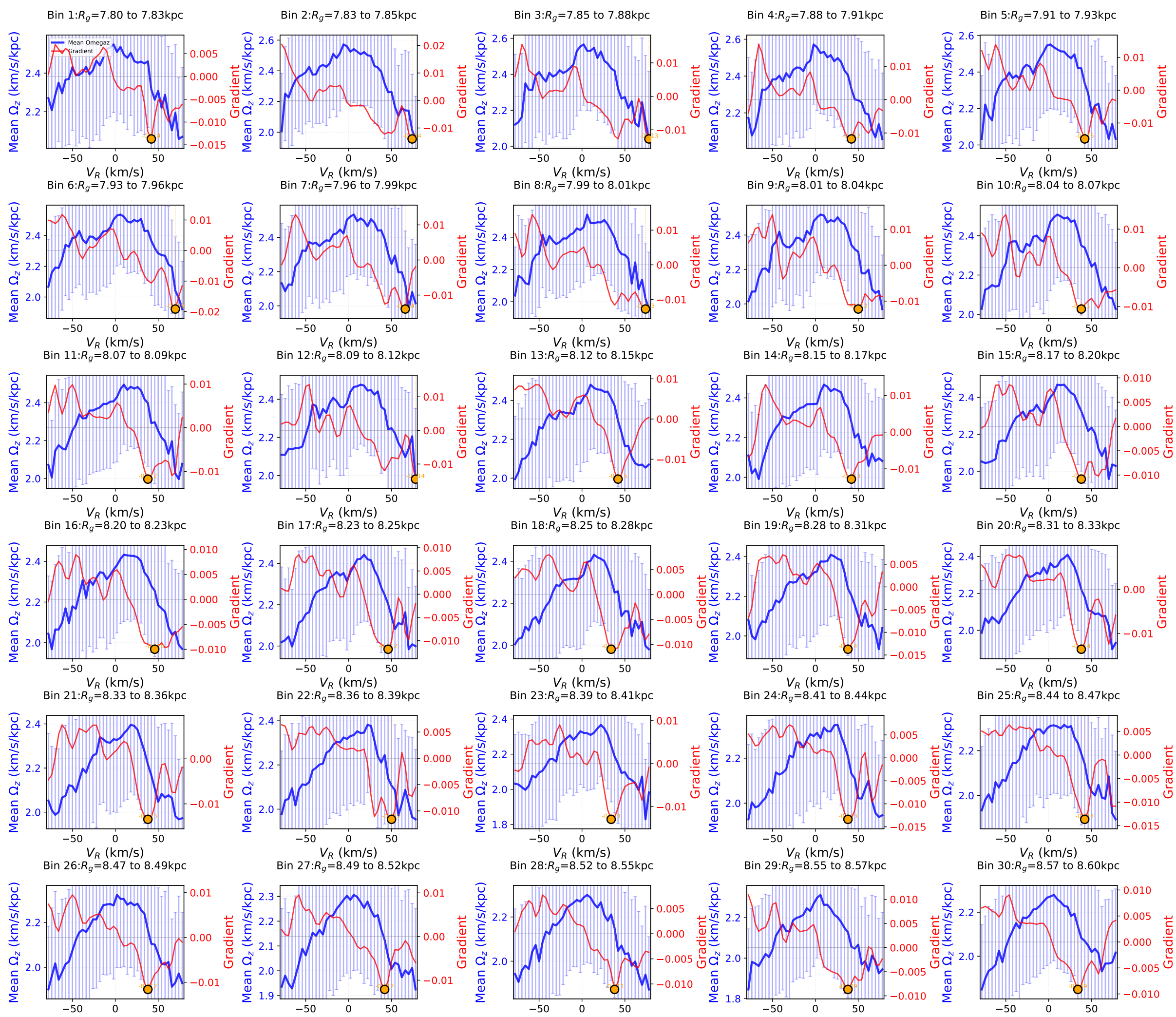

**Figure C4.** Stars with $R_g$ in the range 7.8–8.6 kpc are divided into 30 bins in $R_g$. Each panel shows the mean vertical frequency $\Omega_Z$ as a function of $V_R$ (blue curve) and its gradient with respect to $V_R$ ($d\Omega_Z/dV_R$, red curve). Blue error bars indicate the standard deviation of $\Omega_Z$ within each $V_R$ bin. Orange circles mark the minimum gradient values, and the horizontal gray dashed line indicates the zero-gradient reference. Panel titles give the corresponding $R_g$ bin number and the specific $R_g$ range.

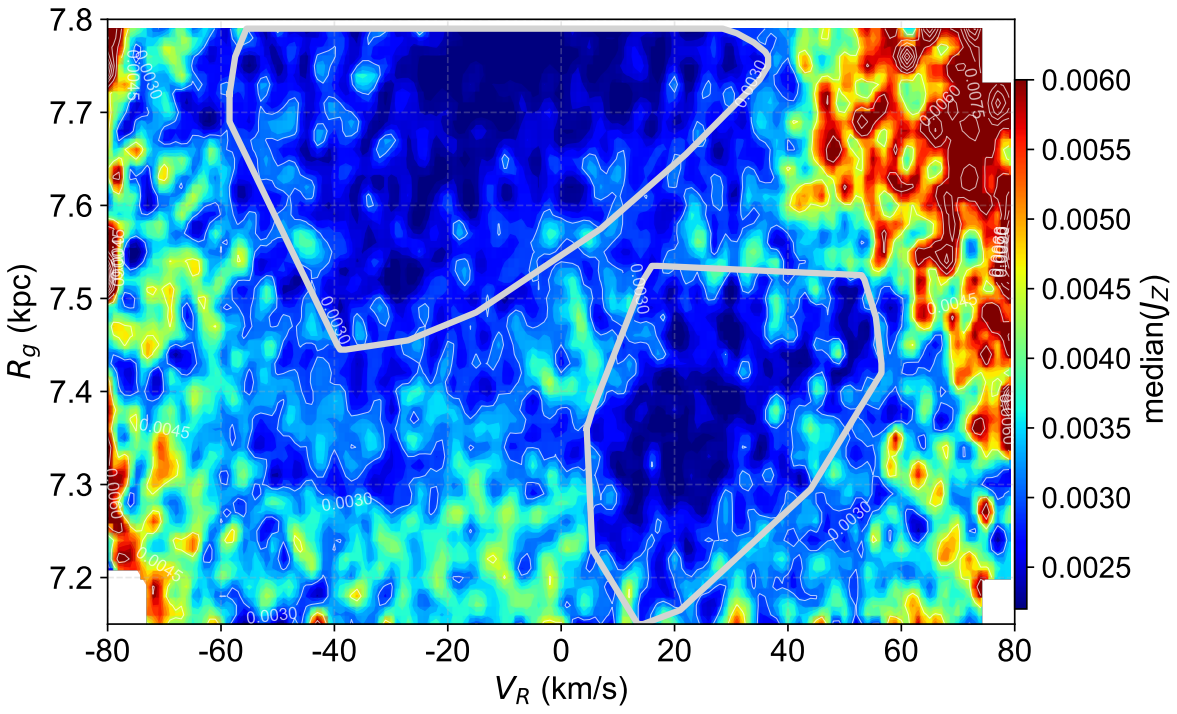

**Figure C5.** Distribution of median $J_Z$ in the ($V_R$, $R_g$) plane for stars with $R_g$ in the range 7.15–7.8 kpc and $V_R$ between −80 and 80 km s$^{-1}$, after Gaussian smoothing with $\sigma = 1$. White contours represent $J_Z$ isopleths (20 automatically determined levels), with selected contours labeled. The two thick light-gray lines outline the boundaries of connected regions where $J_Z < 0.0028$ kpc km s$^{-1}$.

In this work, all analyses of the moving groups are performed by selecting stars within the polygons shown in Figure C3.

## Appendix D Fitting the $V_R$ Phase Spiral with a Logarithmic Spiral in the $Z$–$V_Z$ Plane

In Section 2.7, we examined the $V_R$ distributions in the $Z$–$V_Z$ plane for stars in the moving groups embedded within the P3 and D2 $V_R$ stripes. Stars in the Coma Berenices and Hyades–Pleiades regions of the P3 stripe, as well as those in the Hyades–Pleiades expansion tail and Hercules 1 regions of the D2 stripe, display two-armed and single-armed spirals, respectively. In the $Z$–$V_Z$ plane, the median $V_R$ along these spirals is biased toward zero and positive values relative to the surrounding regions. We define a $V_R$ threshold for the phase-space spiral of each moving group and select bins above the threshold. We then fit an exponential spiral model to the selected bins. To ensure that both coordinates contribute comparably to the fit, we normalize the data by multiplying $Z$ by a factor of 50 before fitting. The model is expressed as

$$r = a \cdot e^{b(\theta - \theta_0)}, \tag{D1}$$

$$x = x_c + r\cos\theta, \quad y = y_c + r\sin\theta \tag{D2}$$

where ($x_c$, $y_c$) represents the spiral center, $a$ is the radius at $\theta = \theta_0$, and $b$ is the growth rate. The initial guesses for the parameters were chosen based on the following considerations.

1. *a (initial radius).* Chosen in the range 10–25, covering the plausible values suggested by the data distribution.
2. *b (growth rate).* Varied from 0.05 to 0.15, encompassing a wide range of spiral morphologies from tightly wound to loosely wound.
3. *$\theta_0$ (angular offset).* Varied between 0.5 and 2.0 rad, allowing for different starting phases of the spiral.
4. *($x_c$, $y_c$) (spiral center).* Allowed to vary within ±10, providing sufficient flexibility for the center to deviate from the origin.

Each set of initial guesses was independently optimized using the full iterative procedure of `curve_fit`, which searches for the optimal solution in the parameter space. The final model was selected as the one yielding the highest $RR^2$ (the coefficient of determination, which quantifies the goodness of fit for each spiral model) among all optimization results. For our two-dimensional data, $RR^2$ is defined as

$$RR^2 = 1 - \frac{SS_{\rm res}}{SS_{\rm tot}}, \tag{D3}$$

where $SS_{\rm res}$ is the residual sum of squares and $SS_{\rm tot}$ is the total sum of squares, defined as

$$SS_{\rm res} = \sum_{i=1}^{n}\left[(Z_i - \hat{Z}_i)^2 + (V_{Z,i} - \hat{V}_{Z,i})^2\right], \tag{D4}$$

$$SS_{\rm tot} = \sum_{i=1}^{n}\left[(Z_i - \bar{Z})^2 + (V_{Z,i} - \bar{V}_Z)^2\right]. \tag{D5}$$

In these expressions, ($Z_i$, $V_{Z,i}$) are the observed coordinates of the $i$th data point, ($\hat{Z}_i$, $\hat{V}_{Z,i}$) are the corresponding coordinates predicted by the spiral model, and ($\bar{Z}$, $\bar{V}_Z$) is the centroid (mean position) of all observed data points in the $Z$–$V_Z$ plane. The $RR^2$ value ranges from zero to one, with values closer to one indicating a better fit.

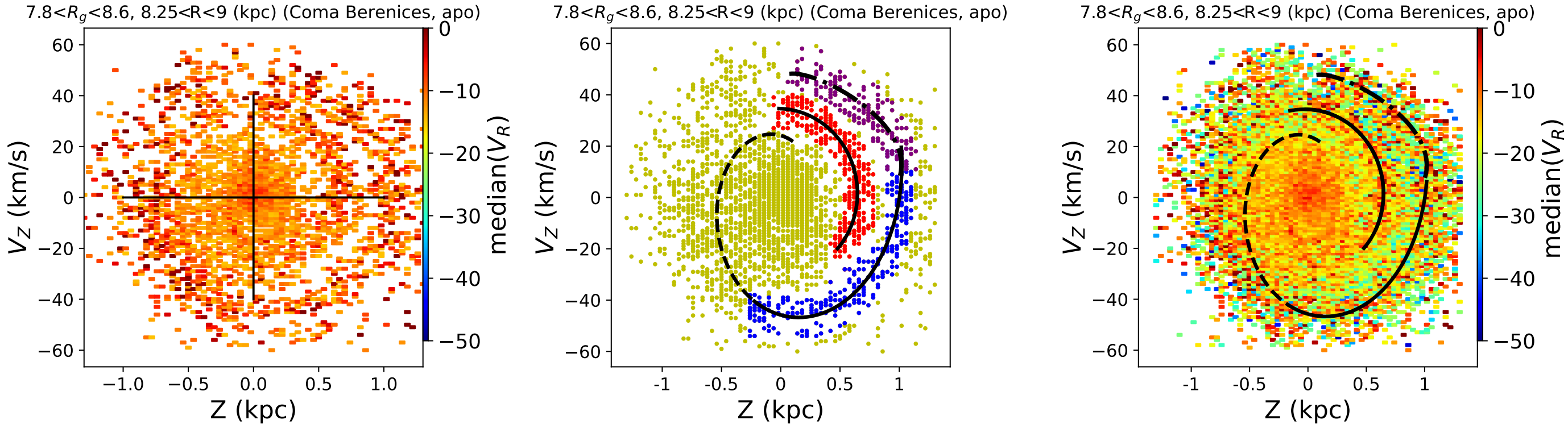

**Figure D1.** Selection and fitting of the spiral arms for the Coma Berenices moving group region in the P3 $V_R$ stripe (7.8 < $R_g$ < 8.6 kpc and 8.25 < $R$ < 9 kpc). Left: bins in the $Z$–$V_Z$ plane where the median $V_R$ exceeds a first threshold ($V_R > -15$ km s$^{-1}$). To reject isolated bins, we additionally require that the median $V_R$ in the four adjacent bins (above, below, left, and right) exceeds a second threshold ($V_R > -20$ km s$^{-1}$). This procedure identifies contiguous regions where $V_R$ is consistently biased toward positive values. Middle: bins selected for fitting. The red and blue bins trace the tails of the right and left arms, respectively; both are fitted with exponential spirals (solid black curves). The black dashed line extends the black curve and closely follows the envelope of the arm. The purple part of this arm forms a sharp turn with the blue part, so they cannot be fitted with a single exponential spiral. The purple bins are instead fitted with a polynomial (dotted–dashed line). Right: the resulting fits overlaid on the $V_R$ distribution in the $Z$–$V_Z$ plane for the same stars.

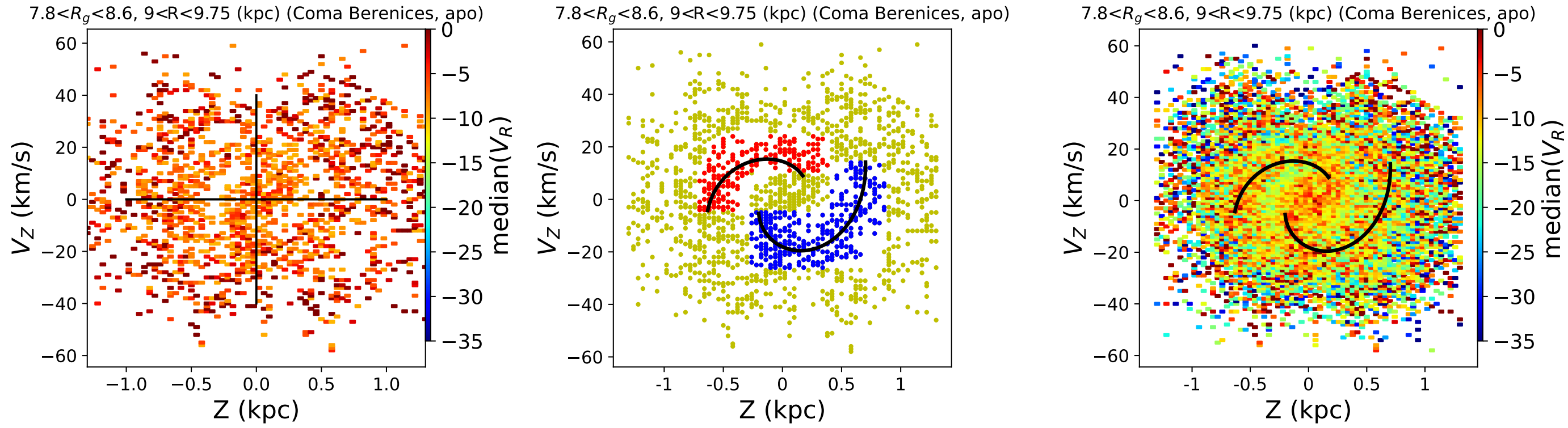

**Figure D2.** Selection and fitting of the spiral arms for the Coma Berenices moving group region in the P3 $V_R$ stripe ($7.8 < R_g < 8.6$ kpc and $9 < R < 9.75$ kpc). Left: bins in the $Z$–$V_Z$ plane where the median $V_R$ exceeds a first threshold ($V_R > -10$ km s$^{-1}$). To reject isolated bins, we additionally require that the median $V_R$ in the four adjacent bins (above, below, left, and right) exceeds a second threshold ($V_R > -15$ km s$^{-1}$), thus identifying contiguous regions with $V_R$ consistently biased toward positive values. Middle: bins selected for fitting; red and blue points mark the right and left arms, respectively, both fitted with exponential spirals (solid black curves). Right: the resulting fits overlaid on the $V_R$ distribution in the $Z$–$V_Z$ plane for the same stars.

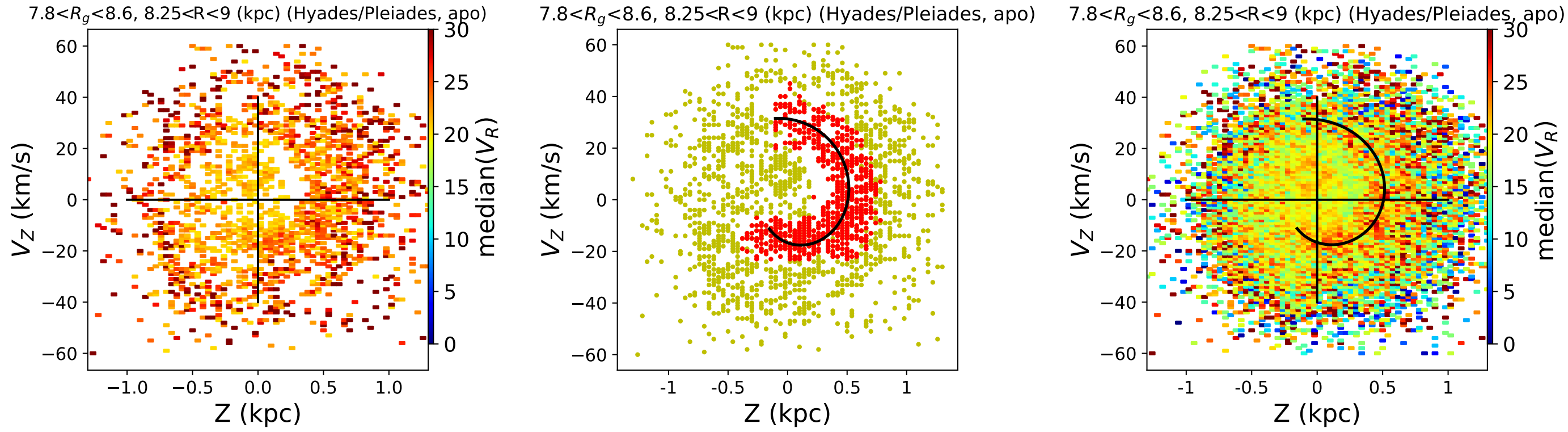

**Figure D3.** Selection and fitting of the spiral arm for the Hyades–Pleiades moving group region in the P3 $V_R$ stripe ($7.8 < R_g < 8.6$ kpc and $8.25 < R < 9$ kpc). Left: bins in the $Z$–$V_Z$ plane where the median $V_R$ exceeds a first threshold ($V_R > 20$ km s$^{-1}$). To reject isolated bins, we additionally require that the median $V_R$ in the four adjacent bins (above, below, left, and right) exceeds a second threshold ($V_R > 18$ km s$^{-1}$), thereby identifying contiguous regions with $V_R$ consistently biased toward positive values. Middle: bins selected for fitting; red points mark the single arm, fitted with an exponential spiral (solid black curve). Right: the resulting fit overlaid on the $V_R$ distribution in the $Z$–$V_Z$ plane for the same stars.

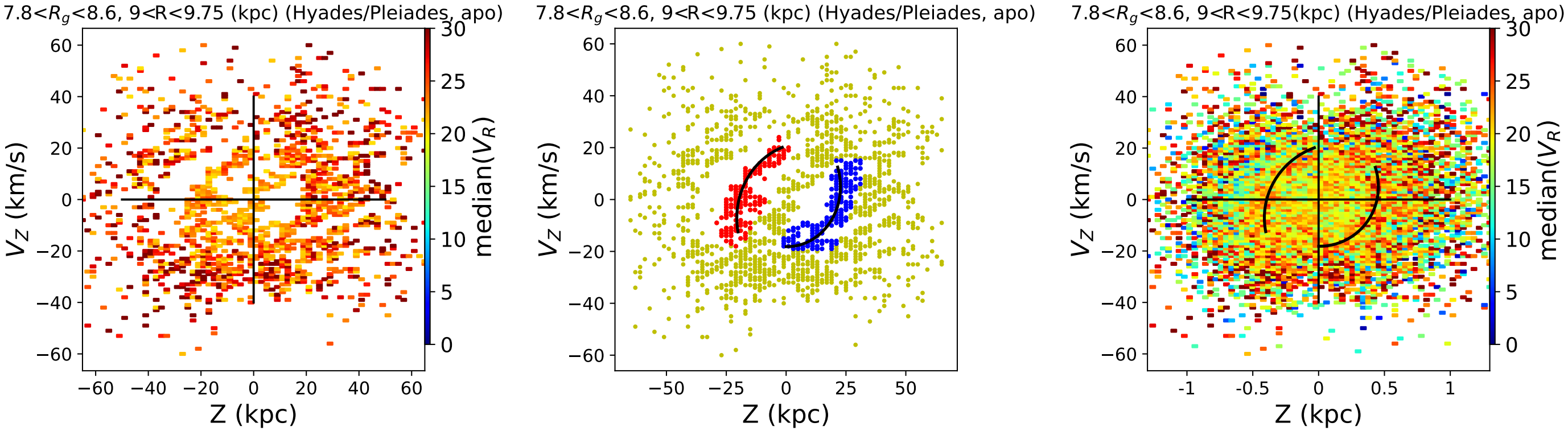

**Figure D4.** Selection and fitting of the spiral arms for the Hyades–Pleiades moving group region in the P3 $V_R$ stripe ($7.8 < R_g < 8.6$ kpc and $9 < R < 9.75$ kpc). Left: bins in the $Z$–$V_Z$ plane where the median $V_R$ exceeds a first threshold ($V_R > 20$ km s$^{-1}$). To reject isolated bins, we additionally require that the median $V_R$ in the four adjacent bins (above, below, left, and right) exceeds a second threshold ($V_R > 18$ km s$^{-1}$), thus selecting contiguous regions with $V_R$ consistently biased toward positive values. Middle: bins used for fitting; red and blue points mark the two arms, both fitted with exponential spirals (solid black curves). Right: the resulting fits overlaid on the $V_R$ distribution in the $Z$–$V_Z$ plane for the same stars.

Figures D1 and D2 present the exponential spiral fitting results for stars in the Coma Berenices moving group region of the P3 $V_R$ stripe, within the radial ranges $8.25 < R < 9$ kpc and $9 < R < 9.75$ kpc, respectively. The left panel of each figure displays the bins that exceed the $V_R$ threshold in the $Z$–$V_Z$ plane. The $V_R$ threshold is determined by comparing the $V_R$ values on and off the spiral arms and then through iterative adjustments, aiming to retain as many on-arm bins as possible while discarding off-arm bins. In the left panel, the selected bins clearly trace a two-armed spiral, with the tails of the two arms well separated. In the middle panel, we select the independent parts of the two arms (red and blue) and fit each with the logarithmic spiral model given by Equations (D1) and (D2) (black curves). The right panel

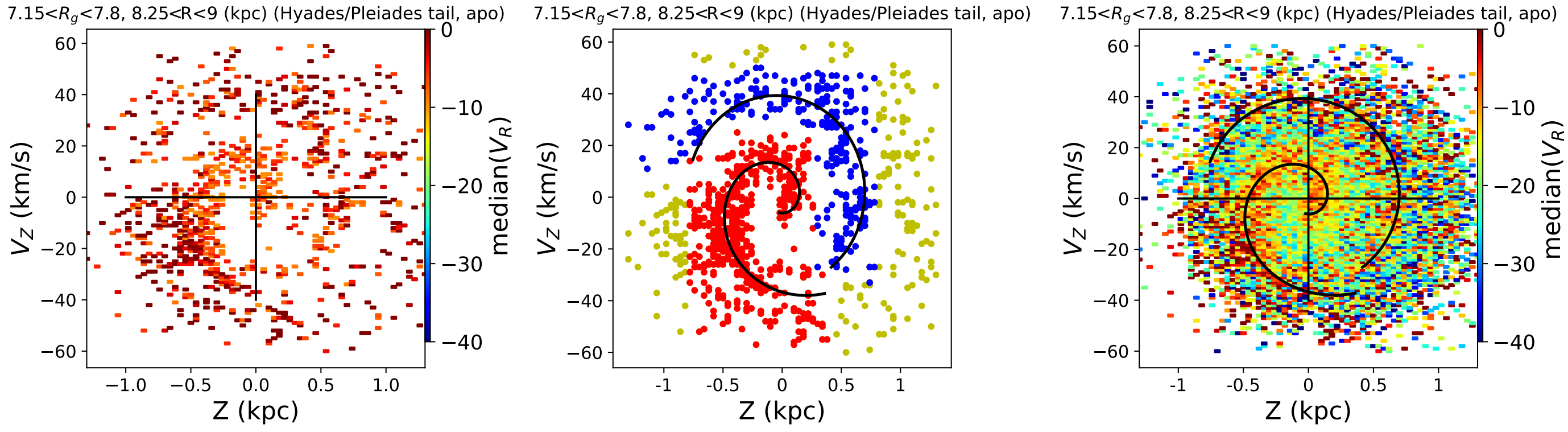

**Figure D5.** Selection and fitting of the spiral arms for the Hyades–Pleiades expansion tail region in the D2 $V_R$ stripe ($7.15 < R_g < 7.8$ kpc and $8.25 < R < 9$ kpc). Left: bins in the $Z$–$V_Z$ plane where the median $V_R$ exceeds a first threshold ($V_R > -10$ km s$^{-1}$). To reject isolated bins, we additionally require that the median $V_R$ in the four adjacent bins (above, below, left, and right) exceeds a second threshold ($V_R > -15$ km s$^{-1}$), thus selecting contiguous regions with $V_R$ consistently biased toward positive values. Middle: bins used for fitting; red and blue points mark the two arms, both fitted with exponential spirals (solid black curves). Right: the resulting fits overlaid on the $V_R$ distribution in the $Z$–$V_Z$ plane for the same stars.

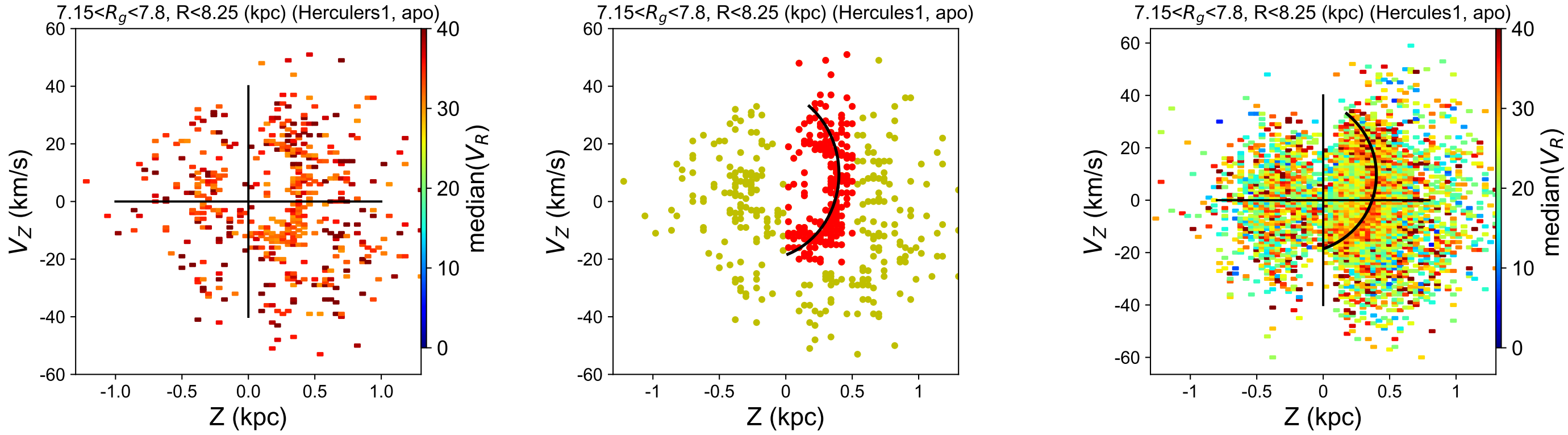

**Figure D6.** Selection and fitting of the short spiral arm for the Hercules 1 moving group region in the D2 $V_R$ stripe ($7.15 < R_g < 7.8$ kpc and $R < 8.25$ kpc). Left: bins in the $Z$–$V_Z$ plane where the median $V_R$ exceeds a first threshold ($V_R > 30$ km s$^{-1}$). To reject isolated bins, we additionally require that the median $V_R$ in the four adjacent bins (above, below, left, and right) exceeds a second threshold ($V_R > 25$ km s$^{-1}$), thus selecting contiguous regions with $V_R$ consistently biased toward positive values. Middle: bins selected for fitting; red points mark the single arm, fitted with an exponential spiral (solid black curve). Right: the resulting fit overlaid on the $V_R$ distribution in the $Z$–$V_Z$ plane for the same stars.

**Table D1**
Parameters of the Fitted Exponential Spirals for the $V_R$ Spiral Arms in the $Z$–$V_Z$ Plane

| Arm | $a$ | $b$ | $\theta_0$ (rad) | $x_c$ | $y_c$ | Pitch Angle (deg) | $RR^2$ |
|---|---|---|---|---|---|---|---|
| The red arm in the middle panel of Figure D1 | 30.704 | 0.0485 | –0.8875 | 0.2728 | 0.0154 | 2.78 | 0.95 |
| Arm of Coma Berenices, $8.25 < R < 9$ kpc | … | … | … | … | … | … | … |
| The blue arm in the middle panel of Figure D1 | 50.3456 | 0.2037 | 0.0531 | 1.0049 | 0.9887 | 11.51 | 0.97 |
| Arm of Coma Berenices, $8.25 < R < 9$ kpc | … | … | … | … | … | … | … |
| The red arm in the middle panel of Figure D2 | 13.4859 | 0.3484 | 1.191 | 1.1382 | 0.94 | 19.21 | 0.88 |
| Arm of Coma Berenices, $9 < R < 9.7$ kpc | … | … | … | … | … | … | … |
| The blue arm in the middle panel of Figure D2 | 12.7466 | 0.378 | –2.461 | 0.503 | –0.441 | 20.7 | 0.917 |
| Arm of Coma Berenices, $9 < R < 9.75$ kpc | … | … | … | … | … | … | … |
| The red arm in the middle panel of Figure D3 | 35.5738 | 0.1845 | 0.6098 | −7.7242 | −7.8941 | 10.45 | 0.7 |
| Arm of Hyades–Pleiades, $8.25 < R < 9$ kpc | … | … | … | … | … | … | … |
| The red arm in the middle panel of Figure D4 | 28.5946 | −0.0311 | 3.1416 | 8 | −8 | 1.78 | 0.85 |
| Arm of Hyades–Pleiades, $9 < R < 9.75$ kpc | … | … | … | … | … | … | … |
| The blue arm in the middle panel of Figure D4 | 28.9043 | 0.0903 | 2.0414 | −1.4286 | 2.7889 | 5.16 | 0.87 |
| Arm of Hyades–Pleiades, $9 < R < 9.75$ kpc | … | … | … | … | … | … | … |
| The red arm in the middle panel of Figure D5 | 7.44 | 0.3073 | 0.382 | −1.44 | −0.7 | 17.08 | 0.79 |
| Arm of Hyades–Pleiades tail, $8.25 < R < 9$ kpc | … | … | … | … | … | … | … |
| The blue arm in the middle panel of Figure D5 | 38.2483 | 0.005 | 5.504 | −2.216 | 1.8741 | 0.29 | 0.92 |
| Arm of Hyades–Pleiades tail, $8.25 < R < 9$ kpc | … | … | … | … | … | … | … |
| The red arm in the middle panel of Figure D6 | 29.8231 | 0.01 | −0.644 | −10 | 9.3452 | 0.57 | 0.95 |
| Arm of Hercules 1, $R < 8.25$ kpc | … | … | … | … | … | … | … |

shows the fitted curves overlaid on the original $V_R$ distribution.

Similarly, Figures D3 and D4 present the fitting results for the $V_R$ spiral in the $Z$–$V_Z$ plane for stars in the Hyades–Pleiades moving group region of the P3 $V_R$ stripe, within the radial ranges $8.25 < R < 9$ kpc and $9 < R < 9.75$ kpc, respectively.

Figure D5 shows the fitting results for the $V_R$ spiral in the $Z$–$V_Z$ plane for stars in the Hyades–Pleiades expansion tail region of the D2 $V_R$ stripe, within the radial range $8.25 < R < 9$ kpc. Figure D6 presents the fitting result for stars in the Hercules 1 moving group region within the radial range $R < 8.25$ kpc.

For the fits shown in Figures D1–D6, the parameters of the fitted exponential spirals and their pitch angles are listed in Table D1. The pitch angle is defined as the angle between the tangent to the spiral and the direction perpendicular to the radial vector.

## ORCID iDs

Yan Xu https://orcid.org/0000-0002-2459-3483
Chao Liu https://orcid.org/0000-0002-1802-6917
Cheng Dong Li https://orcid.org/0000-0001-9949-0625
Heidi Newberg https://orcid.org/0000-0001-8348-0983
Hao Tian https://orcid.org/0000-0003-3347-7596
Hua Jian Wang https://orcid.org/0009-0001-2509-5494
Xiao Dian Chen https://orcid.org/0000-0001-7084-0484
Li Cai Deng https://orcid.org/0000-0001-9073-9914